\documentstyle[12pt]{article}
\begin{document}
\renewcommand{\thefootnote}{\fnsymbol{footnote}}
\newcommand{\leg}{\;\raisebox{-0.4ex}
	{$\stackrel{\scriptstyle >}{\scriptstyle <}$}\;}
\thispagestyle{empty}
\vspace*{-1 cm}
\hspace*{\fill}  \mbox{WUE-ITP-95-024} \\
\vspace*{1 cm}
\begin{center}
{\large \bf Neutralinos and Higgs Bosons in
the Next-To-Minimal Supersymmetric Standard Model
\\ [3 ex] }
{\large F.~Franke\footnote{
email: fabian@physik.uni-wuerzburg.de},
H. Fraas
\\ [2 ex]
Institut f\"ur Theoretische Physik, Universit\"at W\"urzburg \\
D-97074 W\"urzburg, Germany}
\end{center}
\vfill

{\bf Abstract}

The purpose of this paper is to present a complete and consistent list of the
Feynman rules for the vertices of neutralinos and Higgs bosons in the
Next-To-Minimal Supersymmetric Standard Model (NMSSM), which does not yet
exist in the literature. The Feynman rules are derived from the
full expression for the Lagrangian and
the mass matrices of the neutralinos and Higgs bosons in the NMSSM.
Some crucial differences between the vertex functions of
the NMSSM and the
Minimal Supersymmetric Standard Model (MSSM) are discussed.
\vfill
\begin{center}
December 1995
\end{center}
\newpage
\setcounter{page}{1}
\section{Introduction}
With the observation of the top quark \cite{cdf} the particle content of
the electroweak Standard Model (SM) \cite{sm}
is completely experimentally
confirmed with the exception of the Higgs sector \cite{kibble}.
There merely exists
an upper bound for the Higgs mass $m_\Phi$
of about 1 TeV due to unitarity constraints \cite{unitarity}
and a lower bound of about 65 GeV from the so far
unsuccessful experimental Higgs search \cite{alephsm}.

However, in the SM there are some unsolved problems connected with this
minimal Higgs sector. First,
the coupling constants do not meet at one point at
high energies in the simplest Grand Unified Theory (GUT) extensions
of the SM \cite{amaldi}. Also experimental bounds for the proton
decay impose severe constraints on non-supersymmetric GUTs. Further,
the SM does not explain the small ratio
between the energy scale
of the electroweak symmetry breaking and the Planck scale
($m_W^2/m_P^2 \approx 10^{-34}$) \cite{hier} (hierarchy problem)
and does not answer the question
how the large radiative corrections
of the order of the GUT scale to the
Higgs mass are prevented \cite{fine}
(naturalness or fine tuning problem).

Supersymmetric models may provide solutions to these problems without
abandoning the idea of Higgs bosons as elementary particles.
The boson-fermion symmetry stabilizes the Higgs mass against
radiative corrections, electroweak symmetry breaking can be
triggered at the required scale and unification of the
coupling constants at a single point is possible.
The simplest supersymmetric model, the Minimal Supersymmetric
Standard Model (MSSM) \cite{haka},
is characterized by a minimal particle content,
explicit supersymmetry breaking by soft symmetry breaking terms and an
exact symmetry called $R$ parity, which guarantees conservation of baryon and
lepton number. In the MSSM two Higgs doublet fields $H_1$ and $H_2$ with vacuum
expectation values $v_1$ and $v_2$ ($\tan\beta=v_2/v_1$) are needed
in order to avoid anomalies and to give masses to both
up-type and down-type quarks. They lead to
five physical Higgs bosons,
two neutral scalar, one neutral pseudoscalar, and a pair of charged
Higgs particles.

The MSSM predicts for every particle of the SM
a partner with a spin differing by $1/2$,
called gaugino, higgsino, scalar lepton (slepton) and scalar quark
(squark), respectively.
Since "ordinary" and supersymmetric particles obviously have
different masses, supersymmetry must be broken in nature.
In most phenomenological models, this is
simulated by the introduction of explicit soft supersymmetry
breaking terms in the Lagrangian which split the masses
within a supermultiplet. Moreover,
the soft supersymmetry breaking terms
lead to the formation of new mass eigenstates in the supersymmetric sector.
Photinos, zinos and the neutral higgsinos form neutralinos as mass
eigenstates,
the mass eigenstates composed of winos and
charged higgsinos are the charginos, and also the left-handed and
right-handed scalar quarks and leptons mix to new eigenstates.

But the constraint
$\rho \equiv m_W^2/(m_Z^2\cos^2 \theta_W) \approx 1$ also allows
extended supersymmetric models with additional Higgs doublets, singlets or
even triplets. Such nonminimal supersymmetric models gain
more and more attraction since they can evade some of the
constraints of the MSSM and may lead to a variety of
new phenomena with interesting experimental signatures.

In this paper we focus on
the Next-To-Minimal Supersymmetric Standard Model (NMSSM) \cite{drees,ellis},
the minimal extension of the MSSM by a gauge singlet superfield.
The Higgs sector of the NMSSM contains
five physical neutral Higgs bosons, three Higgs scalars $S_a$
($a=1,2,3$) and two pseudoscalars $P_\alpha$ ($\alpha=1,2$), and two
degenerate physical charged Higgs particles $C^\pm$. The neutralino
sector is extended to five neutralinos instead of four in the MSSM,
with masses and eigenstates determined by a $5 \times 5$ mixing
matrix. The remaining particle content is identical with that of the MSSM.

The NMSSM was first developed within the framework of
GUTs and superstring theories \cite{barr,nilsredwy,derendinger}.
It is mainly motivated by its potential to eliminate the
$\mu$ problem of the MSSM \cite{kim}, where the
origin of the
the $\mu$ parameter in the superpotential
\begin{equation}
W_{\mbox{\scriptsize MSSM}}=\mu  H_1 H_2
\end{equation}
is not understood.
For phenomenological reasons it has to be of the order of the electroweak
scale,
while the "natural" mass scale would be
of the order of the GUT or
Planck scale. This problem is evaded in the NMSSM where the
$\mu$ term in the superpotential is dynamically generated through
$\mu = \lambda x$ with a dimensionless coupling $\lambda$ and
the vacuum expectation value $x$ of the Higgs singlet.

As another essential feature of the NMSSM the mass bounds for
the Higgs bosons and neutralinos are weakened.
While in the MSSM experimental data imply lower mass bounds of 18 GeV for the
lightest neutralino \cite{alephneu} (assuming either $\tan\beta > 2$ or
the gluino mass $m_{\tilde{g}} > 100$ GeV), 44 GeV for the
lightest scalar and 21 GeV for the lightest pseudoscalar Higgs boson
\cite{alephhiggs}
very light or massless neutralinos and Higgs bosons are not excluded
in the NMSSM \cite{frankeneu,frankehiggs}. Furthermore the upper
tree level mass bound for the lightest Higgs scalar of the MSSM
\begin{equation}
\label{tlb}
m_h^2 \leq m_Z^2 \cos^22\beta
\end{equation}
is increased to
\begin{equation}
m_{S_1}^2 \le m_Z^2 \cos^2 2\beta+
\lambda^2(v_1^2+v_2^2)
\sin^2 2\beta \, .
\end{equation}
Both bounds are raised by radiative corrections by about 30 GeV
\cite{rad,ellwanger,elliott,pandita}.
Taking into account the weak coupling
of a Higgs scalar of singlet type the NMSSM may still remain a
viable model when the MSSM can be ruled out due to eq.~(\ref{tlb}).

The above arguments make an intensive study of the
NMSSM very desirable. While the
implications for supersymmetric phenomenology have already been studied to
some extent \cite{elliott, ellrs, kimproc, kimneu}, one finds,
however, only incomplete lists of
Feynman rules for the NMSSM in the literature \cite{ellis,higgs,hunter}.
Furthermore, different sign conventions for the
parameters in the superpotential have been established
\cite{ellwanger,elliott,kimproc}. In this paper we
provide the full Lagrangian of the NMSSM and
present a complete list of all Feynman rules
for the neutral Higgs bosons and neutralinos which differ
from those of the MSSM. These differences between NMSSM and MSSM may
arise in two manners: The singlet component of a Higgs boson or
neutralino can explicitly appear in the vertex factor, or the
Feynman rules of NMSSM and MSSM are formally equal differing just by the
mixing of the Higgs bosons or neutralinos.

The production of Higgs bosons or neutralinos is among the
most promising processes suitable for the discovering of minimal or nonminimal
supersymmetric signatures.
Unfortunately, until now, no supersymmetric particles have been observed.
The experimental results at the high energy colliders can be
transformed into mass bounds for the supersymmetric particles and
excluded domains in the parameter space.
We use the derived vertex factors in order to review our previous
analysis of the
allowed parameter space \cite{frankeneu, frankehiggs} in the NMSSM exploring
the meanwhile improved experimental limits from LEP.
For LEP2 and a future linear collider,
neutralino and Higgs production in the NMSSM have been studied
in refs.~\cite{kimproc,
whitehiggs, franke4}.

Comparing the Higgs couplings of NMSSM and MSSM
we point out some fundamental differences between these
models. A crucial test for supersymmetry as well as for the
standard model would be the measurement of
the Higgs self-coupling, e.~g.~at double Higgs production at the NLC
\cite{ilyin, boudjema}. Further significant differences could arise
in Higgs production via gluon fusion or Higgs decays into photons, gluons,
neutralinos or charginos. We emphasize that we do not want to discuss
explicit supersymmetric processes but provide all necessary Feynman rules
for their computation. A detailed study of cross sections and
decay rates would exceed the intention of
this paper by far. This has been either done in other works or
remains as a future challenge.

The outline of this paper is as follows:
First we describe in Sec.~2 the complete Lagrangian
of the NMSSM including all terms for the self-interaction of the
gauge multiplets, the interaction of gauge and matter multiplets
as well as the self-interaction of the matter multiplets.
Explicit expressions for the scalar potential and the soft supersymmetry
breaking potential are given.
Since the additional singlet superfield of the NMSSM leads to extended
Higgs and neutralino sectors, we present the Higgs and neutralino
mixings and review also the chargino and slepton/squark mixings
in order to fix all conventions and to show the influence of all
parameters of the model.
The main part of this paper (Sec.~3) is dedicated to the
Feynman rules of the NMSSM
which are derived from the relevant parts of the Lagrangian.
In Secs.~4 and 5 we illustrate the differences between the
MSSM and NMSSM couplings. Sec.~4 contains a discussion of the
Higgs couplings to gauge bosons and
an analysis of the experimental
constraints on the parameter space and the Higgs and neutralino
masses due to the results at the high
energy colliders. In Sec.~5 we compare in detail the Higgs couplings to
quarks, scalar quarks, neutralinos and charginos and
the trilinear Higgs self-couplings of NMSSM and MSSM and indicate
the phenomenological implications for supersymmetric processes
to be studied in future works.

\section{The Lagrangian of the NMSSM}
The NMSSM is characterized by its superpotential
\begin{eqnarray}
W & = & \lambda \varepsilon_{ij} H_1^i H_2^j N -\frac{1}{3} k N^3
\nonumber \\ & &
+h_u \varepsilon_{ij} \tilde{Q}^i \tilde{U} H_2^j
-h_d \varepsilon_{ij} \tilde{Q}^i \tilde{D} H_1^j
-h_e \varepsilon_{ij} \tilde{L}^i \tilde{R} H_1^j
\end{eqnarray}
where $H_1 = (H_1^0,H^-)$ and $H_2=(H^+,H_2^0)$ are the $SU(2)$ Higgs
doublets with hypercharge $-1/2$ and $1/2$ and
vacuum expectation values $(v_1,0)$, $(0,v_2)$ ($\tan\beta=v_2/v_1$),
respectively, $N$ is the Higgs singlet with hypercharge
0 and vacuum expectation value
$x$, and $\varepsilon_{ij}$ is totally antisymmetric with
$\varepsilon_{12}=-\varepsilon_{21}=1$. The notation of
the squark/slepton doublets and singlets is conventional,
generation indices are understood.
Contrary to the MSSM, the superpotential of the NMSSM consists only of
trilinear terms with dimensionless couplings. Therefore all terms
of the unbroken theory are scale-invariant with
the supersymmetry breaking scale
$m_{\mbox{\scriptsize SUSY}}$
as the only mass scale,
while the MSSM contains as additional
mass scale the parameter $\mu$ which leads to the above
described $\mu$ problem.

The scalar potential
\begin{equation}
V=\frac{1}{2} (D^aD^a+{D'}^2) +F^{\ast}_i F_i,
\end{equation}
is composed of the $D$ and $F$ terms
\begin{eqnarray}
D^a & = & g A_i^{\ast} T^a_{ij} A_j\, ,  \\
D' & = & \frac{1}{2} g' y_i A_i^{\ast} A_i\, , \\
F_i & = & \partial W / \partial A_i\, ,
\end{eqnarray}
with the generators $(T^a_{ij})$ of the weak isospin group $SU(2)$ and the
hypercharge $y_i$.
The $A_i$ are the scalar fields of the theory.

We now give the explicit expressions of the $D$ and $F$ terms
as a function of doublet and singlet fields  and the couplings
$\lambda$, $k$, $h_u$, $h_d$ and $h_e$:
\begin{eqnarray}
\label{vd}
V_D & = & \frac{1}{2} (D^aD^a+{D'}^2)
\nonumber \\ & = &
\frac{1}{8}g^2 \Big[ (H_1^{i\ast} H_1^i)^2 + (H_2^{i\ast} H_2^i )^2
+ (\tilde{Q}^{i\ast} \tilde{Q}^i)^2+
(\tilde{L}^{i\ast} \tilde{L}^i)^2
\nonumber \\ & & \; \; \; \;
+4|H_1^{i\ast}H_2^i|^2 - 2(H_1^{i\ast}H_1^i)(H_2^{j\ast}H_2^j)
+4|H_1^{i\ast}\tilde{Q}^i|^2 - 2(H_1^{i\ast}H_1^i)
(\tilde{Q}^{j\ast}\tilde{Q}^j)
\nonumber \\ & & \;\;\;\;
+4|H_1^{i\ast}\tilde{L}^i|^2 - 2(H_1^{i\ast}H_1^i)
(\tilde{L}^{j\ast}\tilde{L}^j)
+4|H_2^{i\ast}\tilde{Q}^i|^2 - 2(H_2^{i\ast}H_2^i)
(\tilde{Q}^{j\ast}\tilde{Q}^j)
\nonumber \\ & & \;\;\;\;
+4|H_2^{i\ast}\tilde{L}^i|^2 - 2(H_2^{i\ast}H_2^i)
(\tilde{L}^{j\ast}\tilde{L}^j)
+4|\tilde{Q}^{i\ast}\tilde{L}^i|^2 - 2(\tilde{Q}^{i\ast}
\tilde{Q}^i)(\tilde{L}^{j\ast}\tilde{L}^j) \Big]
\nonumber \\ & &
+\frac{1}{8}{g'}^2 \Big[ H_2^{2\ast}H_2^2+H_2^{1\ast}H_2^1
-H_1^{1\ast}H_1^1-H_1^{2\ast}H_1^2
\nonumber \\ & & \;\;\;\;
+y_Q(\tilde{Q}^{1\ast}\tilde{Q}^1+\tilde{Q}^{2\ast}\tilde{Q}^2)
+y_u\tilde{U}^{\ast}\tilde{U}
+y_d\tilde{D}^{\ast}\tilde{D}
\nonumber \\ & & \;\;\;\;
-\tilde{L}^{1\ast}\tilde{L}^1-\tilde{L}^{2\ast}\tilde{L}^2
+2\tilde{R}^{\ast}\tilde{R}\Big] ^2,
\\  & & \mbox{} \nonumber \\
V_F & = & F_i^{\ast} F_i
\nonumber \\ & = &
|\lambda \varepsilon_{ij} H_1^i H_2^j  -kN^2|^2
\nonumber \\ & &
+|\lambda H_2^2 N +h_d\tilde{Q}^2\tilde{D}+
h_e\tilde{L}^2\tilde{R}|^2
+|\lambda H_2^1 N +h_d\tilde{Q}^1\tilde{D}+
h_e\tilde{L}^1\tilde{R}|^2
\nonumber \\ & &
+|\lambda H_1^2N+h_u\tilde{Q}^2\tilde{U}|^2
+|\lambda H_1^1N+h_u\tilde{Q}^1\tilde{U}|^2
\nonumber \\ & &
+|h_uH_2^2\tilde{U}-h_d H_1^2 \tilde{D}|^2
+|-h_u H_2^1 \tilde{U} + h_d H_1^1 \tilde{D}|^2
\nonumber \\ & &
+|h_u \varepsilon_{ij} \tilde{Q}^i H_2^j|^2
+|h_d \varepsilon_{ij} \tilde{Q}^i H_1^j|^2
+|h_e \varepsilon_{ij} \tilde{L}^i H_1^j|^2
\nonumber \\ & &
+|h_e H_1^2 \tilde{R}|^2
+|h_e H_1^1 \tilde{R}|^2.
\label{vf}
\end{eqnarray}
The Yukawa interactions and fermion mass terms arise by the following
part of the Lagrangian:
\begin{equation}
{\cal L}_{\mbox{\scriptsize Yukawa}}=
-\frac{1}{2} [(\partial^2 W/\partial A_i \partial A_j)
\psi_i \psi_j + \mbox{h.c.} ] \; ,
\end{equation}
where the two-component spinors $\psi_i$ are the supersymmetric
partners of the scalar fields $A_i$.

Since supersymmetric particles have not been found in the low-energy
particle spectrum there must exist a mechanism (if SUSY is realized at all)
that breaks SUSY
and splits the masses of the different members
of a supermultiplet. In the NMSSM as well as in the MSSM one simulates
the supersymmetry breaking by adding explicit soft supersymmetry
breaking terms to the Lagrangian.
The most general supersymmetry breaking potential \cite{gira} can be
written as
\begin{eqnarray}
\label{vs}
V_{\mbox{soft}} & = &
m_1^2 |H_1|^2 + m_2^2 |H_2|^2+m_3^2 |N|^2
\nonumber \\ & &
+m_Q^2 |\tilde{Q}|^2 + m_U^2 |\tilde{U}|^2 + m_D^2 |
\tilde{D}|^2
\nonumber \\ & &
+m_L^2 |\tilde{L}|^2 + m_E^2 |\tilde{R}|^2
\nonumber \\  & &
-(\lambda A_\lambda \varepsilon_{ij} H_1^i H_2^j N + \mbox{h.c.})
-(\frac{1}{3}kA_k N^3 + \mbox{h.c.})
\nonumber \\  & &
+(h_u A_U \varepsilon_{ij} \tilde{Q}^i \tilde{U} H_2^j
-h_d A_D \varepsilon_{ij} \tilde{Q}^i \tilde{D} H_1^j
-h_e A_E \varepsilon_{ij} \tilde{L}^i \tilde{R} H_1^j
+\mbox{h.c.})
\nonumber \\ & &
+\frac{1}{2}M \lambda^a \lambda^a
+\frac{1}{2}M' \lambda ' \lambda '.
\end{eqnarray}
The scalar potential as well as the soft supersymmetry breaking potential
appear with negative signs in the Lagrangian of the NMSSM.

Finally, we give
for completeness the parts of the Lagrangian responsible
for the self-interaction of the gauge multiplet and for the interaction
of gauge and matter multiplets.
The gauge field interacts with itself and the gauginos via
\begin{eqnarray}
{\cal L}_{V} & = & -\frac{1}{4} V^a_{\mu\nu} V^{a\mu\nu}
-\frac{1}{4}(\partial_\mu V'_\nu-\partial_\nu V'_\mu)^2
\nonumber \\ & &
-i\bar{\lambda}^a(\delta_{ab}\sigma^\mu\partial_\mu-
f_{abc}\sigma^\mu V_\mu^c) \lambda^b,
\label{vym}
\end{eqnarray}
where
\begin{equation}
\label{vmunu}
V^a_{\mu\nu}=\partial_\mu V^a_{\nu} -\partial_\nu V^a_{\mu}
+gf_{abc}V^a_{\mu}V^a_{\nu}
\end{equation}
and $f_{abc}$ are the structure constants of the nonabelian gauge group
$SU(2)$. The Pauli matrices are denoted by $\sigma^\mu=(1,\vec{\sigma})$.
The interaction between gauge and matter multiplets is described by
\begin{eqnarray}
{\cal L}_M & = &
(-gT^a_{ij}V_\mu^a-
\frac{1}{2}g'y_i\delta_{ij}V'_\mu)
(\bar{\psi}_i\bar{\sigma}^\mu
\psi_j+iA_i^\ast \stackrel{\leftrightarrow}{\partial}\!{}^\mu A_j)
\nonumber \\ & &
+ig\sqrt{2}T^a_{ij}(\lambda^a\psi_jA_i^\ast-\bar{\lambda}^a
\bar{\psi}_i A_j)
+\frac{ig'}{\sqrt{2}}y_i(\lambda'\psi_iA_i^\ast-\bar{\lambda}'
\bar{\psi}_i A_i)
\nonumber \\ & &
+A_i^\ast A_j(gT^a_{ik}V_\mu^a+\frac{1}{2}g'y_i\delta_{ik} V'_\mu)
(gT^b_{kj}V^{\mu b}+\frac{1}{2}g'y_j\delta_{kj} {V'}^\mu)
\label{lf}
\end{eqnarray}
Since the additional Higgs singlet field
has hypercharge 0, it does not interact with gauge and matter fields.
Therefore, this part of the Lagrangian is unchanged compared to the MSSM.
In eqs.~(\ref{vym}) -- (\ref{lf}), $V_\mu^a$ and
$V'_\mu$ denote the $SU(2)$ and $U(1)$ gauge fields of the model,
respectively, $\lambda^a$ and $\lambda '$ are their gaugino fermionic partners.

Now the interaction Lagrangian of the NMSSM is complete.
As free parameters appear
the ratio of the doublet vacuum expectation values,
$\tan\beta$, the singlet vacuum expectation value $x$,
the couplings in the superpotential $\lambda$ and $k$, the parameters
$A_\lambda$, $A_k$, as well as $A_U$, $A_D$, $A_E$ (for all three generations)
in the supersymmetry breaking potential, the gaugino mass
parameters $M$ and $M'$, and the scalar mass parameters
for the Higgs bosons $m_{1,2,3}$, squarks $m_{Q,U,D}$ and sleptons
$m_{L,E}$.
Note that the sign convention for the gaugino mass parameters in the NMSSM is
normally chosen to be opposite to that of the MSSM in order to recover
the minimal model with $\mu=\lambda x$ in the limit $x \rightarrow
\infty$, with $\lambda x , kx$ fixed.

This low-energy Lagrangian of the NMSSM obviously contains more free
parameters than the MSSM, the couplings $\lambda$ and $k$ as well
as the parameters $A_\lambda$ and $A_k$ are absent in the minimal model.
Within the framework of
supergravity theories, however, the total number of parameters
remains the same in both models. At the GUT scale, the couplings
$\lambda$ and $k$ have to be fixed in the NMSSM instead of $\mu$ and
$B$ (the coefficient of the quadratic Higgs term in the soft supersymmetry
breaking potential) in the MSSM. Then all other parameters at the
electroweak scale follow by renormalization group equations \cite{renor}.
The mass spectrum of such a constrained NMSSM has been studied
in refs.~\cite{ellwanger, ekwuni, kwconstr}.

Even without imposing unification constraints there exist some
restrictions for the low-energy parameters: Explicit CP violation
in the scalar sector is avoided by choosing the parameters
$\lambda$, $A_\lambda$, $k$ and $A_k$ to be real. Further, a
sufficient condition for the vacuum to conserve CP is to assume
$\lambda$ and $k$ to be positive which allows a choice of vacuum with
$v_1,v_2,x >0$ \cite{ellis}.

In the next step
we will study the mixing in the Higgs, neutralino/chargino and
slepton/squark sector before deriving the Feynman rules.
\subsection{The Higgs sector}
The $10 \times 10$ Higgs mass squared matrix decouples
in two $3\times 3$ blocks for the CP-even scalar and CP-odd pseudoscalar
Higgs bosons, respectively, and two $2\times 2$ blocks for the charged
Higgs particles. One eigenvalue of the CP-odd and charged Higgs
matrices vanishes and corresponds to an unphysical Goldstone mode.
The minimization conditions for the scalar potential
$\partial V / \partial v_{1,2}
=0$, $\partial V / \partial x =0$ eliminate
three parameters of the Higgs sector
which are normally chosen to be $m_1^2$, $m_2^2$ and
$m_3^2$.
Then at tree level the elements of the symmetric CP-even mass squared
matrix ${\cal M}_S^2=({M_{ij}^S}^2)$ become in the basis
$(H_1,H_2,N)$
\begin{eqnarray}
{M_{11}^S}^2 & = & \frac{1}{2} v_1^2 ({g'}^2+g^2) +
\lambda x\tan\beta (A_\lambda + kx), \\
{M_{12}^S}^2 & = & -\lambda x (A_\lambda + kx)
+v_1v_2(2\lambda^2-\frac{1}{2} {g'}^2 - \frac{1}{2} g^2 )\\
{M_{13}^S}^2 & = & 2 \lambda ^2 v_1 x -2\lambda kxv_2 -
\lambda A_\lambda v_2, \\
{M_{22}^S}^2 & = & \frac{1}{2} v_2^2 ({g'}^2+g^2) +
\lambda x\cot\beta (A_\lambda + kx), \\
{M_{23}^S}^2 & = & 2 \lambda ^2 v_2 x -2\lambda kxv_1 -
\lambda A_\lambda v_1, \\
{M_{33}^S}^2 & = & 4k^2x^2-kA_kx+\frac{\lambda A_\lambda v_1 v_2}{x}.
\end{eqnarray}
In the same way one finds for the elements of
the CP-odd matrix ${\cal M}_P^2$
\begin{eqnarray}
{M_{11}^P}^2 & = &
\lambda x(A_\lambda + kx) \tan\beta ,\\
{M_{12}^P}^2 & = &
\lambda x(A_\lambda + kx)  ,\\
{M_{13}^P}^2 & = &  \lambda v_2 (A_\lambda - 2kx), \\
{M_{22}^P}^2 & = &
\lambda x(A_\lambda + kx) \cot\beta ,\\
{M_{23}^P}^2 & = &  \lambda v_1 (A_\lambda - 2kx), \\
{M_{33}^P}^2 & = & \lambda A_\lambda \frac{v_1v_2}{x} + 4\lambda k v_1 v_2
+3kA_kx ,
\end{eqnarray}
and for the charged Higgs matrix one obtains
\begin{equation}
{\cal M}_c^2 = \left( \lambda A_\lambda x + \lambda k x^2 - v_1 v_2
\left( \lambda -\frac{g^2}{2} \right) \right)
\left( \begin{array}{cc} \tan\beta & 1 \\ 1 & \cot\beta
\end{array} \right).
\end{equation}
All Higgs mass matrices obtain radiative corrections from loops
of the heavy quarks, scalar quarks, Higgs particles, higgsinos,
gauge bosons and gauginos. Appropriate formulae can be found in
refs.~\cite{ellwanger,elliott,pandita}.

Assuming CP conservation in the Higgs sector, the Higgs matrices
are diagonalized by the real orthogonal $3\times 3$ matrices $U^S$ and $U^P$,
respectively,
\begin{eqnarray}
\mbox{Diag}(m_{S_1}^2,m_{S_2}^2,m_{S_3}^2) & = & {U^S}^T {\cal M}_S^2
U^S, \\
\mbox{Diag}(m_{P_1}^2,m_{P_2}^2,0) & = & {U^P}^T {\cal M}_P^2 U^P,
\end{eqnarray}
where $m_{S_1} < m_{S_2} < m_{S_3}$ and $m_{P_1}<m_{P_2}$ denote the
Higgs masses in ascending order. The mass eigenstates
$S_a$ $(a=1,2,3)$ of the neutral scalar
Higgs bosons, $P_\alpha$ $(\alpha = 1,2)$ of the physical neutral
pseudoscalar Higgs particles and $C^\pm$ of the physical charged Higgs
boson are obtained by the transformations
\begin{eqnarray}
\label{shiggs}
\left( \begin{array}{c} S_1 \\ S_2 \\ S_3 \end{array} \right) & = &
\sqrt{2} {U^S}  \left[
\left( \begin{array}{c} \mbox{Re} H_1^0 \\
\mbox{Re} H_2^0 \\ \mbox{Re} N \end{array} \right) -
\left( \begin{array}{c} v_1 \\ v_2 \\ x \end{array} \right) \right] \; ,
\\ & & \nonumber \\
\label{phiggs}
\left( \begin{array}{c} P_1 \\ P_2  \\ P_G \end{array} \right) & = &
\sqrt{2}
U^P
\left( \begin{array}{c} \mbox{Im} H_1^0 \\
\mbox{Im} H_2^0 \\ \mbox{Im} N \end{array} \right) \; , \\
\label{chiggs}
C^+ & = & \cos\beta H_2^1 + \sin\beta H_1^{2 \ast} \; .
\end{eqnarray}
Note that only the upper $ 3 \times 2$ matrix of
$U^P$ is physically relevant since the eigenstate $P_G$ of the CP-odd
matrix corresponds to an unphysical Goldstone mode.
Since the diagonalization matrices are normally to be found numerically,
we do not bother about analytical results.
For the rest of this paper, indices of the scalar Higgs bosons are
denoted by latin characters which can take values from 1
to 3 while
greek letters with possible values
1 or 2 are used for the pseudoscalar Higgs bosons.

The phenomenology of the NMSSM
Higgs sector has been studied in ref.~\cite{elliott}.
At tree level, the masses and mixings of the Higgs bosons
depend on six parameters: the couplings in the superpotential
$\lambda$ and $k$, the ratio of the vacuum expectation values
of the Higgs doublets
$\tan\beta$, the vacuum expectation value $x$
of the singlet and the parameters in the
supersymmetry breaking potential $A_\lambda$ and $A_k$.
In addition, the radiative corrections are influenced by the
squark masses and $A$-terms in the
supersymmetry breaking potential as well as by the parameters of the
gaugino/higgsino sector.

For comparison with the MSSM
we also quote the corresponding results for the Higgs sector of
the minimal supersymmetric model. In the MSSM there exist two neutral
scalar Higgs bosons $h$ and $H$ ($m_h < m_H$), one physical neutral
pseudoscalar
Higgs particle $A$ and a pair of degenerate charged Higgs bosons.
Their masses and mixings are conventionally parameterized in terms
of $m_A$ and $\tan\beta$. Then the
tree level mass squared matrix of the Higgs scalars can be written as
\begin{equation}
{\cal M}^2 = \left( \begin{array}{cc} m_Z^2 \cos^2\beta+m_A^2\sin^2\beta &
-(m_Z^2+m_A^2) \sin\beta\cos\beta \\ -(m_Z^2+m_A^2) \sin\beta\cos\beta &
m_Z^2 \sin2\beta+m_A^2\cos^2\beta \end{array} \right).
\end{equation}
The tree level masses are
\begin{equation}
m^2_{h,H}=\frac{1}{2} \left( (m_Z^2+m_A^2) \mp \sqrt{(m_Z^2+m_A^2)^2-
4m_A^2m_Z^2\cos^22\beta} \right),
\end{equation}
they correspond to the physical Higgs mass eigenstates
\begin{eqnarray}
\label{shiggsmssm}
\left( \begin{array}{c} h \\ H  \end{array} \right) & = &
\sqrt{2} \left( \begin{array}{cc} -\sin\alpha & \cos\alpha \\
\cos\alpha & \sin\alpha \end{array} \right) \left[
\left( \begin{array}{c} \mbox{Re} H_1^0 \\
\mbox{Re} H_2^0 \end{array} \right) -
\left( \begin{array}{c} v_1 \\ v_2  \end{array} \right) \right] \; ,
\\ & & \nonumber \\
A & = &
\sqrt{2}
\left( \begin{array}{c} \sin\beta \\ \cos\beta \end{array} \right)
\left( \begin{array}{c} \mbox{Im} H_1^0 \\
\mbox{Im} H_2^0  \end{array} \right) \; .
\label{phiggsmssm}
\end{eqnarray}
The mixing angle $\alpha$ is determined by
\begin{equation}
\tan 2\alpha = \tan 2\beta \frac{m_A^2+m_Z^2}{m_A^2-m_Z^2}.
\end{equation}

\subsection{The neutralino sector}
Although the neutralino sector does not depend on the parameters
$A_\lambda$ and $A_k$ of the supersymmetry breaking potential,
neutralino and Higgs sector of the NMSSM are nevertheless
strongly correlated contrary  to the MSSM.
With fixed parameters of the Higgs sector
the masses and mixings of the neutralinos are determined by the two
further parameters $M$ and $M'$ of the Lagrangian
\begin{eqnarray}
{\cal L}_{m_\chi^0} & = & \frac{1}{\sqrt{2}}ig\lambda^3(v_1\psi_{H_1}^1-
v_2\psi_{H_2}^2)-\frac{1}{\sqrt{2}}ig'\lambda '
(v_1 \psi_{H_1}^1-v_2 \psi_{H_2}^2) \nonumber \\
& & - \frac{1}{2} M \lambda^3 \lambda^3
-\frac{1}{2} M' \lambda' \lambda' \nonumber \\
& & -\lambda x \psi_{H_1}^1 \psi_{H_2}^2
-\lambda v_1 \psi_{H_2}^2 \psi_N
-\lambda v_2 \psi_{H_1}^1 \psi_N
+kx\psi_N^2 \nonumber \\
& & \mbox{} \mbox{} + \mbox{h.c.} \;\;\; .
\end{eqnarray}
In the basis
\begin{equation}
(\psi^0)^T=(-i\lambda_{\gamma},-i\lambda_Z,\psi_H^a,\psi_H^b,
\psi_N) ,
\end{equation}
with
\begin{eqnarray}
\psi_H^a & = &
\psi_{H_1}^1 \cos \beta - \psi_{H_2}^2 \sin\beta ,
\nonumber \\
\psi_H^b & = &
\psi_{H_1}^1 \sin\beta + \psi_{H_2}^2 \cos\beta ,
\end{eqnarray}
the mass term of the Lagrangian reads
\begin{equation}
{\cal L}_{m_{\chi^0}} = -\frac{1}{2} (\psi^0)^T Y \psi^0 + \mbox{h.c.} \;\; .
\end{equation}
The symmetric neutralino mixing matrix
\begin{equation}
Y= \left(
\begin{array}{ccccc}
-M s^2_W-M' c^2_W &
(M'-M) s_W c_W &
0 & 0 & 0 \\
(M'-M) s_W c_W &
-M c^2_W-M' s^2_W &
m_Z & 0 & 0 \\
0 & m_Z &
-\lambda x \sin2\beta &
\lambda x \cos2\beta &
0 \\
0 & 0 &
\lambda x \cos2\beta &
\lambda x \sin2\beta &
\lambda v \\
0 & 0 & 0 &
\lambda v &
-2kx
\end{array}
\right)
\end{equation}
can be diagonalized by a unitary $5 \times 5$ matrix $N$
\begin{equation}
m_{\chi^0_i}\delta_{ij} = N^\ast_{im}Y_{mn}N_{jn} \; ,
\end{equation}
with real and positive mass eigenvalues $m_{\chi^0_i}$.
If one tolerates negative eigenvalues, the diagonalization matrix
can be chosen to be real. Then
the absolute values of $m_{\chi^0_i} \leg 0$ are the physical
neutralino masses.
The upper $4\times 4$ matrix of $Y$ represents the neutralino mixing
matrix of the MSSM with $\mu=\lambda x$. The negative sign of
the parameters $M$ and $M'$ and $\mu$ in $Y$ opposite to the
convention in ref.~\cite{bartlneu} leaves all neutralino physics
unchanged.

As in the MSSM, from the two-component mass eigenstates
\begin{equation}
\chi_i^0 = N_{ij} \psi^0_j , \hspace*{1cm}
(i,j = 1,\ldots , 5),
\end{equation}
the four-component
Majorana spinors are formed by
\begin{equation}
\tilde{\chi}^0_i = \left( \begin{array}{c}
\chi_i^0 \\ \bar{\chi}_i^0
\end{array} \right) , \hspace{1cm}
(i=1,\dots , 5) .
\end{equation}
The Feynman rules in Sec.~\ref{sec:hnn}
are derived using these four-component spinors.
\subsection{The chargino sector}
Although the NMSSM does not extend the chargino sector of the MSSM,
we include the most important definitions in order to fix our
conventions.
With the basis
\begin{equation}
\psi^+ = \left( -i\lambda^+, \psi^1_{H_2} \right), \; \; \;
\psi^- = \left( -i\lambda^-, \psi^2_{H_1} \right),
\end{equation}
the chargino mass term in the Lagrangian is
\begin{equation}
{\cal L}_{m_{\chi^\pm}} = -\frac{1}{2} (\psi^+,\psi^-)
\left( \begin{array}{cc} 0 & X^T \\ X & 0 \end{array} \right)
\left( \begin{array}{c} \psi^+ \\ \psi^- \end{array} \right)  .
\end{equation}
The chargino mass matrix
\begin{equation}
\label{charmatrix}
X=\left( \begin{array}{cc} -M & \sqrt{2}m_W\sin\beta \\
\sqrt{2}m_W \cos\beta & -\lambda x \end{array} \right)
\end{equation}
can be diagonalized by two unitary matrices
$U$ and $V$, so that the mass eigenvalues $m_{\chi^\pm_i}$ become
\begin{equation}
m_{\chi^\pm_i} \delta_{ij} = U^\ast_{im} X_{mn} V_{jn} .
\end{equation}
{}From the two-component mass eigenstates
\begin{equation}
\tilde{\chi}_i^+ = V_{ij} \psi_j^+ , \; \; \;
\tilde{\chi}_i^- = U_{ij} \psi_j^- ,
\end{equation}
one constructs the four-component chargino fields
\begin{equation}
\label{charmass}
\tilde{\chi}_i^+=\left( \begin{array}{c} \chi^+_i \\
\bar{\chi}_i^- \end{array} \right), \hspace{0.5cm} i=1,2,
\end{equation}
which will be used in our Feynman rules.
Analytic expressions for the chargino masses and mixings are given in
ref.~\cite{bartlchar}. Note that we again use a different sign convention
in eq.~(\ref{charmatrix}),
which, however, does not alter the physical results.
\subsection{The slepton/squark sector}
\label{squark}
Likewise, the mixings between the left and right-handed sleptons and
squarks remain
unchanged compared to the MSSM. The mass term of the Lagrangian
contained in the $D$ and $F$ terms of the scalar potential and
in the soft supersymmetry breaking potential reads
\begin{equation}
{\cal L} _{m_{\tilde{u}}} = -\left( \tilde{u}_L^\ast,\tilde{u}_R^\ast  \right)
\left( \begin{array}{cc} a & b \\ b & c \end{array} \right)
\left( \begin{array}{c} \tilde{u}_L \\ \tilde{u}_R \end{array}
\right)
\end{equation}
with
\begin{eqnarray}
a & = & m_Q^2+m_u^2+m_Z^2 \cos 2\beta \left( \frac{1}{2}
-e_u \sin^2\theta_W \right), \\
b & = & m_u \left( \lambda x \cot\beta +A_U\right), \\
c & = & m_U^2+m_u^2+e_um_Z^2\cos 2\beta\sin^2\theta_W,
\end{eqnarray}
for scalar up-type quarks (or sleptons with isospin $+1/2$),
and
\begin{equation}
{\cal L}_ {m_{\tilde{d}}} = -\left( \tilde{d}_L^\ast ,\tilde{d}_R^\ast \right)
\left( \begin{array}{cc} a' & b' \\ b' & c' \end{array} \right)
\left( \begin{array}{c} \tilde{d}_L \\ \tilde{d}_R \end{array}
\right)
\end{equation}
with
\begin{eqnarray}
a' & = & m_Q^2+m_d^2-m_Z^2 \cos 2\beta \left( \frac{1}{2}
+e_d \sin^2\theta_W \right), \\
b' & = & m_d \left( \lambda x \tan\beta +A_D\right), \\
c' & = & m_D^2+m_d^2+e_dm_Z^2\cos 2\beta\sin^2\theta_W,
\end{eqnarray}
for scalar down-type quarks (or sleptons with isospin $-1/2$).
The mass eigenvalues are
\begin{equation}
m^2_{\tilde{q}_{1,2}} = \frac{1}{2}\left( a^{(')}+c^{(')} \mp \sqrt{(a^{(')}
-c^{(')})^2+4b^{(')2}} \right),
\end{equation}
and the mass eigenstates $\tilde{q}_1$ and $\tilde{q}_2$ read
\begin{equation}
\label{squarkmix}
\left( \begin{array}{c} \tilde{q}_L \\ \tilde{q}_R
\end{array} \right) =
\left( \begin{array}{cr} \cos\theta^{(')} & -\sin\theta^{(')} \\
\sin\theta^{(')} & \cos\theta^{(')} \end{array} \right)
\left( \begin{array}{c} \tilde{q}_1 \\ \tilde{q}_2
\end{array} \right)
\end{equation}
with
\begin{equation}
\sin\theta^{(')} = \frac{1}{\sqrt{1+d^{(')2}}}, \; \; \;
\cos\theta^{(')} = -d^{(')}\sin\theta^{(')},
\end{equation}
and
\begin{equation}
d^{(')}=\frac{b^{(')}}{a^{(')}-m_{\tilde{q}_1}^2}.
\end{equation}
\section{Feynman rules}
In this section we give a complete set of Feynman rules for all
vertex factors in the NMSSM that
differ from those of the MSSM.
Therefore it contains all Feynman rules with scalar or
pseudoscalar Higgs particles as well as the
$\tilde{\chi}^0 \tilde{\chi}^\pm C^\mp$ coupling.
Since the charged Higgs sector of the NMSSM is not enlarged,
all further vertex functions with charged Higgs bosons but
without neutral Higgs particles remain unchanged in the NMSSM.
The corresponding
Feynman rules for the MSSM can be found e.~g.~in refs.~\cite{higgs,hunter}.

Differences between the vertex functions of NMSSM and MSSM can
arise because of two reasons: First, the Feynman rules can be
formally identical, but the Higgs mixings are different. Generally,
in this case the vertex factor is suppressed in the NMSSM when
the Higgs bosons have a significant singlet component.
Second, the Feynman rules can differ explicitly if they contain terms
with the Higgs singlet components $U^S_{a3}$, $U^P_{\alpha 3}$, or
certain terms proportional to $\lambda $ or $k$. Then the vertex factor
can be suppressed or enhanced depending on the choice of the parameters.

In the NMSSM as well as in the MSSM, Higgs bosons interact with gauge bosons,
quarks, leptons, other Higgs bosons and their supersymmetric partners.
We first give those pieces of the Lagrangian that are responsible for
the coupling and then derive the Feynman rules for the respective
vertex functions in the unitary gauge.
\subsection{Interaction of two Higgs bosons with one gauge boson}
\label{sec:hhv}
\begin{figure}[p]
\begin{center}
\begin{picture}(12,4)
\put(0.8,2.3){$Z^0$}
\put(2.8,0.5){$P_{\alpha}$}
\put(2.8,3.0){$S_a$}
\put(3.3,2.4){$p$}
\put(3.3,1.4){$p'$}
\put(0,-0.8){\includegraphics{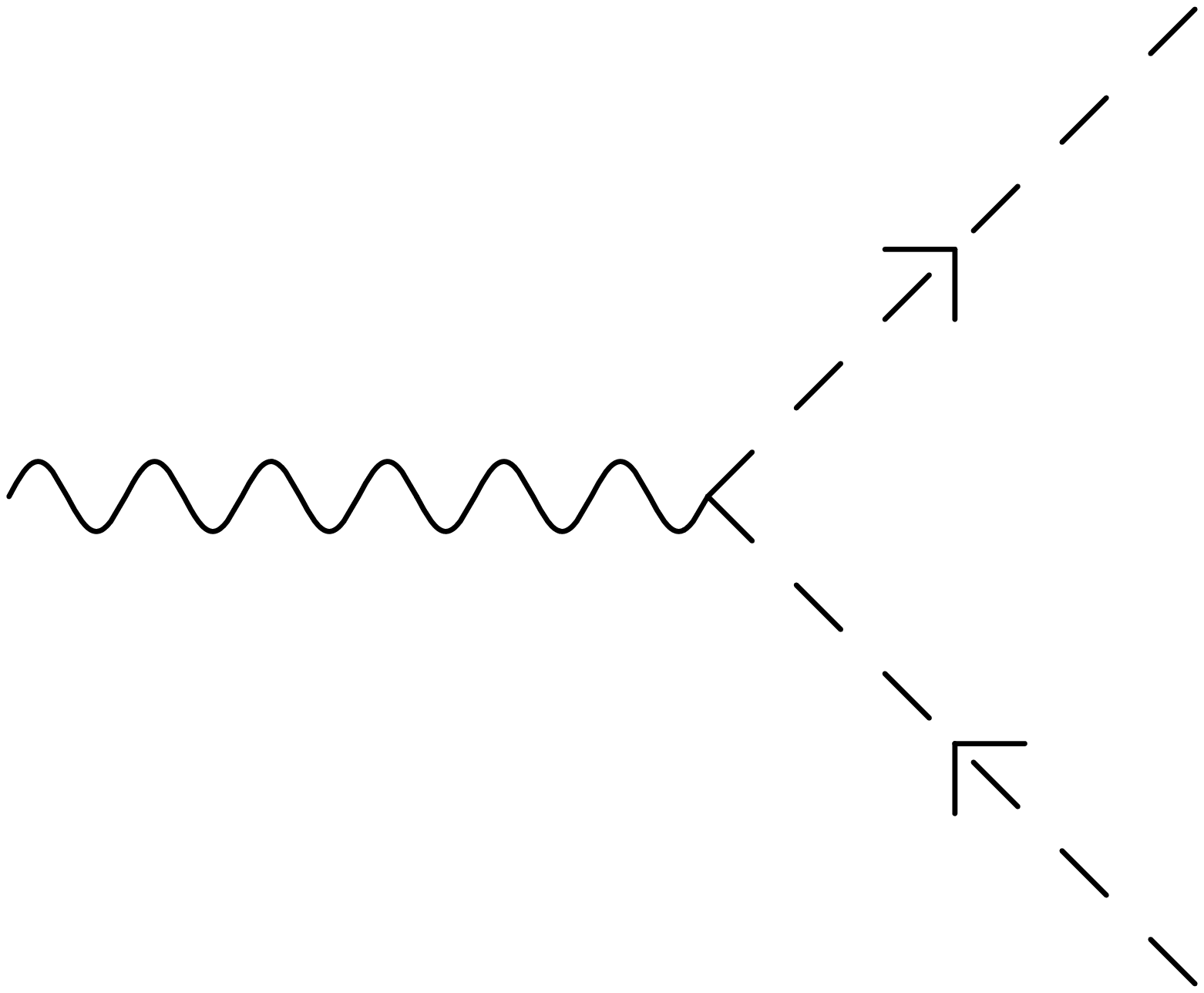}}
\put(7,2){\mbox{$\frac{g}{2\cos\theta_W}
(U^S_{a1}U^P_{\alpha 1}-U^S_{a2}U^P_{\alpha 2})(p+p')^{\mu}$}}
\end{picture}
\begin{picture}(12,4)
\put(0.8,2.3){$W^+$}
\put(2.8,0.5){$S_a$}
\put(2.8,3.0){$C^+$}
\put(3.3,2.4){$p$}
\put(3.3,1.4){$p'$}
\put(0,-0.8){\includegraphics{hhv.ps}}
\put(7,2){\mbox{$\frac{ig}{2}
(\sin\beta U^S_{a1}-\cos\beta U^S_{a2})(p+p')^{\mu}$}}
\end{picture}
\begin{picture}(12,4)
\put(0.8,2.3){$W^+$}
\put(2.8,0.5){$P_{\alpha}$}
\put(2.8,3.0){$C^+$}
\put(3.3,2.4){$p$}
\put(3.3,1.4){$p'$}
\put(0,-0.8){\includegraphics{hhv.ps}}
\put(7,2){\mbox{$\frac{g}{2}
(\sin\beta U^P_{\alpha 1}+\cos\beta U^P_{\alpha 2})(p+p')^{\mu}$}}
\end{picture}
\end{center}
\caption{Feynman rules for the $ZS_aP_\alpha$, $W^+C^+S_a$ and
$W^+C^+P_\alpha$ vertices.}
\label{fighhv}
\end{figure}

The relevant part of the Lagrangian is
\begin{eqnarray}
{\cal L}_{HHV} & = & -\frac{ig}{\sqrt{2}}
[W^+_\mu(H_1^{1\ast} \stackrel{\leftrightarrow}{\partial}\!{}^\mu
H_1^2+H_2^{1\ast} \stackrel{\leftrightarrow}{\partial}\!{}^\mu
H_2^2) + \mbox{h.c.} ]
\nonumber \\ & &
-\frac{ig}{2\cos\theta_W} Z_\mu
[(H_1^{1\ast} \stackrel{\leftrightarrow}{\partial}\!{}^\mu
H_1^1-H_2^{2\ast} \stackrel{\leftrightarrow}{\partial}\!{}^\mu
H_2^2) +
\nonumber \\ & & \hspace*{1.5cm}
(-1+2\sin\theta_W)
(H_1^{2\ast} \stackrel{\leftrightarrow}{\partial}\!{}^\mu
H_1^2+H_2^{2\ast} \stackrel{\leftrightarrow}{\partial}\!{}^\mu
H_2^2)]
\nonumber \\ & &
+ieA_\mu
(H_1^{2\ast} \stackrel{\leftrightarrow}{\partial}\!{}^\mu
H_1^2+H_2^{1\ast} \stackrel{\leftrightarrow}{\partial}\!{}^\mu
H_2^1).
\label{lhvv}
\end{eqnarray}
{}From eq.~(\ref{lhvv}) we get the Lagrangian for the interaction
between
\begin{enumerate}
\item one scalar and one pseudoscalar Higgs boson and one $Z$ boson
\begin{eqnarray}
\label{frspz}
{\cal L}_{S_aP_{\alpha}Z} & =
\frac{g}{2\cos\theta_W} Z_\mu &
(U^S_{a1} S_a \stackrel{\leftrightarrow}{\partial}\!{}^\mu
U^P_{\alpha 1} P_\alpha -
U^S_{a2} S_a \stackrel{\leftrightarrow}{\partial}\!{}^\mu
U^P_{\alpha 2} P_\alpha ) \; ;
\end{eqnarray}
\item one neutral scalar and one charged Higgs boson and one $W$ boson
\begin{eqnarray}
\label{frscw}
{\cal L}_{S_aCW} & = &
-\frac{ig}{2} W^+_\mu
(\sin\beta \,U^S_{a1}S_a \stackrel{\leftrightarrow}{\partial}\!{}^\mu
C^- -
\cos\beta \, U^S_{a2}S_a \stackrel{\leftrightarrow}{\partial}\!{}^\mu
C^- )+\mbox{h.c.}\; ;
\end{eqnarray}
\item one neutral pseudoscalar and one charged Higgs boson and one $W$ boson
\begin{eqnarray}
\label{frpcw}
{\cal L}_{P_\alpha CW} & = &
-\frac{g}{2}
(\sin\beta \, U^P_{\alpha 1}P_\alpha \stackrel{\leftrightarrow}{\partial}
\! {}^\mu
C^- +
\cos\beta \, U^P_{\alpha 2}P_\alpha \stackrel{\leftrightarrow}{\partial}
\! {}^\mu
C^- )+\mbox{h.c.}\; \; .
\end{eqnarray}
\end{enumerate}
As in the MSSM, Bose symmetry forbids the coupling of the $Z$ boson to two
identical Higgs bosons, while CP invariance forbids $ZS_aS_b$ ($a \neq b$)
and $ZP_\alpha P_\beta$ ($\alpha \neq \beta$) vertices.
The Feynman rules for the vertices with two Higgs bosons and one
gauge boson are given in Fig.~\ref{fighhv}.
\subsection{Interaction of Higgs bosons with two gauge bosons}
\label{sec:hvv}
\begin{figure}[p]
\begin{center}
\begin{picture}(12,4)
\put(0.8,2.3){$S_a$}
\put(2.8,0.5){$Z^0$}
\put(2.8,3.2){$Z^0$}
\put(0,-0.8){\includegraphics{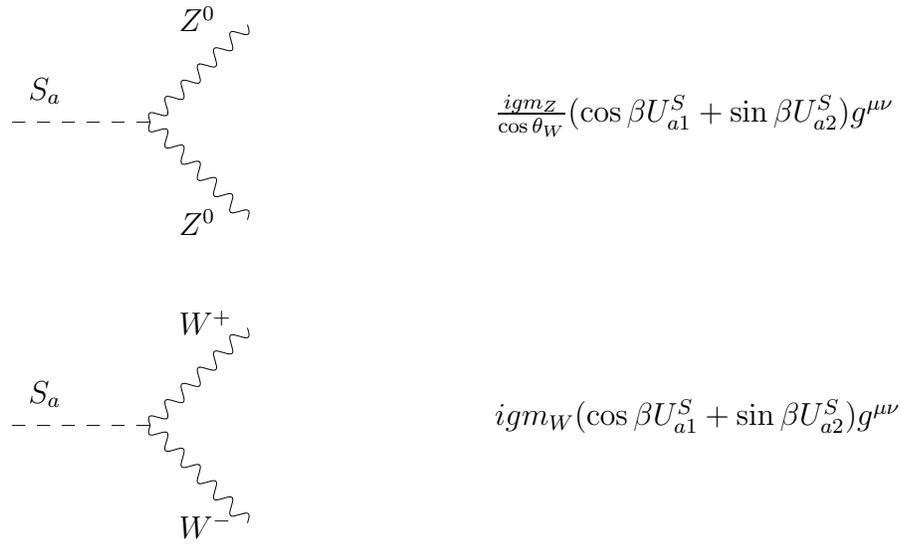}}
\put(7,2){\mbox{$\frac{igm_Z}{\cos\theta_W}
(\cos\beta U^S_{a1}+\sin\beta U^S_{a2})g^{\mu\nu}$}}
\end{picture}
\begin{picture}(12,4)
\put(0.8,2.3){$S_a$}
\put(2.8,0.5){$W^-$}
\put(2.8,3.2){$W^+$}
\put(0,-0.8){\includegraphics{hvv.ps}}
\put(7,2){\mbox{$igm_W
(\cos\beta U^S_{a1}+\sin\beta U^S_{a2})g^{\mu\nu}$}}
\end{picture}
\end{center}
\caption{Feynman rules for the vertices with two gauge bosons and one
neutral scalar Higgs boson.}
\label{fighvv}
\end{figure}

These interactions are contained in
\begin{eqnarray}
{\cal L}_{H(H)VV} & = &
\frac{g^2}{2} W^{\mu+}W_\mu^-
(|H_1^1|^2+|H_1^2|^2+|H_2^1|^2+|H_2^2|^2)
\nonumber \\ & &
-\frac{g}{\sqrt{2}}(eA^\mu-\frac{g\sin^2\theta_W}{\cos\theta_W}
Z^\mu)[W_\mu^+(H_1^{1\ast}H_1^2 -H_2^{1\ast} H_2^2)
+\mbox{h.c.}]
\nonumber \\ & &
+eA_\mu A^\mu (|H_1^2|^2+|H_2^1|^2)
\nonumber \\ & &
+\frac{g^2}{4\cos^2\theta_W} Z_\mu Z^\mu
(|H_1^1|^2+|H_2^2|^2+\cos^22\theta_W
(|H_1^2|^2+|H_2^1|^2))
\nonumber \\ & &
+\frac{eg}{\cos\theta_W} A_\mu Z^\mu \cos 2\theta_W
(|H_1^2|^2+|H_2^1|^2).
\label{hhvv}
\end{eqnarray}
Substituting the Higgs mass eigenstates of eqs.~(\ref{shiggs}) and
(\ref{phiggs})
one obtains for the trilinear
interaction between
\begin{enumerate}
\item one scalar Higgs boson and two $Z$ bosons
\begin{eqnarray}
\label{frszz}
{\cal L}_{S_aZZ} & = &
\frac{gm_Z}{2\cos\theta_W} Z_\mu Z^\mu
(\cos\beta U^S_{a1}+\sin\beta U^S_{a2}) S_a;
\end{eqnarray}
\item one scalar Higgs boson and two $W$ bosons
\begin{eqnarray}
\label{frsww}
{\cal L}_{S_aWW} & = &
gm_WW^{+\mu}W^-_\mu
(\cos\beta U^S_{a1}+\sin\beta U^S_{a2}) S_a.
\end{eqnarray}
\end{enumerate}
The corresponding Feynman rules are shown in Fig.~\ref{fighvv}.
All other trilinear $HVV$ couplings vanish at tree level.
Since the singlet Higgs field does not couple to gauge bosons, the
Feynman rules for the trilinear couplings
of Secs.~\ref{sec:hhv} and \ref{sec:hvv}
differ from those of the minimal model
only by the mixings of the
Higgs bosons. For the case of a
Higgs with a large singlet component, the coupling
to gauge bosons may become so small that production of a singlet
like scalar or pseudoscalar Higgs boson via the $ZS_aP_\alpha$ or $ZZS_a$
vertex is suppressed. We discuss the implications for the Higgs search
in Sec.~\ref{sec:con}.

\begin{figure}[p]
\begin{center}
\begin{picture}(12,4)
\put(2.8,0.5){$S_b$}
\put(2.8,3.2){$S_a$}
\put(1.6,0.5){$W^-$}
\put(1.6,3.2){$W^+$}
\put(0,-0.8){\includegraphics{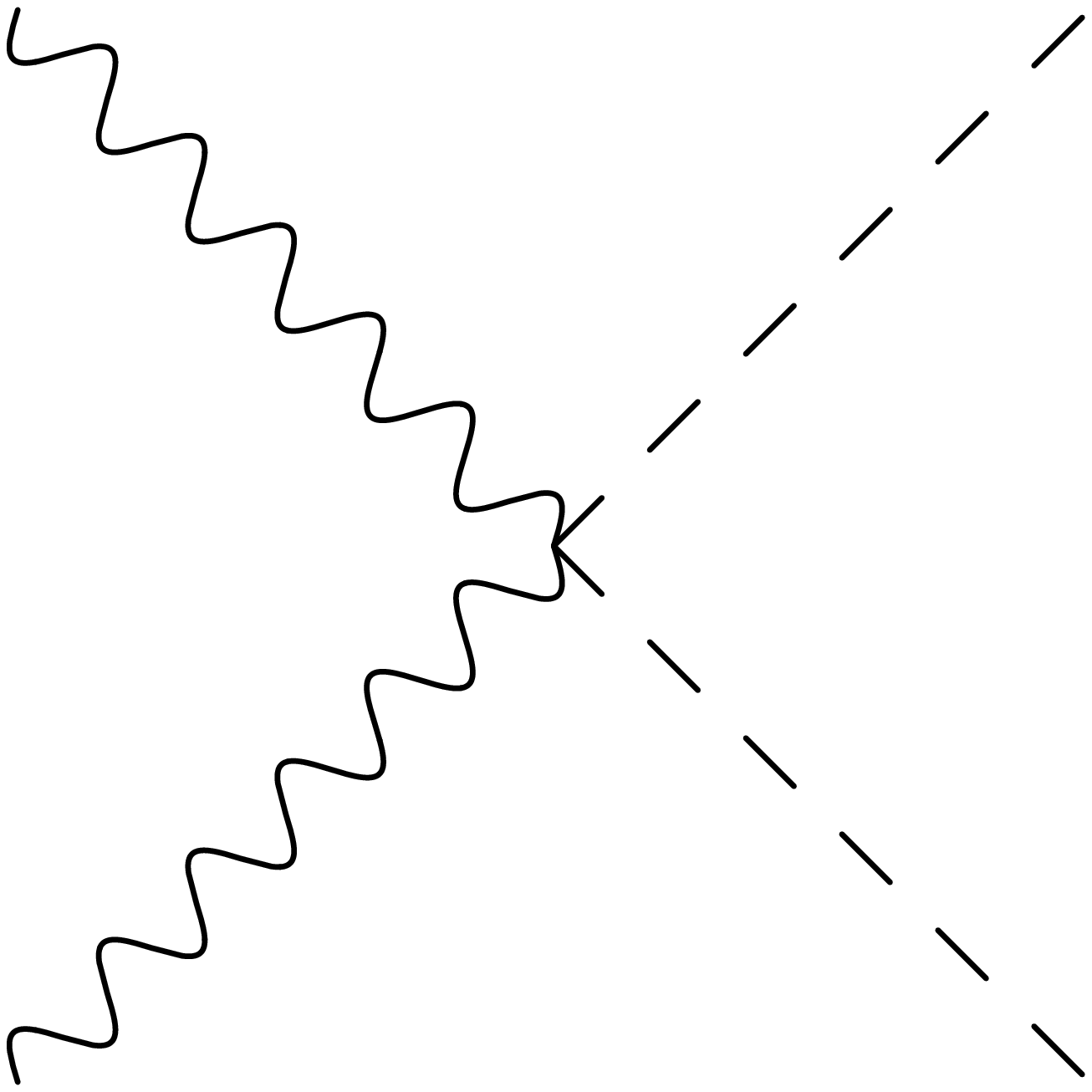}}
\put(7,2){\mbox{$\frac{ig^2}{2}
(U^S_{a1}U^S_{b1}+U^S_{a2}U^S_{b2})g^{\mu\nu}$}}
\end{picture}
\begin{picture}(12,4)
\put(2.8,0.5){$S_b$}
\put(2.8,3.2){$S_a$}
\put(1.6,0.5){$Z^0$}
\put(1.6,3.2){$Z^0$}
\put(0,-0.8){\includegraphics{hhvv.ps}}
\put(7,2){\mbox{$\frac{ig^2}{2\cos^2\theta_W}
(U^S_{a1}U^S_{b1}+U^S_{a2}U^S_{b2})g^{\mu\nu}$}}
\end{picture}
\begin{picture}(12,4)
\put(2.8,0.5){$S_a$}
\put(2.8,3.2){$C^{\pm}$}
\put(1.6,0.5){$Z^0$}
\put(1.6,3.2){$W^{\pm}$}
\put(0,-0.8){\includegraphics{hhvv.ps}}
\put(7,2){\mbox{$\frac{ig^2\sin^2\theta_W}{2\cos^2\theta_W}g^{\mu\nu}
(U^S_{a1}\sin\beta-U^S_{a2}\cos\beta)$}}
\end{picture}
\begin{picture}(12,4)
\put(2.8,0.5){$S_a$}
\put(2.8,3.2){$C^{\pm}$}
\put(1.6,0.5){$\gamma$}
\put(1.6,3.2){$W^{\pm}$}
\put(0,-0.8){\includegraphics{hhvv.ps}}
\put(7,2){\mbox{$\frac{-ieg}{2}g^{\mu\nu}
(U^S_{a 1}\sin\beta-U^S_{a 2}\cos\beta)$}}
\end{picture}
\end{center}
\caption{Feynman rules for the quartic couplings of scalar Higgs bosons
to gauge bosons.}
\label{figssvv}
\end{figure}

\begin{figure}[p]
\begin{center}
\begin{picture}(12,4)
\put(2.8,0.5){$P_{\beta}$}
\put(2.8,3.2){$P_{\alpha}$}
\put(1.6,0.5){$W^-$}
\put(1.6,3.2){$W^+$}
\put(0,-0.8){\includegraphics{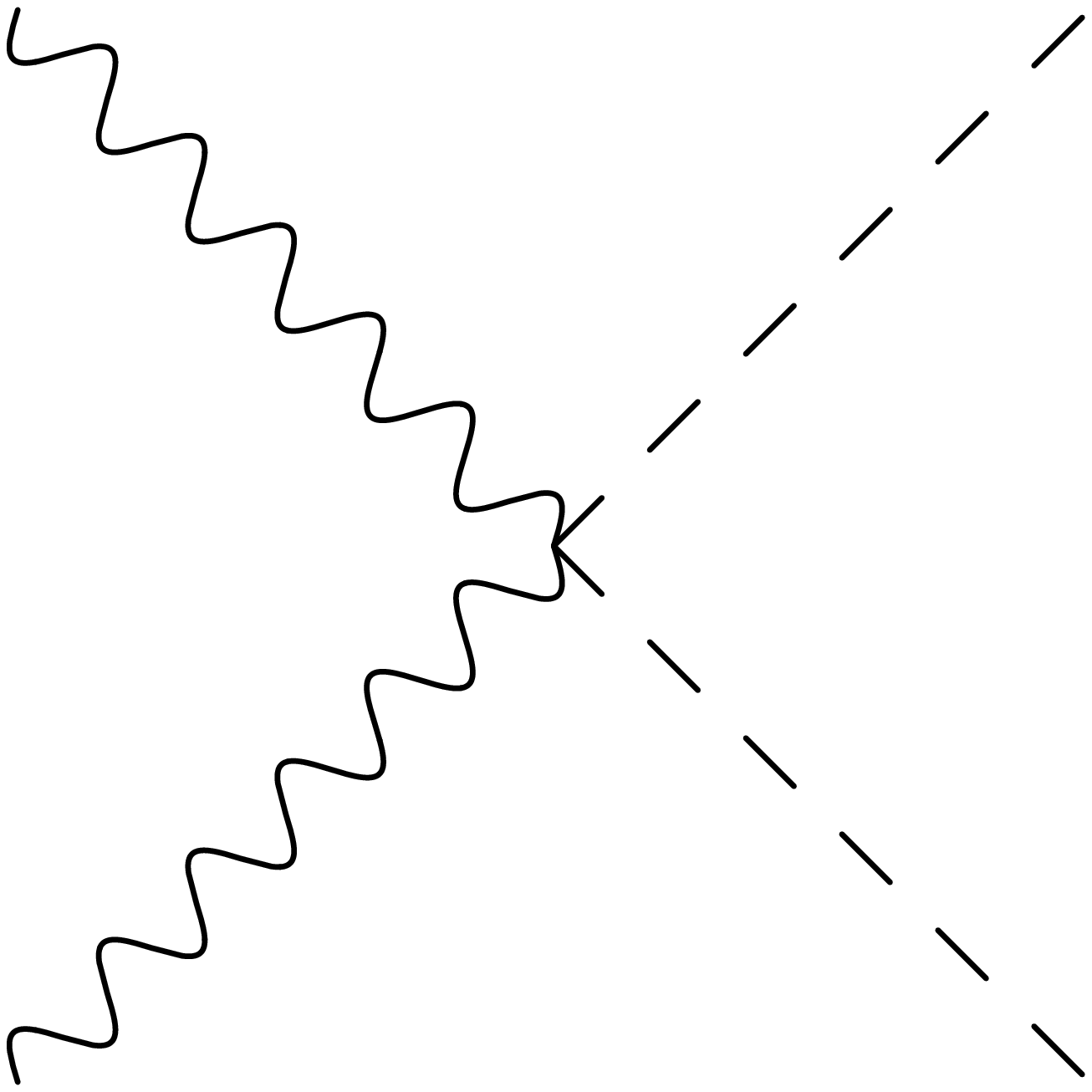}}
\put(7,2){\mbox{$\frac{ig^2}{2}
(U^P_{\alpha 1}U^P_{\beta 1}+U^P_{\alpha 2}U^P_{\beta 2})g^{\mu\nu}$}}
\end{picture}
\begin{picture}(12,4)
\put(2.8,0.5){$P_{\beta}$}
\put(2.8,3.2){$P_{\alpha}$}
\put(1.6,0.5){$Z^0$}
\put(1.6,3.2){$Z^0$}
\put(0,-0.8){\includegraphics{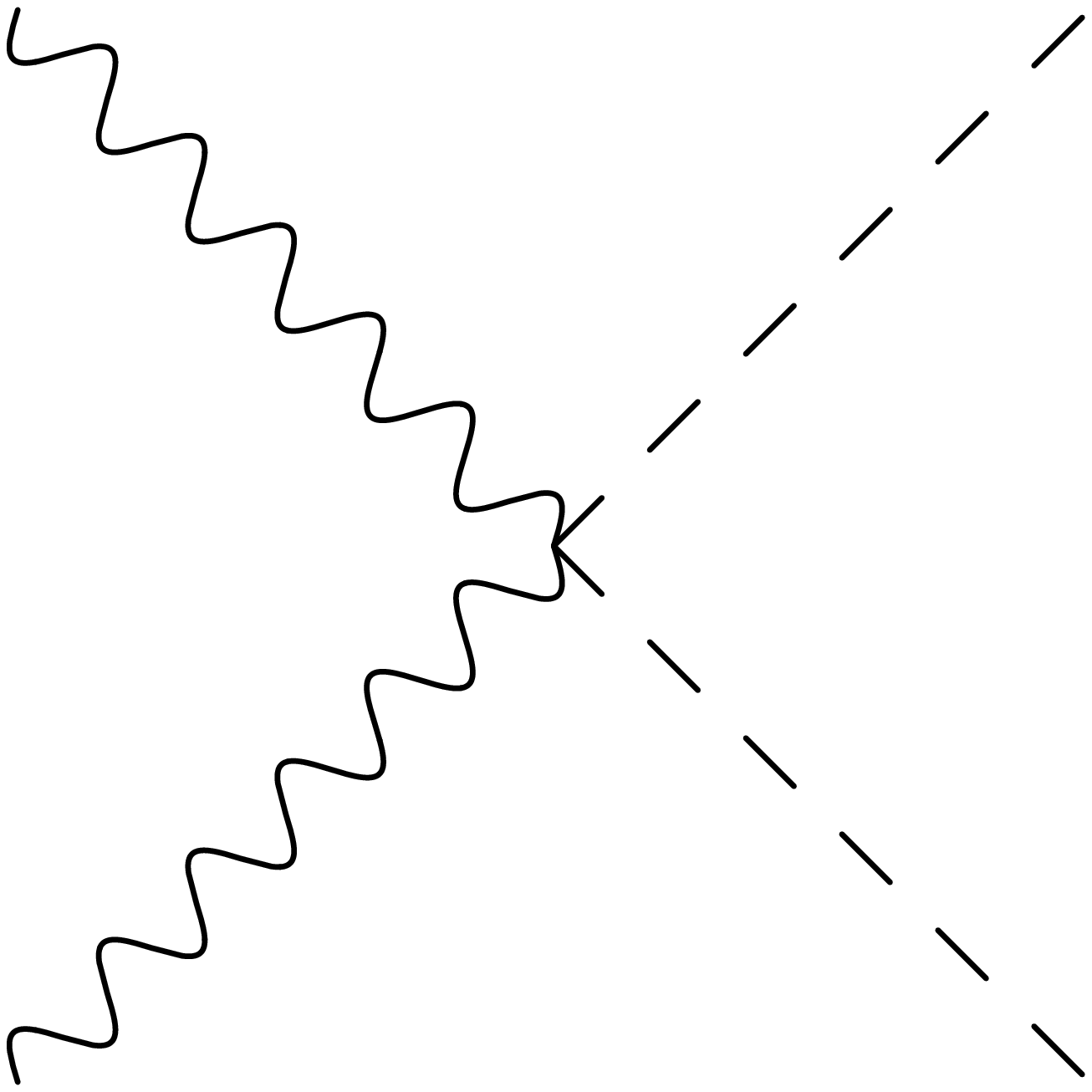}}
\put(7,2){\mbox{$\frac{ig^2}{2\cos^2\theta_W}
(U^P_{\alpha 1}U^P_{\beta 1}+U^P_{\alpha 2}U^P_{\beta2})g^{\mu\nu}$}}
\end{picture}
\begin{picture}(12,4)
\put(2.8,0.5){$P_{\alpha}$}
\put(2.8,3.2){$C^{\pm}$}
\put(1.6,0.5){$Z^0$}
\put(1.6,3.2){$W^{\pm}$}
\put(0,-0.8){\includegraphics{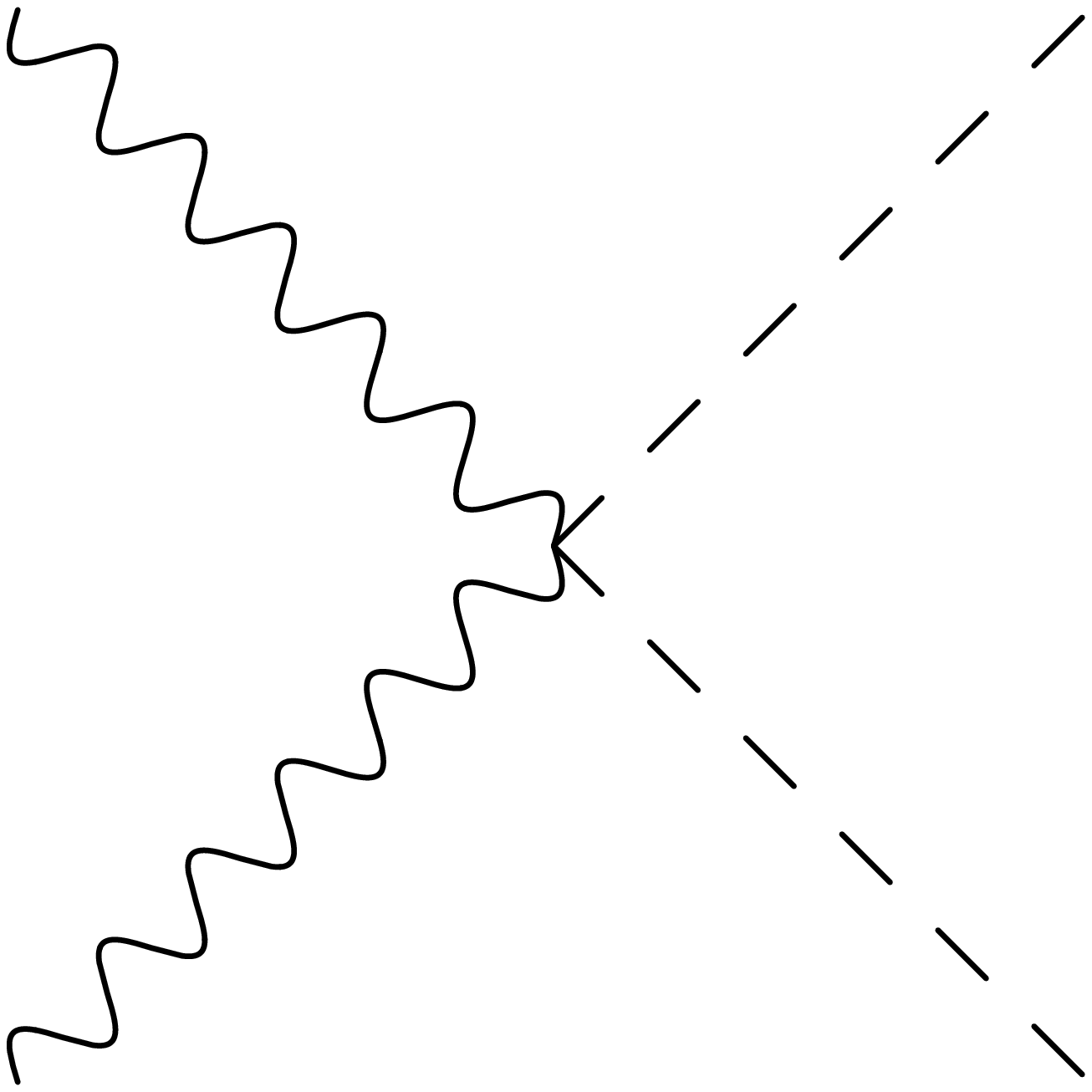}}
\put(7,2){\mbox{$\pm\frac{g^2\sin^2\theta_W}{2\cos^2\theta_W}g^{\mu\nu}
(U^P_{\alpha 1}\sin\beta+U^P_{\alpha 2}\cos\beta)$}}
\end{picture}
\begin{picture}(12,4)
\put(2.8,0.5){$P_{\alpha}$}
\put(2.8,3.2){$C^{\pm}$}
\put(1.6,0.5){$\gamma$}
\put(1.6,3.2){$W^{\pm}$}
\put(0,-0.8){\includegraphics{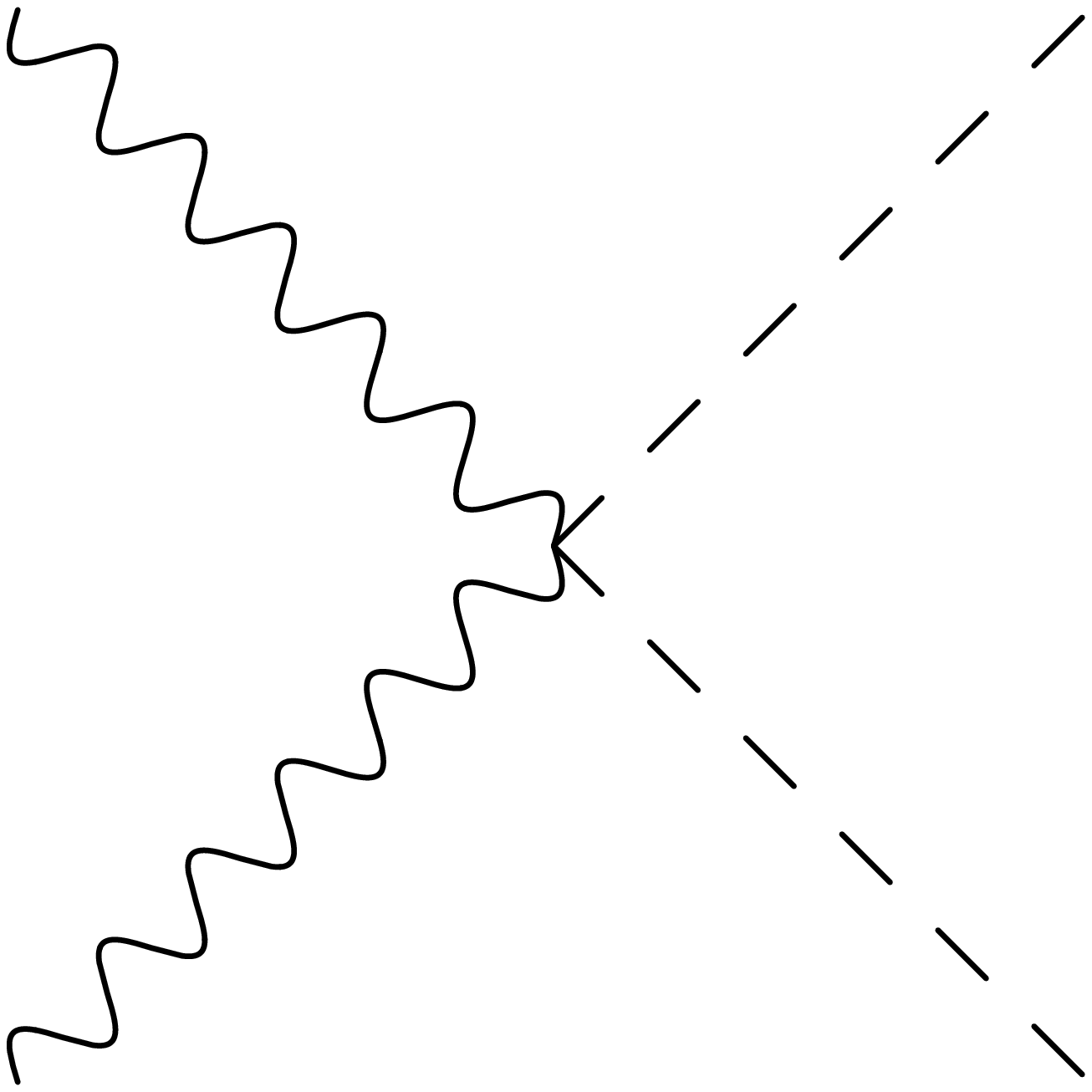}}
\put(7,2){\mbox{$\mp\frac{eg}{2}g^{\mu\nu}
(U^P_{\alpha 1}\sin\beta+U^P_{\alpha 2}\cos\beta)$}}
\end{picture}
\end{center}
\caption{Feynman rules for the quartic couplings of pseudoscalar Higgs bosons
to gauge bosons.}
\label{figppvv}
\end{figure}

The quartic couplings, however, show significant differences between
the minimal and nonminimal model.
{}From the interaction Lagrangian eq.~(\ref{hhvv}) one derives for
the interaction of
\begin{enumerate}
\item two neutral scalar Higgs bosons and two $W$ bosons
\begin{equation}
\label{wwss}
{\cal L}_{S_aS_bWW} =
\frac{g^2}{4} (2-\delta_{ab}) (U^S_{a1} U^S_{b1} +
U^S_{a2} U^S_{b2}) W^{\mu +}W^-_\mu S_a S_b ;
\end{equation}
\item two neutral pseudoscalar Higgs bosons and two $W$ bosons
\begin{equation}
\label{wwsp}
{\cal L}_{P_\alpha P_\beta WW} =
\frac{g^2}{4} (2-\delta_{ab}) (U^P_{\alpha 1} U^P_{\beta 1} +
U^P_{\alpha 2} U^P_{\beta 2}) W^{\mu +}W^-_\mu P_\alpha P_\beta ;
\end{equation}
\item two neutral scalar Higgs bosons and two $Z$ bosons
\begin{equation}
\label{zzss}
{\cal L}_{S_aS_bZZ} =
\frac{g^2}{8\cos^2\theta_W} (2-\delta_{ab}) (U^S_{a1} U^S_{b1} +
U^S_{a2} U^S_{b2}) Z^{\mu}Z_\mu S_a S_b ;
\end{equation}
\item two neutral pseudoscalar Higgs bosons and two $Z$ bosons
\begin{equation}
\label{zzpp}
{\cal L}_{P_\alpha P_\beta ZZ} =
\frac{g^2}{8\cos^2\theta_W} (2-\delta_{ab}) (U^P_{\alpha 1} U^P_{\beta 1} +
U^P_{\alpha 2} U^P_{\beta 2}) Z^{\mu}Z_\mu P_\alpha P_\beta ;
\end{equation}
\item one neutral scalar Higgs boson, one charged Higgs particle,
one $Z$ and one $W$ boson
\begin{equation}
{\cal L}_{S_aCZW} =
+\frac{g^2}{2} \frac{\sin^2\theta_W}{\cos\theta_W}
(U^S_{a1}\sin\beta-U^S_{a2}\cos\beta)Z^\mu W^+_\mu C^- S_a
+\mbox{h.c.};
\end{equation}
\item one neutral pseudoscalar Higgs boson, one charged Higgs particle,
one $Z$ and one $W$ boson
\begin{equation}
{\cal L}_{P_\alpha CZW} =
-\frac{ig^2}{2} \frac{\sin^2\theta_W}{\cos\theta_W}
(U^P_{\alpha 1}\sin\beta+U^P_{\alpha 2}\cos\beta)Z^\mu W^+_\mu C^-
P_\alpha+\mbox{h.c.};
\end{equation}
\item one neutral scalar Higgs boson, one charged Higgs particle,
one photon and one $W$ boson
\begin{equation}
{\cal L}_{S_aC\gamma W} =
-\frac{eg}{2}
(U^S_{a1}\sin\beta-U^S_{a2}\cos\beta)A^\mu W^+_\mu C^- S_a
+\mbox{h.c.};
\end{equation}
\item one neutral pseudoscalar Higgs boson, one charged Higgs particle,
one photon and one $W$ boson
\begin{equation}
{\cal L}_{P_\alpha C\gamma W} =
\frac{ieg}{2}
(U^P_{\alpha 1}\sin\beta+U^P_{\alpha 2}\cos\beta)A^\mu W^+_\mu C^-
P_\alpha+\mbox{h.c.}\; \; .
\end{equation}
\end{enumerate}
The Feynman rules for these vertex function can be found in
Figs.~\ref{figssvv} and \ref{figppvv}.
Contrary to the NMSSM, the vertices with two different neutral
scalar Higgs bosons
and two gauge bosons eqs.~(\ref{wwss}) and (\ref{zzss})
vanish in the MSSM due to the orthogonality of the $2\times 2$ MSSM
diagonalization matrix.
\subsection{Interaction of Higgs bosons with quarks and leptons}
\begin{figure}[p]
\begin{center}
\begin{picture}(12,4)
\put(0.8,2.3){$S_a$}
\put(2.8,0.5){$u$}
\put(2.8,3.0){$u$}
\put(0,-0.8){\includegraphics{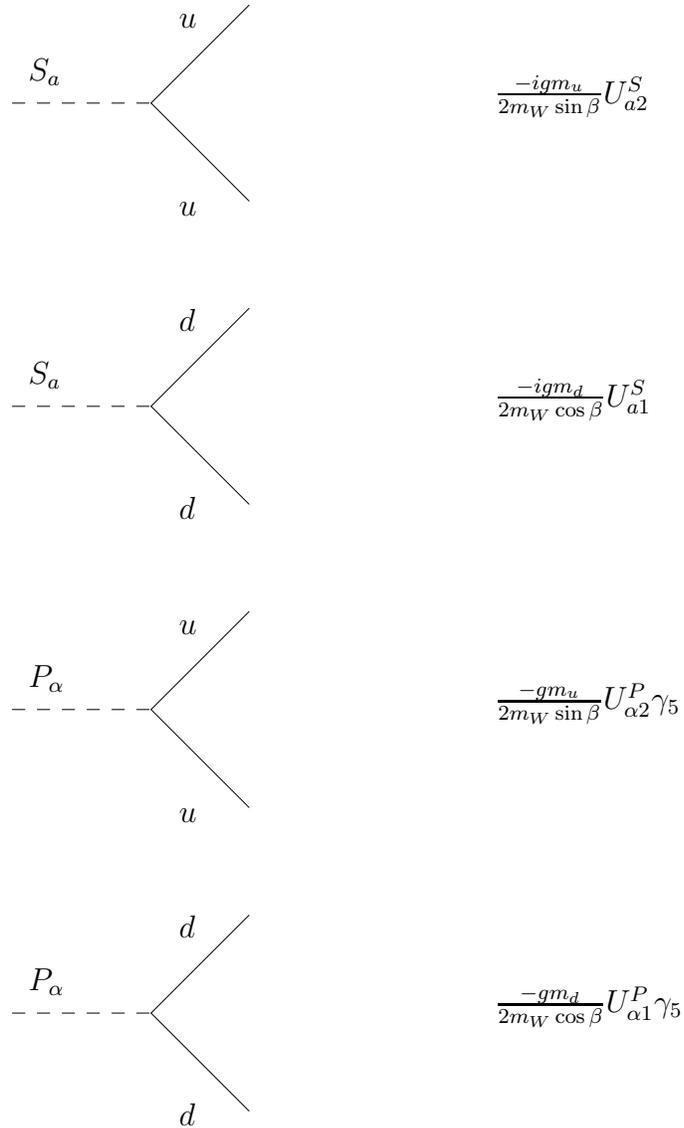}}
\put(7,2){\mbox{$\frac{-igm_u}{2m_W\sin\beta}U_{a2}^S$}}
\end{picture}
\begin{picture}(12,4)
\put(0.8,2.3){$S_a$}
\put(2.8,0.5){$d$}
\put(2.8,3.0){$d$}
\put(0,-0.8){\includegraphics{hqqo.ps}}
\put(7,2){\mbox{$\frac{-igm_d}{2m_W\cos\beta}U_{a1}^S$}}
\end{picture}
\begin{picture}(12,4)
\put(0.8,2.3){$P_{\alpha}$}
\put(2.8,0.5){$u$}
\put(2.8,3.0){$u$}
\put(0,-0.8){\includegraphics{hqqo.ps}}
\put(7,2){\mbox{$\frac{-gm_u}{2m_W\sin\beta}U_{\alpha 2}^P\gamma_5$}}
\end{picture}
\begin{picture}(12,4)
\put(0.8,2.3){$P_{\alpha}$}
\put(2.8,0.5){$d$}
\put(2.8,3.0){$d$}
\put(0,-0.8){\includegraphics{hqqo.ps}}
\put(7,2){\mbox{$\frac{-gm_d}{2m_W\cos\beta}U_{\alpha 1}^P\gamma_5$}}
\end{picture}
\end{center}
\caption{Feynman rules for the couplings of neutral Higgs bosons to
quarks.}
\label{fighqq}
\end{figure}

The part of the Lagrangian containing the terms responsible for
the masses of the quarks and their couplings to Higgs bosons
reads in two-component notation
\begin{eqnarray}
{\cal L} _{Hq\bar{q}} & = & -h_d (H_1^1Q^2D-H_1^2Q^1D)
\nonumber \\ & &
-h_u(H_2^2Q^1U-H_2^1Q^2U) + \mbox{h.c.} \;\; .
\end{eqnarray}
Introducing four-component spinors
\begin{equation}
u=\left( \begin{array}{c} Q^1 \\ U \end{array} \right),
\hspace*{1cm}
d=\left( \begin{array}{c} Q^2 \\ D \end{array} \right),
\end{equation}
one finds besides the relation between the coupling parameters $h_{u,d}$
and the quark masses $m_{u,d}$
\begin{eqnarray}
h_u & = & \frac{gm_u}{\sqrt{2}m_W\sin\beta}, \\
h_d & = & \frac{gm_d}{\sqrt{2}m_W\cos\beta},
\end{eqnarray}
the trilinear interaction terms
\begin{eqnarray}
{\cal L} _{S_au\bar{u}} & = &
-\frac{gm_u}{2m_W\sin\beta}U^S_{a2}S_a\bar{u}u, \\
{\cal L} _{S_ad\bar{d}} & = &
-\frac{gm_d}{2m_W\cos\beta}U^S_{a1}S_a\bar{d}d, \\
{\cal L} _{P_\alpha u\bar{u}} & = &
\frac{igm_u}{2m_W\sin\beta}U^P_{\alpha 2}P_\alpha\bar{u}\gamma_5 u, \\
{\cal L} _{P_\alpha d\bar{d}} & = &
\frac{igm_d}{2m_W\cos\beta}U^P_{\alpha 1}P_\alpha \bar{d}\gamma_5 d.
\end{eqnarray}
As in the MSSM, up-type quarks couple to the
$H_2$ component of the respective Higgs boson, while the
coupling to down-type quarks contains the $H_1$ component.
The relevant Feynman rules are displayed in Fig.~\ref{fighqq}.
The interactions with leptons are obtained by the replacement
$(u,d)\longrightarrow (\nu,e)$. The generalization to three
generations proceeds in the same way as in the MSSM \cite{higgs}.
\subsection{Interaction of Higgs bosons with scalar quarks}
\label{sec:hsqsq}
These interactions arise from the following $D$ and $F$ terms of the scalar
potential as well as from the supersymmetry breaking terms:
\begin{eqnarray}
\label{ldhsqsq}
{\cal L}^D_{H\tilde{q}\tilde{q}} & = &
(H_2^{2\ast}H_2^2-H_1^{1\ast}H_1^1)
\Big\{ \frac{g^2}{4}(\tilde{Q}^{1\ast} \tilde{Q}^1 -
\tilde{Q}^{2\ast} \tilde{Q}^2)
\nonumber \\ & & \; \; \; \; \;
-\frac{{g'}^2}{4}[(y_q (\tilde{Q}^{1\ast} \tilde{Q}^1
+\tilde{Q}^{2\ast} \tilde{Q}^2)+y_u \tilde{U}^{\ast} \tilde{U}
+y_d \tilde{D}^{\ast} \tilde{D})]\Big\}
\nonumber \\ & &
-\frac{g^2}{2} [H_1^1 \tilde{Q}^{1\ast} H_1^{2\ast}
\tilde{Q}^2 +
H_2^1 \tilde{Q}^{1\ast} H_2^{2\ast}
\tilde{Q}^2 + \mbox{h.c.} ], \\ [0.5em]
\label{lfhsqsq}
{\cal L}^F_{H\tilde{q}\tilde{q}} & = &
-|h_u\tilde{Q}^1H_2^2|^2-|h_u\tilde{U}H_2^2|^2
-|h_d\tilde{Q}^2H_1^1|^2-|h_d\tilde{D}H_1^1|^2
\nonumber \\ & &
+\{ -\lambda H_2^2N h_d \tilde{Q}^{2\ast} \tilde{D}^\ast -
\lambda H_1^1N h_u \tilde{Q}^{1\ast} \tilde{U}^\ast
\nonumber \\ & & \; \;
- \lambda H_2^1 N h_d \tilde{Q}^{1\ast} \tilde{D}^{\ast}
- \lambda H_1^{2\ast} N^{\ast} h_u \tilde{Q}^2 \tilde{U}
\nonumber \\ & & \; \;
+h_u h_d (H_1^{2\ast} H_2^2 \tilde{U} \tilde{D}^{\ast}
+ H_2^1 H_1^{1\ast} \tilde{U} \tilde{D}^{\ast})
\nonumber \\ & & \; \;
+h_u^2 H_2^1 H_2^{2\ast}\tilde{Q}^{1\ast}  \tilde{Q}^2
+h_d^2 H_1^{2\ast}  H_1^1 \tilde{Q}^{1\ast} \tilde{Q}^2
+ \mbox{h.c.}\}, \\ [0.5em]
\label{lshsqsq}
{\cal L}^{\mbox{\scriptsize soft}}_{H\tilde{q}\tilde{q}} & = &
-h_uA_U\tilde{Q}^1 \tilde{U} H_2^2 -
h_dA_D\tilde{Q}^2 \tilde{D} H_1^1
\nonumber \\ & &
+(h_uA_U H_2^1 \tilde{Q}^2 \tilde{U}
+h_d A_D H_1^{2\ast} H_1^1 \tilde{Q}^{1\ast} \tilde{D}^{\ast}
+\mbox{h.c.}) \; \; .
\end{eqnarray}
We list in the following the trilinear and quartic interactions
of Higgs bosons with the weak interaction eigenstates $\tilde{q}_L$,
$\tilde{q}_R$.
The mass eigenstates of the squarks and
sleptons are obtained with the
transformation eq.~(\ref{squarkmix}),
so that the vertex functions can be converted
with
\begin{equation}
\label{v}
V(\tilde{q}_i \tilde{q}'_j) = \sum_{k,l=L,R} C_{ijkl}
V(\tilde{q}_k \tilde{q}'_l).
\end{equation}
In eq.~(\ref{v}), $V(\tilde{q}_i \tilde{q}'_j)$ denotes any vertex function
with two scalar quarks $\tilde{q}_i$ and $\tilde{q}'_j$ $(i,j=1,2)$.
The coefficients $C_{ijkl}$
are given in Table 1.
\begin{table}[t]
\begin{center}
$\begin{array}{ccccc}
\hline
& \tilde{q}_L \tilde{q}'_L & \tilde{q}_L \tilde{q}'_R &
\tilde{q}'_L \tilde{q}_R & \tilde{q}_R \tilde{q}'_R \\
\hline
\tilde{q}_1 \tilde{q}'_1 & \cos\theta \cos\theta' &
\cos\theta \sin\theta' & \cos\theta' \sin\theta &
\sin\theta \sin\theta' \\
\tilde{q}_1 \tilde{q}'_2 & -\cos\theta \sin\theta' &
\cos\theta \cos\theta' & -\sin\theta \sin\theta' &
\cos\theta' \sin\theta \\
\tilde{q}_2 \tilde{q}'_2 & \sin\theta \sin\theta' &
-\sin\theta \cos\theta' & -\sin\theta' \cos\theta' &
\cos\theta \cos\theta' \\ \hline
\end{array}$
\end{center}
\caption{Coefficients that convert Feynman rules for the vertex
functions with two scalar
quarks from the $\tilde{q}_L - \tilde{q}_R$ basis to the
$\tilde{q}_1 - \tilde{q}_2$ basis. The squark mixing angles
$\theta$ are defined in Sec.~\ref{squark}.}
\end{table}

\begin{figure}[p]
\begin{center}
\begin{picture}(12,4)
\put(0.8,2.3){$S_a$}
\put(2.8,0.5){$\tilde{u}_L$}
\put(2.8,3.0){$\tilde{u}_L$}
\put(0,-0.8){\includegraphics{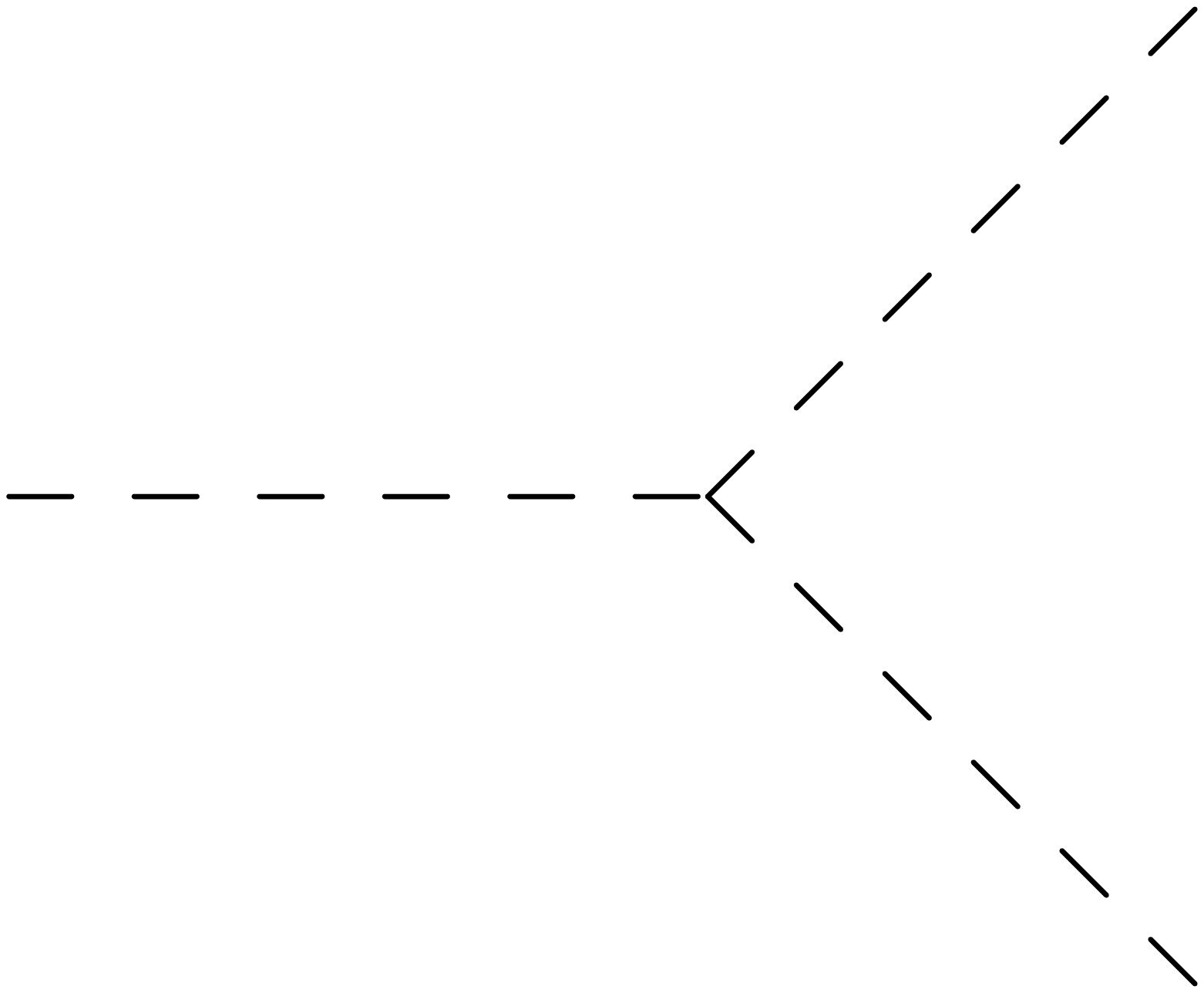}}
\put(5,2){\mbox{$\begin{array}{l}
\frac{-igm_u^2}{m_W\sin\beta}U_{a2}^S \\
+\frac{ig}{2}\frac{m_Z}{\cos\theta_W}(1-2e_u\sin^2\theta_W)
(U^S_{a2}\sin\beta -U^S_{a1}\cos\beta )
\end{array}$}}
\end{picture}
\begin{picture}(12,4)
\put(0.8,2.3){$S_a$}
\put(2.8,0.5){$\tilde{u}_R$}
\put(2.8,3.0){$\tilde{u}_R$}
\put(0,-0.8){\includegraphics{hhh.ps}}
\put(5,2){\mbox{$\begin{array}{l}
\frac{-igm_u^2}{m_W\sin\beta}U_{a2}^S \\
+igm_We_u\tan^2\theta_W
(U^S_{a2}\sin\beta -U^S_{a1}\cos\beta )
\end{array}$}}
\end{picture}
\begin{picture}(12,4)
\put(0.8,2.3){$S_a$}
\put(2.8,0.5){$\tilde{u}_R$}
\put(2.8,3.0){$\tilde{u}_L$}
\put(0,-0.8){\includegraphics{hhh.ps}}
\put(5,2){\mbox{$
\frac{-igm_u}{2m_W\sin\beta}(\lambda (v_1U_{a3}^S
+xU^S_{a1})+A_UU^S_{a2} )$}}
\end{picture}
\end{center}
\caption{Feynman rules for the trilinear vertices with one neutral
scalar Higgs boson and two scalar up-type quarks.}
\label{figssusu}
\end{figure}

\begin{figure}[p]
\begin{center}
\begin{picture}(12,4)
\put(0.8,2.3){$S_a$}
\put(2.8,0.5){$\tilde{d}_L$}
\put(2.8,3.0){$\tilde{d}_L$}
\put(0,-0.8){\includegraphics{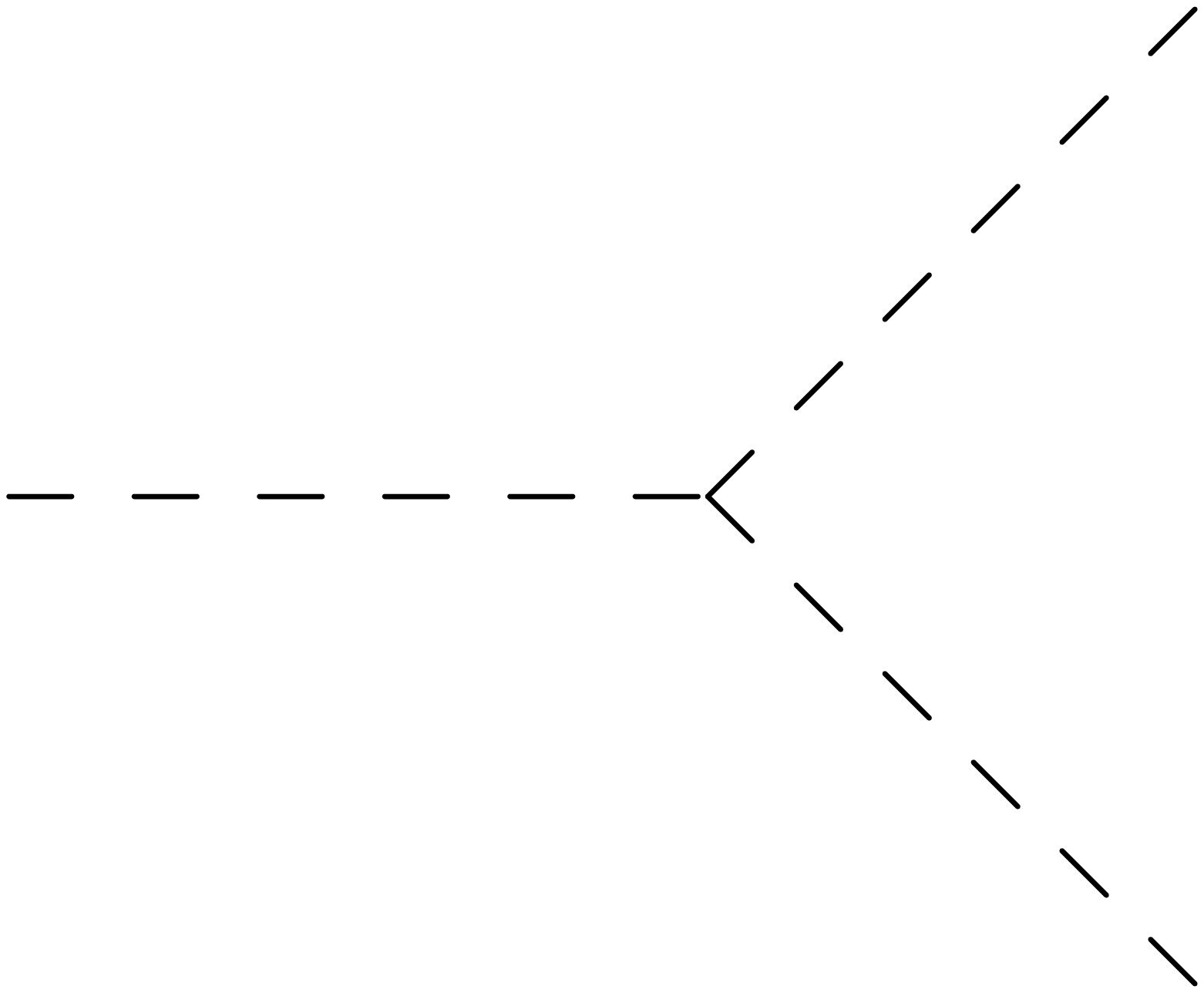}}
\put(5,2){\mbox{$\begin{array}{l}
\frac{-igm_d^2}{m_W\cos\beta}U_{a1}^S \\
-\frac{ig}{2}\frac{m_Z}{\cos\theta_W}(1+2e_d\sin^2\theta_W)
(U^S_{a2}\sin\beta -U^S_{a1}\cos\beta )
\end{array}$}}
\end{picture}
\begin{picture}(12,4)
\put(0.8,2.3){$S_a$}
\put(2.8,0.5){$\tilde{d}_R$}
\put(2.8,3.0){$\tilde{d}_R$}
\put(0,-0.8){\includegraphics{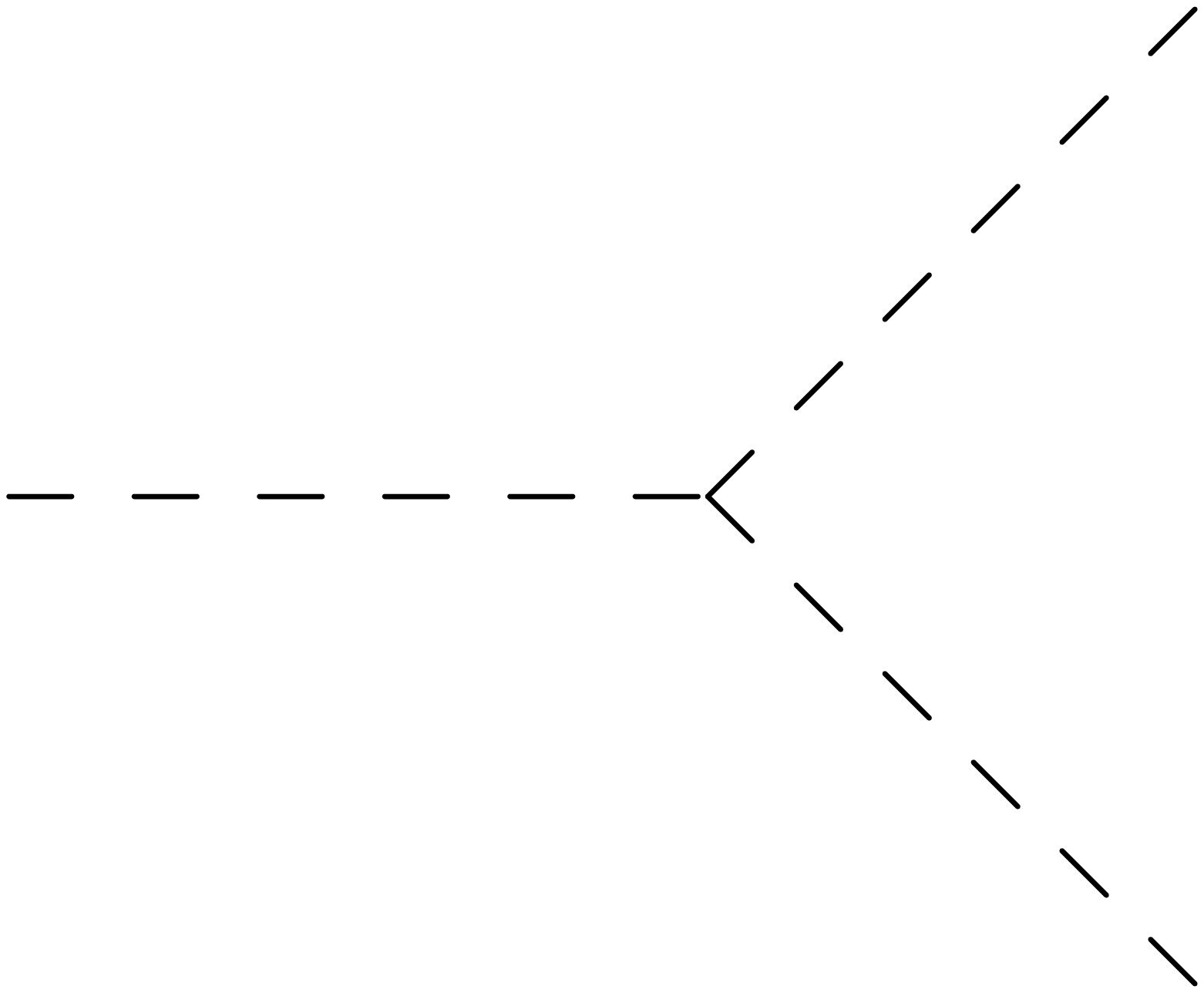}}
\put(5,2){\mbox{$\begin{array}{l}
\frac{-igm_d^2}{m_W\cos\beta}U_{a1}^S \\
+igm_We_d\tan^2\theta_W
(U^S_{a2}\sin\beta -U^S_{a1}\cos\beta )
\end{array}$}}
\end{picture}
\begin{picture}(12,4)
\put(0.8,2.3){$S_a$}
\put(2.8,0.5){$\tilde{d}_R$}
\put(2.8,3.0){$\tilde{d}_L$}
\put(0,-0.8){\includegraphics{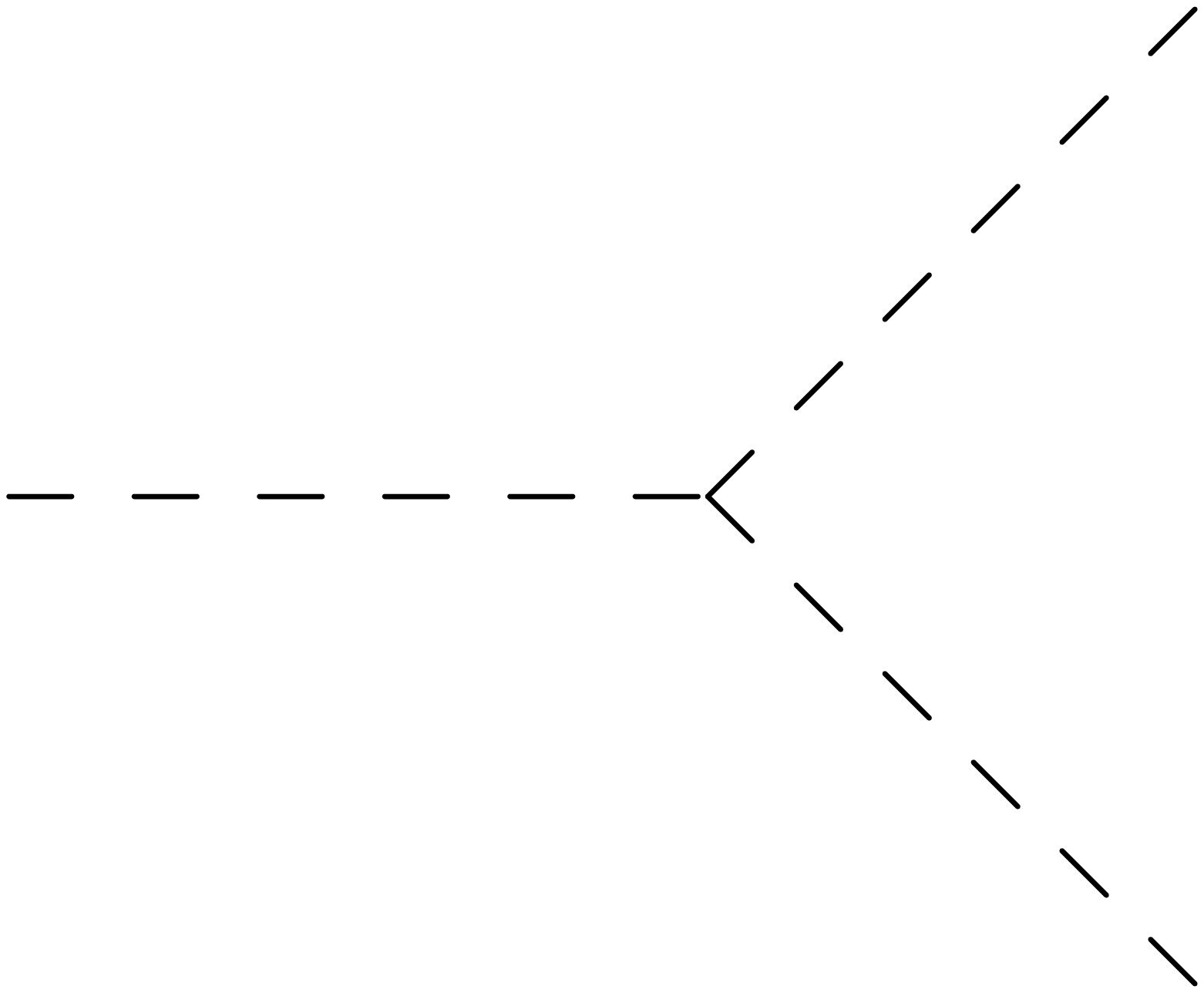}}
\put(5,2){\mbox{$
\frac{-igm_d}{2m_W\cos\beta}(\lambda (v_2U_{a3}^S
+xU^S_{a2})+A_DU^S_{a1} )$}}
\end{picture}
\end{center}
\caption{Feynman rules for the trilinear vertices with one neutral
scalar Higgs boson and two scalar down-type quarks.}
\label{figssdsd}
\end{figure}

\begin{figure}[p]
\begin{center}
\begin{picture}(12,4)
\put(0.8,2.3){$P_{\alpha}$}
\put(2.8,0.5){$\tilde{u}_R$}
\put(2.8,3.0){$\tilde{u}_L$}
\put(0,-0.8){\includegraphics{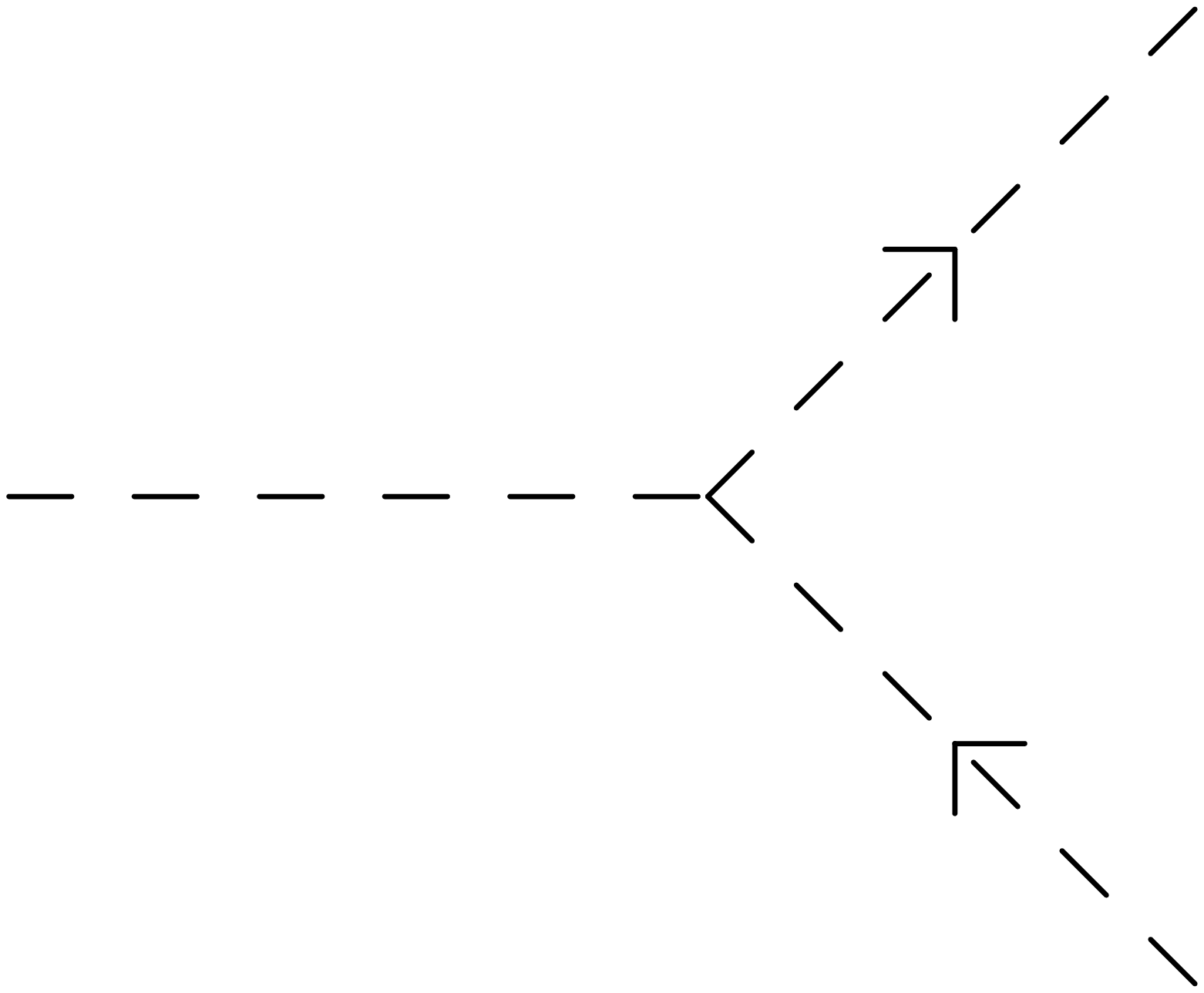}}
\put(7,2){\mbox{$
\frac{gm_u}{2m_W\sin\beta}(\lambda (v_1U_{\alpha 3}^P
+xU^P_{\alpha 1})-A_UU^P_{\alpha 2} )$}}
\end{picture}
\begin{picture}(12,4)
\put(0.8,2.3){$P_{\alpha}$}
\put(2.8,0.5){$\tilde{d}_R$}
\put(2.8,3.0){$\tilde{d}_L$}
\put(0,-0.8){\includegraphics{hosqsq.ps}}
\put(7,2){\mbox{$
\frac{gm_d}{2m_W\cos\beta}(\lambda (v_2U_{\alpha 3}^P
+xU^P_{\alpha 2})-A_DU^P_{\alpha 1} )$}}
\end{picture}
\end{center}
\caption{Feynman rules for the trilinear vertices with one neutral
pseudoscalar Higgs boson and two scalar quarks.}
\label{figpsqsq}
\end{figure}

In the $q_L - q_R$ basis we derive from eqs.~(\ref{ldhsqsq}) --
(\ref{lshsqsq}) the trilinear interactions for
\begin{enumerate}
\item one scalar Higgs boson and two left-handed up-type squarks
\begin{eqnarray}
{\cal L}_{S_a\tilde{u}_L \tilde{u}_L} & = &
[-\frac{gm_u^2}{m_W\sin\beta} U^S_{a2}
\nonumber \\ & & \;
+\frac{g}{2} \frac{m_Z}{\cos\theta_W}(1-2e_u\sin^2\theta_W)
(\sin\beta U^S_{a2}-\cos\beta U^S_{a1})]
S_a \tilde{u}_L^\ast \tilde{u}_L  ;
\end{eqnarray}
\item one scalar Higgs boson and two left-handed down-type squarks
\begin{eqnarray}
{\cal L}_{S_a\tilde{d}_L \tilde{d}_L} & = &
[-\frac{gm_d^2}{m_W\cos\beta} U^S_{a1}
\nonumber \\ & & \;
-\frac{g}{2}\frac{m_Z}{\cos\theta_W}(1+2e_d\sin^2\theta_W)
(\sin\beta U^S_{a2}-\cos\beta U^S_{a1})]
S_a \tilde{d}_L^\ast \tilde{d}_L ;
\end{eqnarray}
\item one scalar Higgs boson and two right-handed up-type squarks
\begin{eqnarray}
{\cal L}_{S_a\tilde{u}_R \tilde{u}_R} & = &
[-\frac{gm_u^2}{m_W\sin\beta} U^S_{a2}
\nonumber \\ & & \;
+gm_We_u\tan^2\theta_W
(\sin\beta U^S_{a2}-\cos\beta U^S_{a1})]
S_a \tilde{u}_R^\ast \tilde{u}_R ;
\end{eqnarray}
\item one scalar Higgs boson and two right-handed down-type squarks
\begin{eqnarray}
{\cal L}_{S_a\tilde{d}_R \tilde{d}_R} & = &
[-\frac{gm_d^2}{m_W\cos\beta} U^S_{a1}
\nonumber \\ & & \;
+gm_We_d\tan^2\theta_W
(\sin\beta U^S_{a2}-\cos\beta U^S_{a1})]
S_a \tilde{d}_R^\ast \tilde{d}_R ;
\end{eqnarray}
\item one scalar Higgs boson and one left-handed and one right-handed
up-type squark
\begin{eqnarray}
\label{sulur}
{\cal L}_{S_a\tilde{u}_L \tilde{u}_R} & = &
-\frac{gm_u}{2m_W\sin\beta} \left(\lambda \left( v_1 U^S_{a3} +
xU^S_{a1}\right) + A_U U_{a2}^S \right)
S_a \tilde{u}_R^\ast \tilde{u}_L
+\mbox{h.c.} \; ;
\end{eqnarray}
\item one scalar Higgs boson and one left-handed and one right-handed
down-type squark
\begin{eqnarray}
{\cal L}_{S_a\tilde{d}_L \tilde{d}_R} & = &
-\frac{gm_d}{2m_W\cos\beta}\left( \lambda \left( v_2 U^S_{a3} +
xU^S_{a2}\right) + A_D U_{a1}^S \right)
S_a \tilde{d}_R^\ast \tilde{d}_L
+\mbox{h.c.} \; ;
\end{eqnarray}
\item one pseudoscalar Higgs boson and one left-handed and one
right-handed up-type squark
\begin{eqnarray}
{\cal L}_{P_\alpha\tilde{u}_L \tilde{u}_R} & = &
\frac{igm_u}{2m_W\sin\beta}\left( \lambda \left( v_1 U^P_{\alpha 3} +
xU^P_{\alpha 1}\right) - A_U U_{\alpha 2}^P \right)
P_\alpha \tilde{u}_R^\ast \tilde{u}_L
+\mbox{h.c.} \; ;
\end{eqnarray}
\item one pseudoscalar Higgs boson and one left-handed and one
right-handed down-type squark
\begin{eqnarray}
\label{pdldr}
{\cal L}_{P_\alpha\tilde{d}_L \tilde{d}_R} & = &
\frac{igm_d}{2m_W\cos\beta}\left( \lambda \left( v_2 U^P_{\alpha 3} +
xU^P_{\alpha 2}\right) - A_D U_{\alpha 1}^P \right)
P_\alpha \tilde{d}_R^\ast \tilde{d}_L
+\mbox{h.c.} \; \; .
\end{eqnarray}
\end{enumerate}
The corresponding
Feynman rules are shown in Figs.~\ref{figssusu} -- \ref{figpsqsq}.
The vertex factors with one right-handed and one left-handed scalar quark in
eqs.~(\ref{sulur}) -- (\ref{pdldr})
explicitly depend on
the singlet components $U^S_{a3}$, $U^P_{\alpha 3}$ of the
neutral Higgs bosons. Therefore these couplings could be enhanced in the
NMSSM compared to the minimal model.

\begin{figure}[p]
\begin{center}
\begin{picture}(12,4)
\put(1.8,0.5){$\tilde{u}_L$}
\put(1.8,3.2){$\tilde{u}_L$}
\put(0.3,0.5){$S_b$}
\put(0.3,3.2){$S_a$}
\put(-1.0,-0.8){\includegraphics{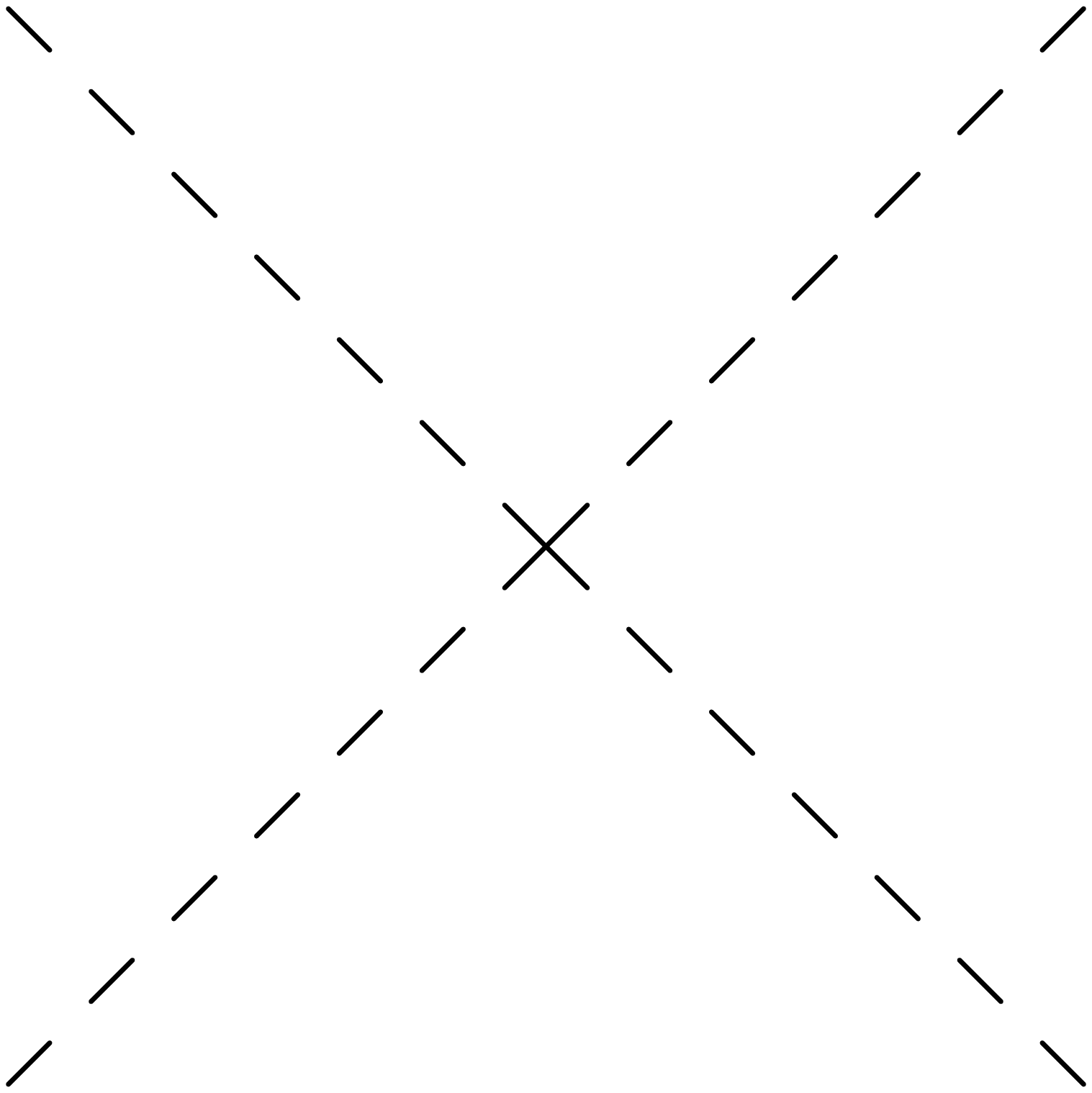}}
\put(5,2.2){$\frac{ig^2}{4}\left[ \left( \frac{1}{\cos^2\theta_W}-
2e_u\tan^2\theta_W \right)
(U^S_{a2}U^S_{b2} - U^S_{a1}U^S_{b1}) \right.$}
\put(5.7,1.4){$
-2\frac{m_u^2}{m_W^2\sin^2\beta} U^S_{a2}U^S_{b2} \; \; \; \big]
$}
\end{picture}
\begin{picture}(12,4)
\put(1.8,0.5){$\tilde{d}_L$}
\put(1.8,3.2){$\tilde{d}_L$}
\put(0.3,0.5){$S_b$}
\put(0.3,3.2){$S_a$}
\put(-1.0,-0.8){\includegraphics{hhhh.ps}}
\put(5,2.2){$\frac{-ig^2}{4}\left[ \left( \frac{1}{\cos^2\theta_W}+
2e_d\tan^2\theta_W \right)
(U^S_{a2}U^S_{b2} - U^S_{a1}U^S_{b1}) \right.$ }
\put(5.7,1.4){$
+2\frac{m_d^2}{m_W^2\cos^2\beta} U^S_{a1}U^S_{b1} \; \; \; \big]$}
\end{picture}
\begin{picture}(12,4)
\put(1.8,0.5){$\tilde{u}_R$}
\put(1.8,3.2){$\tilde{u}_R$}
\put(0.3,0.5){$S_b$}
\put(0.3,3.2){$S_a$}
\put(-1.0,-0.8){\includegraphics{hhhh.ps}}
\put(5,2.2){$\frac{ig^2}{2}[
e_u\tan^2\theta_W
(U^S_{a2}U^S_{b2} - U^S_{a1}U^S_{b1}) $ }
\put(5.7,1.4){
$ -\frac{m_u^2}{m_W^2\sin^2\beta} U^S_{a2}U^S_{b2} \; \; \; ]$}
\end{picture}
\begin{picture}(12,4)
\put(1.8,0.5){$\tilde{d}_R$}
\put(1.8,3.2){$\tilde{d}_R$}
\put(0.3,0.5){$S_b$}
\put(0.3,3.2){$S_a$}
\put(-1.0,-0.8){\includegraphics{hhhh.ps}}
\put(5,2.2){$\frac{ig^2}{2}[
e_d\tan^2\theta_W
(U^S_{a2}U^S_{b2} - U^S_{a1}U^S_{b1}) $ }
\put(5.7,1.4){
$ -\frac{m_d^2}{m_W^2\cos^2\beta} U^S_{a1}U^S_{b1} \; \; \; ]$}
\end{picture}
\end{center}
\caption{Feynman rules for the quartic interactions of two scalar
Higgs bosons and two left-handed or two right-handed scalar quarks.}
\label{figsssqsq}
\end{figure}

\begin{figure}[p]
\begin{center}
\begin{picture}(12,4)
\put(1.8,0.5){$\tilde{u}_L$}
\put(1.8,3.2){$\tilde{u}_L$}
\put(0.3,3.2){$P_{\alpha}$}
\put(0.3,0.5){$P_{\beta}$}
\put(-1.0,-0.8){\includegraphics{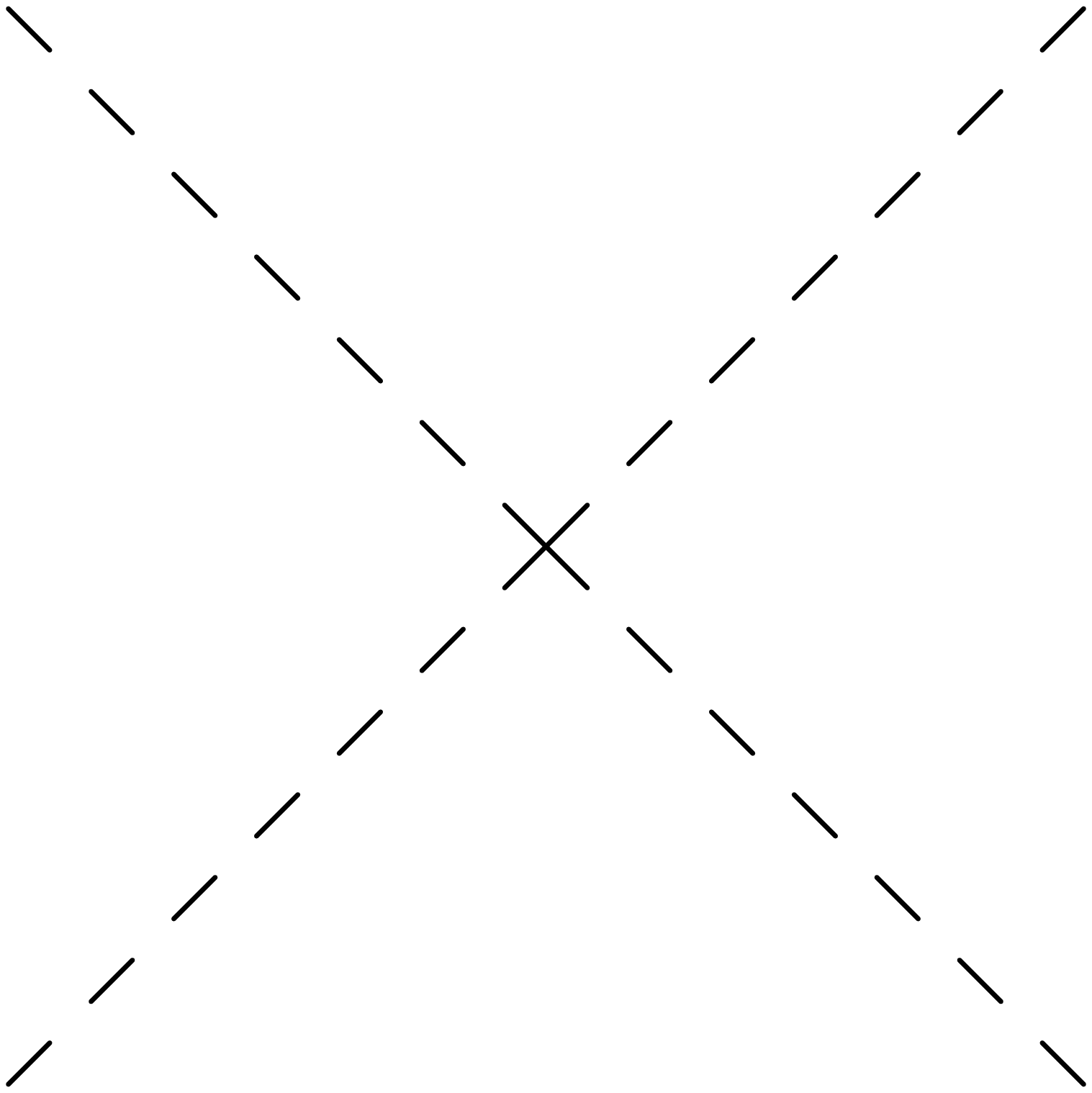}}
\put(5,2.2){$\frac{ig^2}{4}\left[\left( \frac{1}{\cos^2\theta_W}-
2e_u\tan^2\theta_W \right)
(U^P_{\alpha 2}U^P_{\beta 2} - U^P_{\alpha 1}U^P_{\beta 1}) \right. $}
\put(5.7,1.4){$
-2\frac{m_u^2}{m_W^2\sin^2\beta} U^P_{\alpha 2}U^P_{\beta 2} \; \; \; \big]$}
\end{picture}
\begin{picture}(12,4)
\put(1.8,0.5){$\tilde{d}_L$}
\put(1.8,3.2){$\tilde{d}_L$}
\put(0.3,3.2){$P_{\alpha}$}
\put(0.3,0.5){$P_{\beta}$}
\put(-1.0,-0.8){\includegraphics{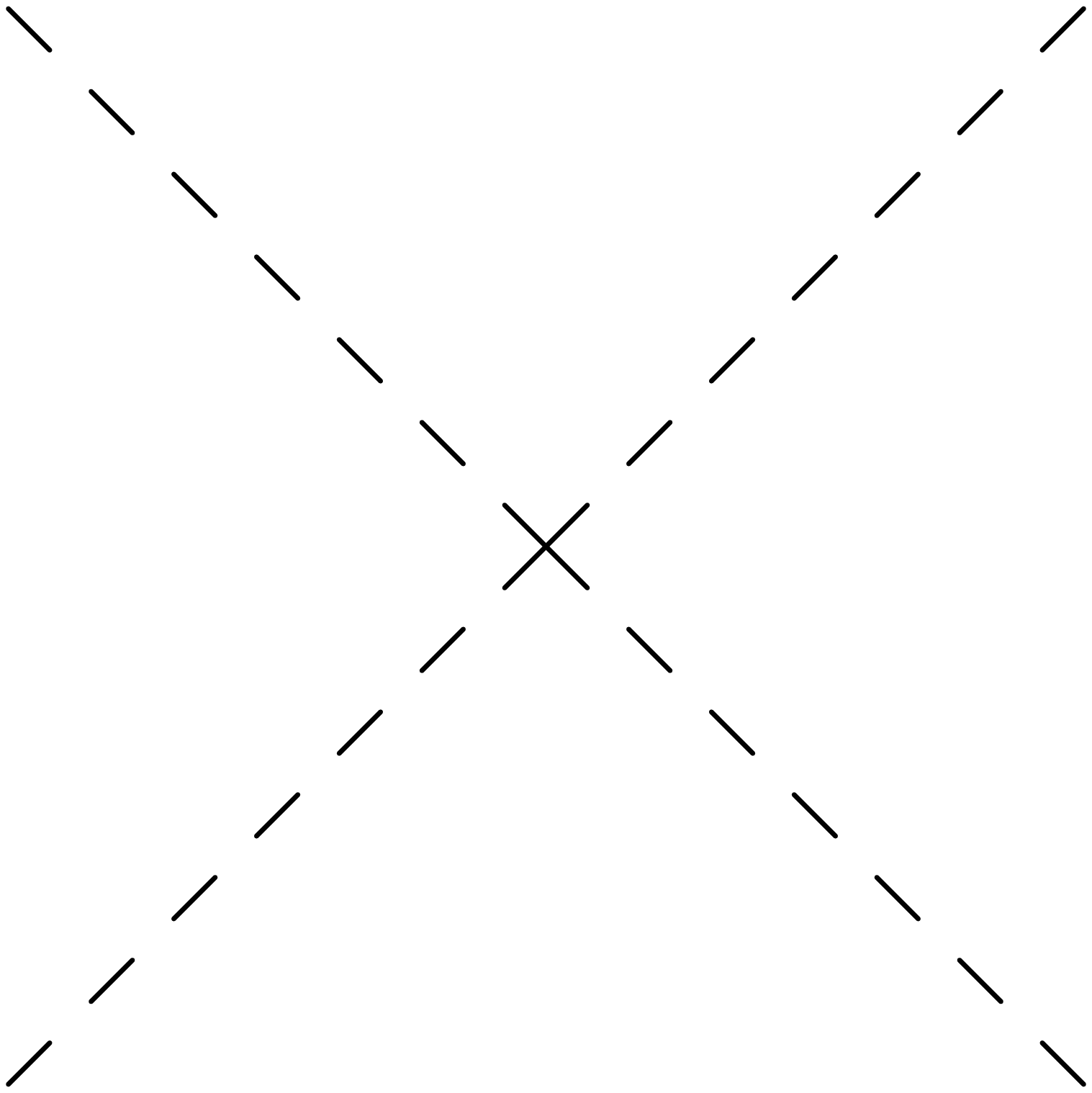}}
\put(5,2.2){$\frac{ig^2}{4}\left[ \left( \frac{1}{\cos^2\theta_W}+
2e_d\tan^2\theta_W \right)
(U^P_{\alpha 2}U^P_{\beta 2} - U^P_{\alpha 1}U^P_{\beta 1}) \right.$}
\put(5.7,1.4){$
-2\frac{m_d^2}{m_W^2\cos^2\beta} U^P_{\alpha 1}U^P_{\beta 1} \; \; \; \big]$}
\end{picture}
\begin{picture}(12,4)
\put(1.8,0.5){$\tilde{u}_R$}
\put(1.8,3.2){$\tilde{u}_R$}
\put(0.3,3.2){$P_{\alpha}$}
\put(0.3,0.5){$P_{\beta}$}
\put(-1.0,-0.8){\includegraphics{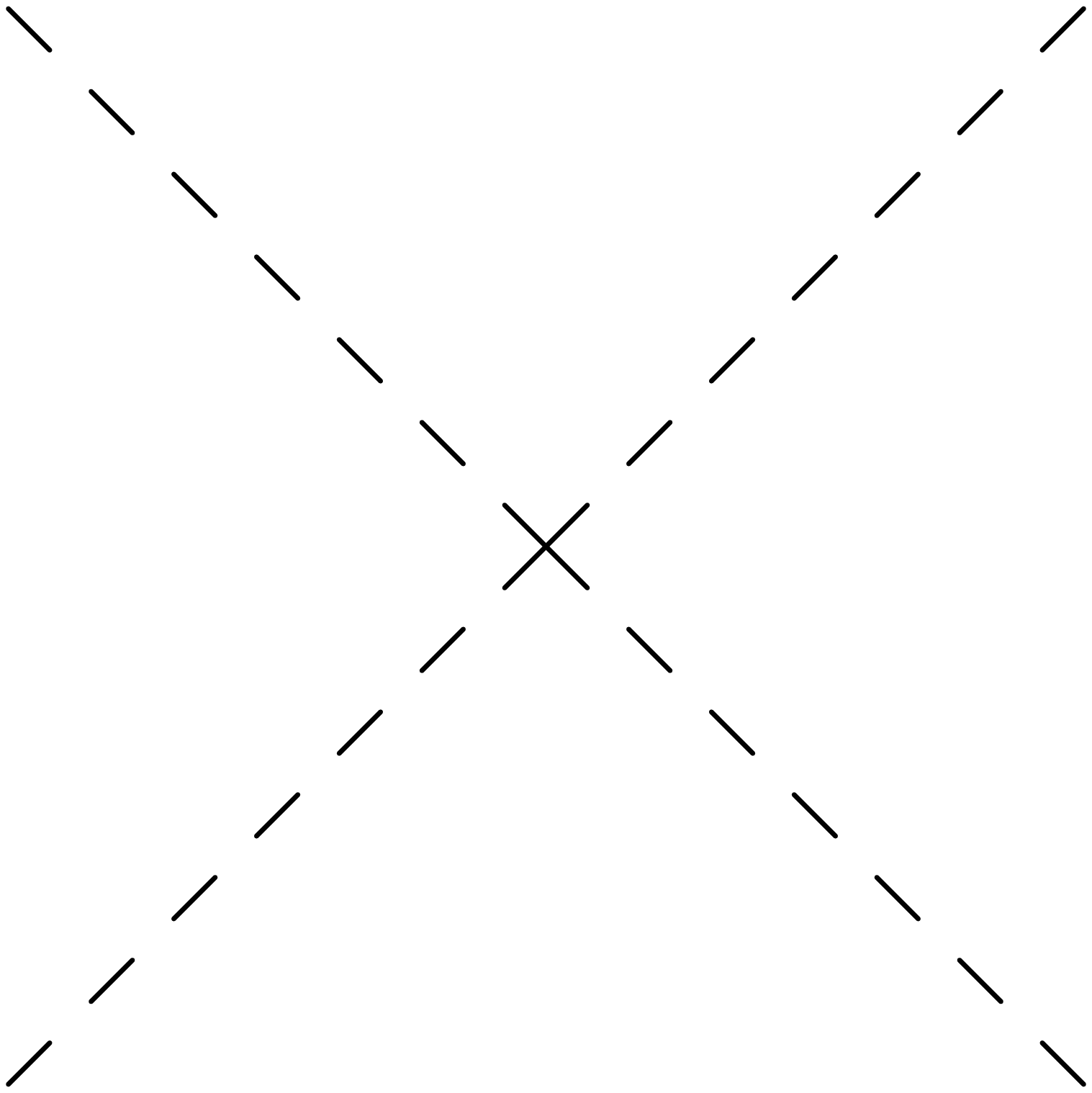}}
\put(5,2.2){$\frac{ig^2}{2}[
e_u\tan^2\theta_W
(U^P_{\alpha 2}U^P_{\beta 2} - U^P_{\alpha 1}U^P_{\beta 1}) $ }
\put(5.7,1.4){
$ -\frac{m_u^2}{m_W^2\sin^2\beta} U^P_{\alpha 2}U^P_{\beta 2} \; \; \; ]$}
\end{picture}
\begin{picture}(12,4)
\put(1.8,0.5){$\tilde{d}_R$}
\put(1.8,3.2){$\tilde{d}_R$}
\put(0.3,3.2){$P_{\alpha}$}
\put(0.3,0.5){$P_{\beta}$}
\put(-1.0,-0.8){\includegraphics{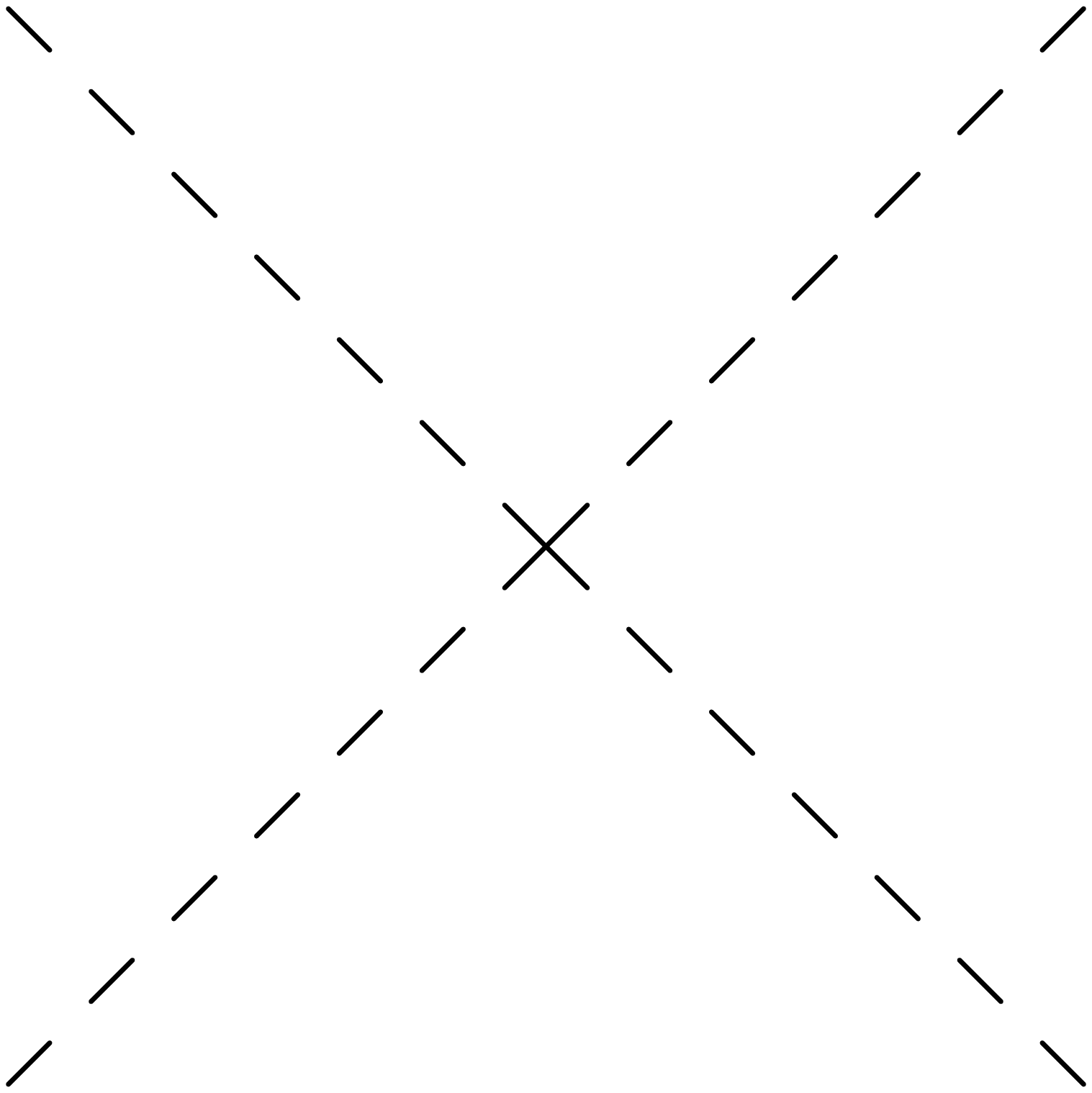}}
\put(5,2){$\frac{ig^2}{2}[
e_d\tan^2\theta_W
(U^P_{\alpha 2}U^P_{\beta 2} - U^P_{\alpha 1}U^P_{\beta 1}) $}
\put(5.7,1.2){
$ -\frac{m_d^2}{m_W^2\cos^2\beta} U^P_{\alpha 1}U^P_{\beta 1} \; \; \; ]$}
\end{picture}
\end{center}
\caption{Feynman rules for the quartic interactions of two pseudoscalar
Higgs bosons and two left-handed or two right-handed scalar quarks.}
\label{figppsqsq}
\end{figure}

\begin{figure}[p]
\begin{center}
\begin{picture}(12,4)
\put(1.8,0.5){$\tilde{u}_R$}
\put(1.8,3.2){$\tilde{u}_L$}
\put(0.3,0.5){$S_b$}
\put(0.3,3.2){$S_a$}
\put(-1.0,-0.8){\includegraphics{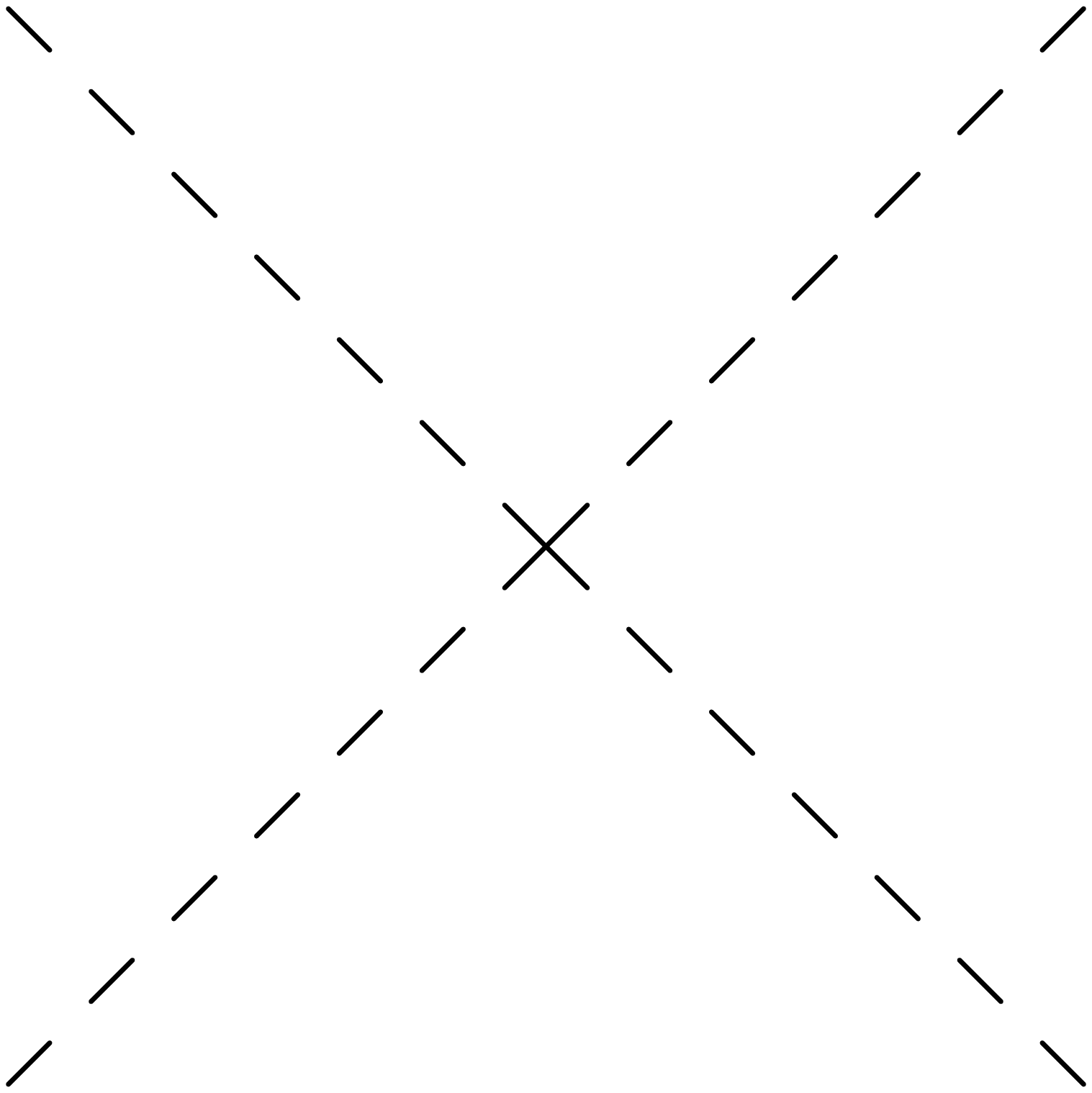}}
\put(7,2){\mbox{
$\frac{-ig\lambda m_u}{2\sqrt{2}m_W\sin\beta}
(U^S_{a1}U^S_{b3}+U^S_{a3}U^S_{b1})$}}
\end{picture}
\begin{picture}(12,4)
\put(1.8,0.5){$\tilde{d}_R$}
\put(1.8,3.2){$\tilde{d}_L$}
\put(0.3,0.5){$S_b$}
\put(0.3,3.2){$S_a$}
\put(-1.0,-0.8){\includegraphics{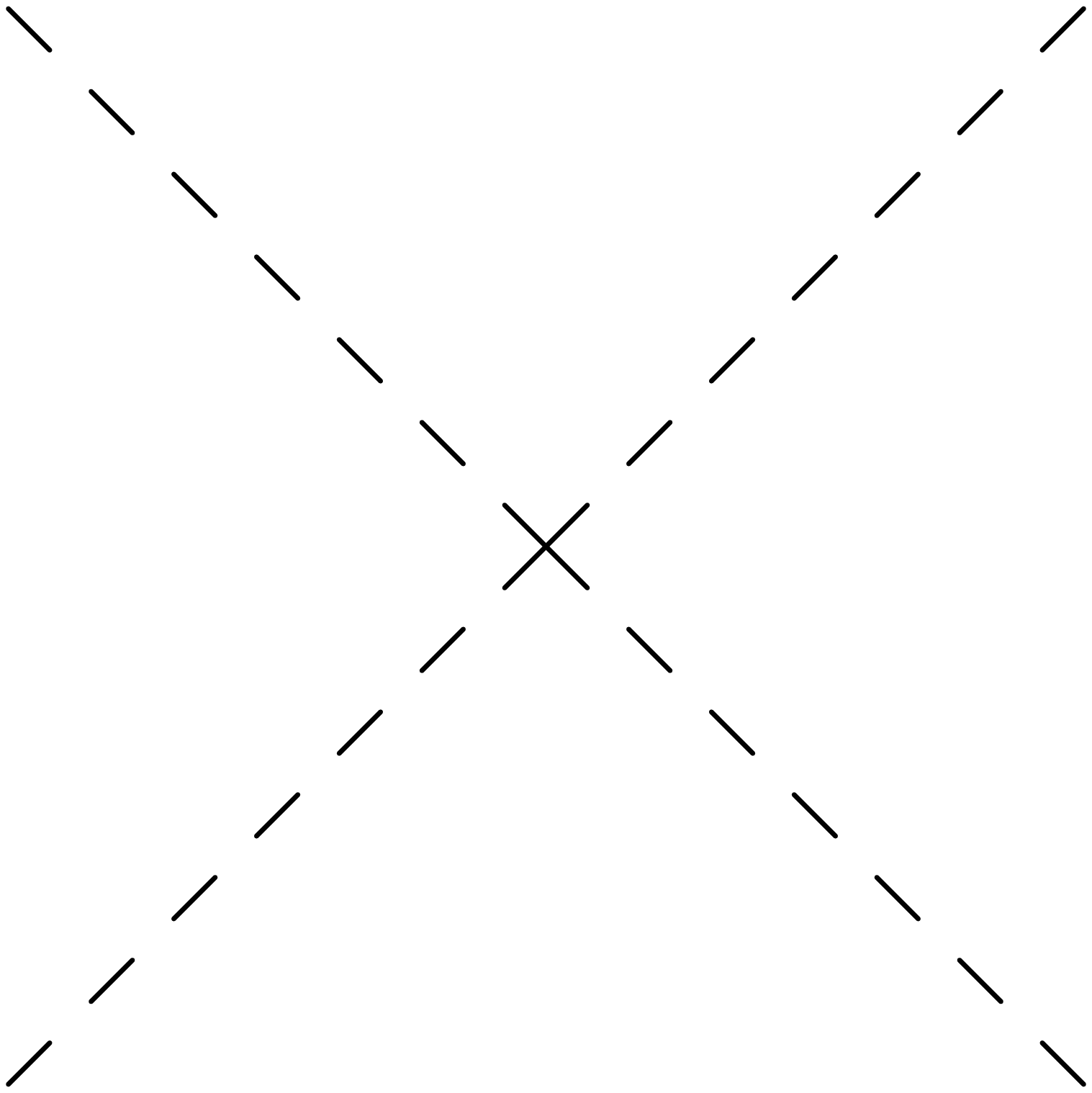}}
\put(7,2){\mbox{
$\frac{-ig\lambda m_d}{2\sqrt{2}m_W\cos\beta}
(U^S_{a2}U^S_{b3}+U^S_{a3}U^S_{b2})$}}
\end{picture}
\end{center}
\caption{Feynman rules for the quartic interactions of two scalar
Higgs bosons and one left-handed and one right-handed scalar quark.}
\label{figssslsr}
\end{figure}

\begin{figure}[p]
\begin{center}
\begin{picture}(12,4)
\put(2.8,0.5){$\tilde{u}_R$}
\put(2.8,3.2){$\tilde{u}_L$}
\put(1.3,3.2){$P_{\alpha}$}
\put(1.3,0.5){$P_{\beta}$}
\put(0,-0.8){\includegraphics{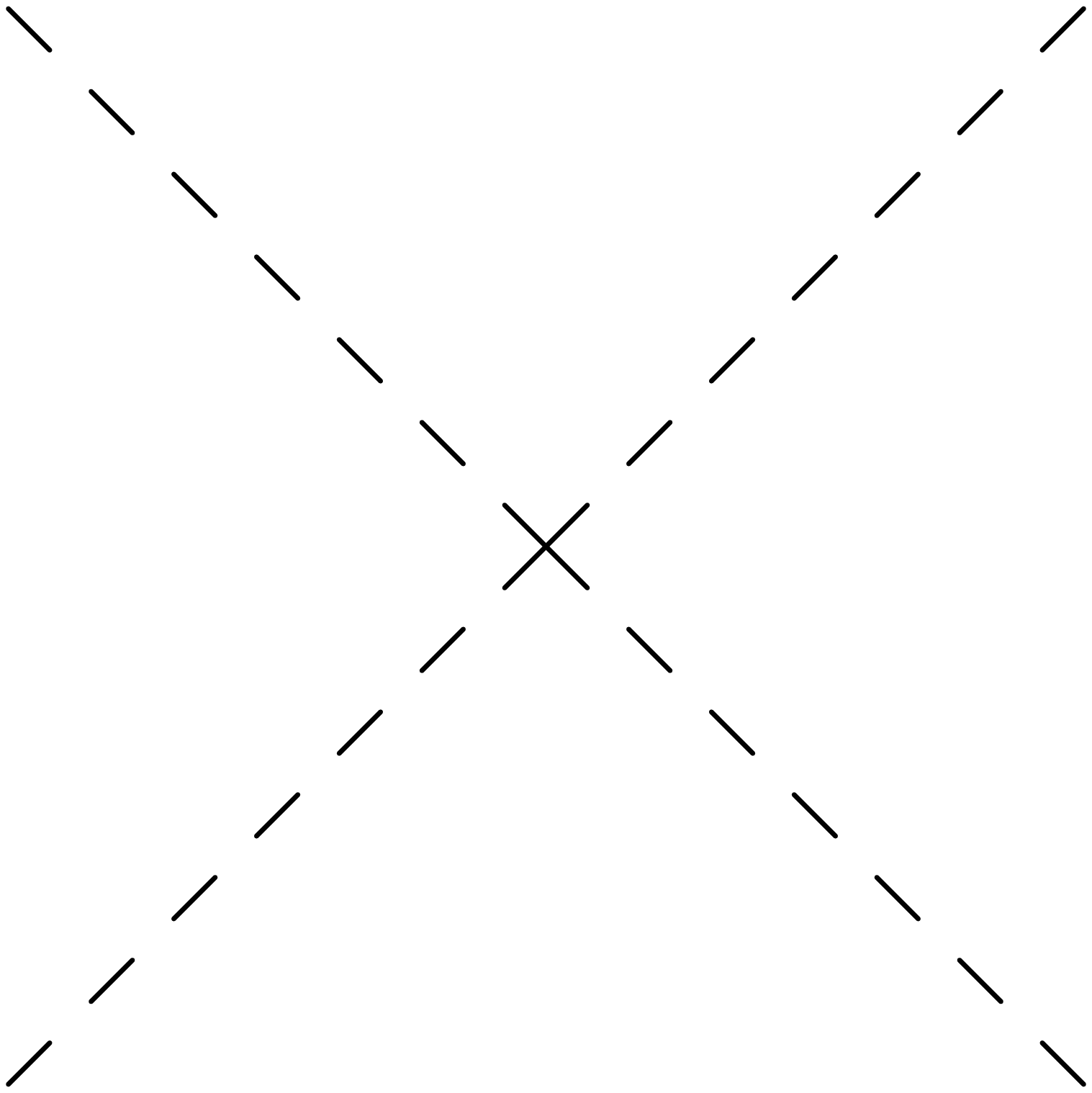}}
\put(7,2){\mbox{
$\frac{ig\lambda m_u}{2\sqrt{2}m_W\sin\beta}
(U^P_{\alpha 1}U^P_{\beta 3}+U^P_{\alpha 3}U^P_{\beta 1})$}}
\end{picture}
\begin{picture}(12,4)
\put(2.8,0.5){$\tilde{d}_R$}
\put(2.8,3.2){$\tilde{d}_L$}
\put(1.3,3.2){$P_{\alpha}$}
\put(1.3,0.5){$P_{\beta}$}
\put(0,-0.8){\includegraphics{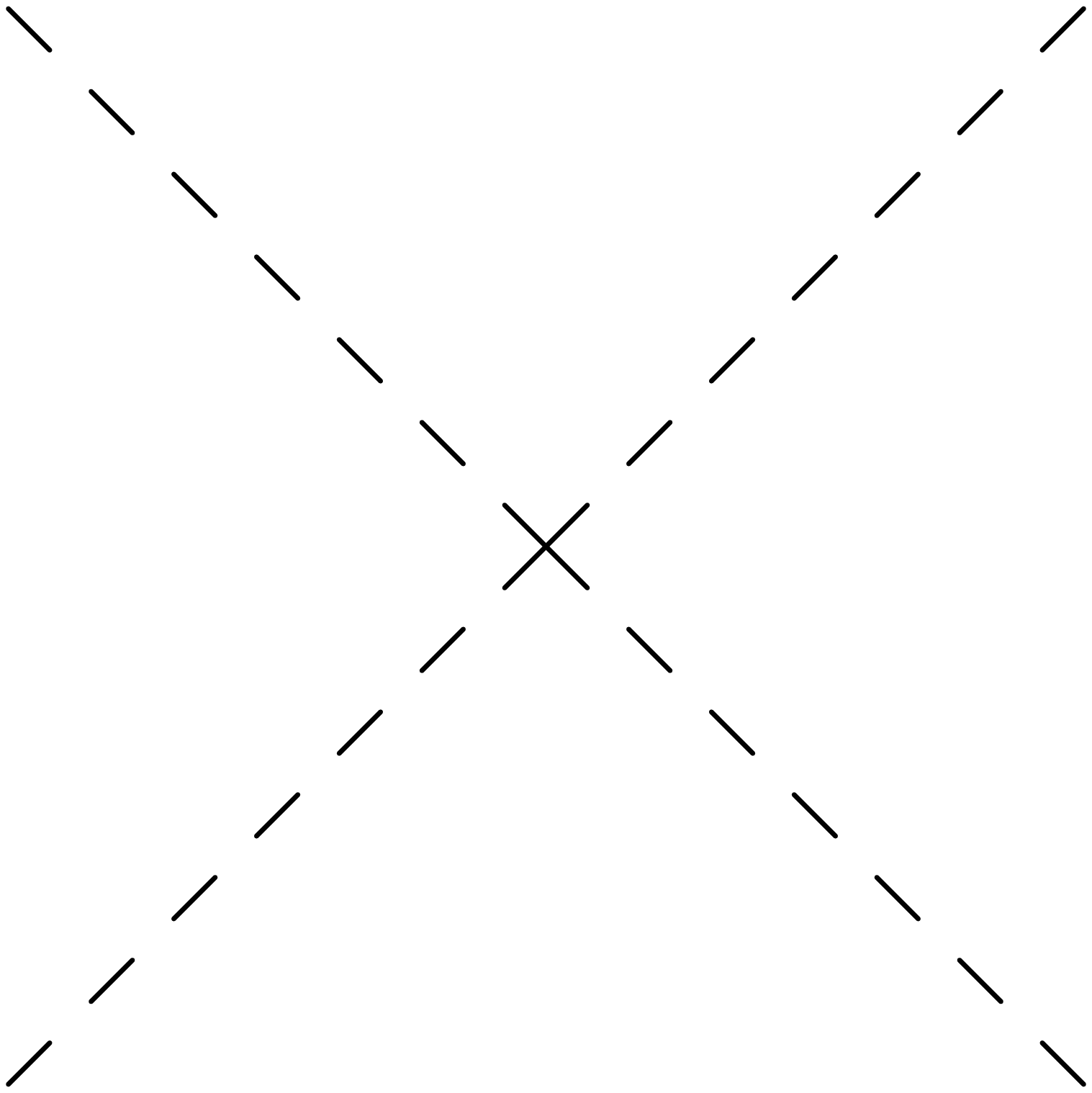}}
\put(7,2){\mbox{
$\frac{ig\lambda m_d}{2\sqrt{2}m_W\cos\beta}
(U^P_{\alpha 2}U^P_{\beta 3}+U^P_{\alpha 3}U^P_{\beta 2})$}}
\end{picture}
\end{center}
\caption{Feynman rules for the quartic interactions of two pseudoscalar
Higgs bosons and one left-handed and one right-handed scalar quark.}
\label{figppslsr}
\end{figure}

\begin{figure}[p]
\begin{center}
\begin{picture}(12,4)
\put(2.8,0.5){$\tilde{u}_R$}
\put(2.8,3.2){$\tilde{u}_L$}
\put(1.3,0.5){$P_{\alpha}$}
\put(1.3,3.2){$S_a$}
\put(0,-0.8){\includegraphics{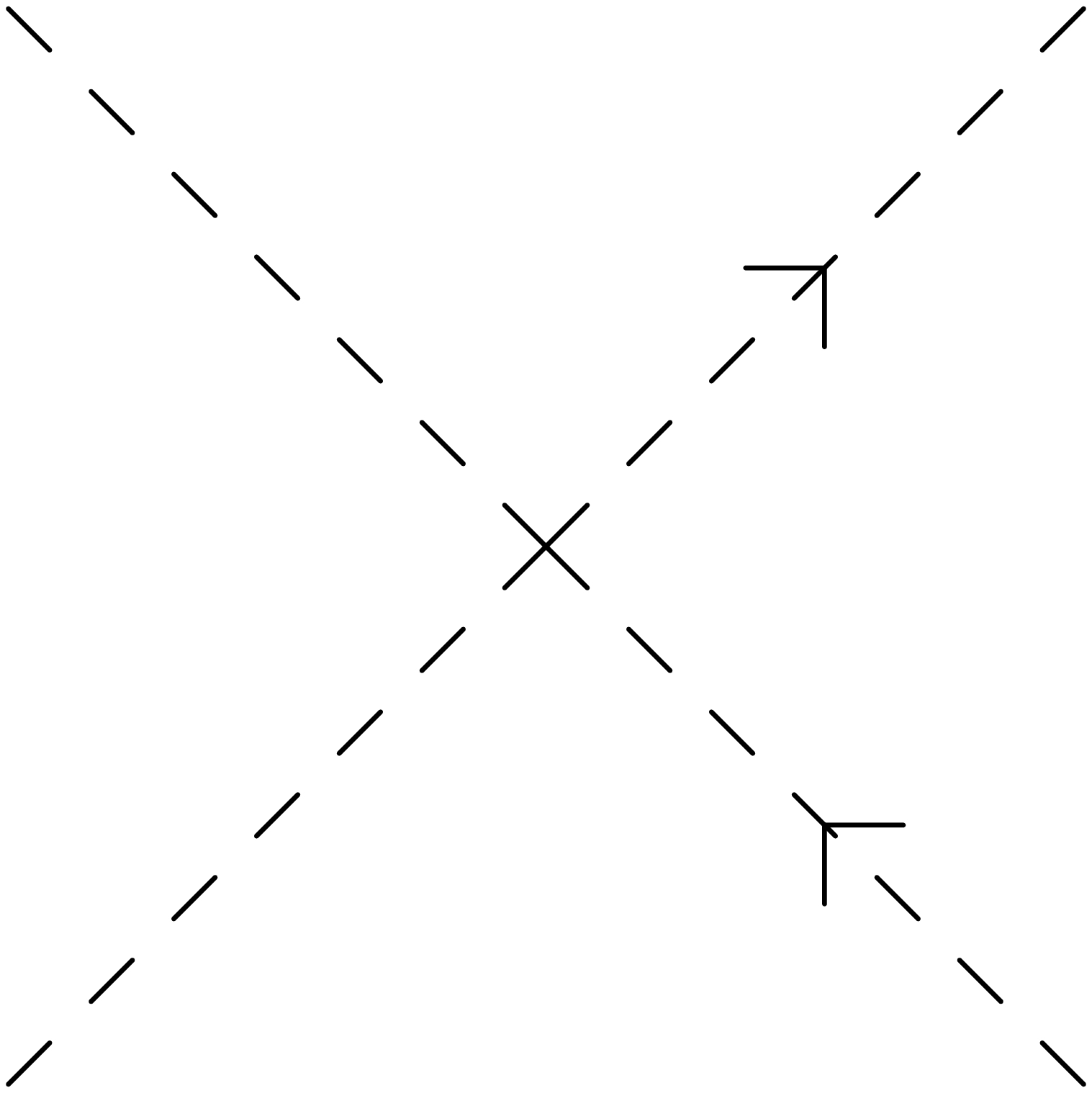}}
\put(7,2){\mbox{
$\frac{g\lambda m_u}{2\sqrt{2}m_W\sin\beta}
(U^S_{a1}U^P_{\alpha 3}+U^S_{a3}U^P_{\alpha 1})$}}
\end{picture}
\begin{picture}(12,4)
\put(2.8,0.5){$\tilde{d}_R$}
\put(2.8,3.2){$\tilde{d}_L$}
\put(1.3,0.5){$P_{\alpha}$}
\put(1.3,3.2){$S_a$}
\put(0,-0.8){\includegraphics{hhhh2.ps}}
\put(7,2){\mbox{
$\frac{g\lambda m_d}{2\sqrt{2}m_W\cos\beta}
(U^S_{a2}U^P_{\alpha 3}+U^S_{a3}U^P_{\alpha 2})$}}
\end{picture}
\end{center}
\caption{Feynman rules for the quartic interactions of one scalar and one
pseudoscalar Higgs boson and two scalar quarks.}
\label{figspslsr}
\end{figure}

\begin{figure}[p]
\begin{center}
\begin{picture}(12,4)
\put(2.8,0.5){$\tilde{d}_L$}
\put(2.8,3.2){$\tilde{u}_L$}
\put(1.3,0.5){$C^+$}
\put(1.3,3.2){$S_a$}
\put(0,-0.8){\includegraphics{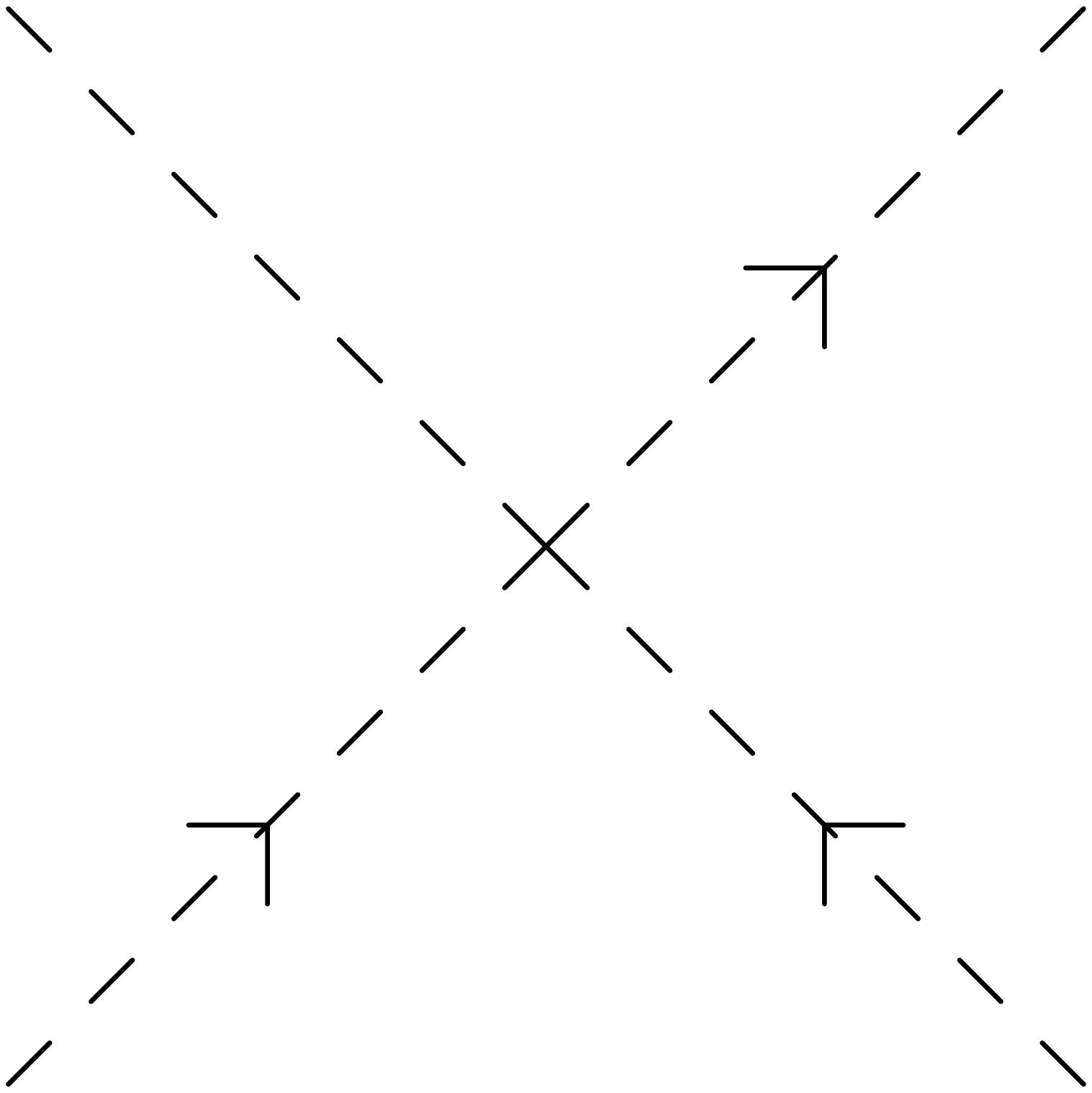}}
\put(7,2){$
\frac{-ig^2}{2\sqrt{2}} (
U^S_{a1} \sin\beta + U^S_{a2} \cos\beta$}
\put(8,1.4){$
-\frac{m_u^2}{m_W^2} \frac{\cos\beta}{\sin^2\beta}
U^S_{a2}
-\frac{m_d^2}{m_W^2} \frac{\sin\beta}{\cos^2\beta}
U^S_{a1})$}
\end{picture}
\begin{picture}(12,4)
\put(2.8,0.5){$\tilde{d}_R$}
\put(2.8,3.2){$\tilde{u}_R$}
\put(1.3,0.5){$C^+$}
\put(1.3,3.2){$S_a$}
\put(0,-0.8){\includegraphics{hhhh3.ps}}
\put(7,2){\mbox{$
\frac{ig^2m_um_d}{\sqrt{2}m_W^2\sin 2\beta} (
U^S_{a2} \sin\beta + U^S_{a1} \cos\beta)$}}
\end{picture}
\begin{picture}(12,4)
\put(2.8,0.5){$\tilde{d}_R$}
\put(2.8,3.2){$\tilde{u}_L$}
\put(1.3,0.5){$C^+$}
\put(1.3,3.2){$S_a$}
\put(0,-0.8){\includegraphics{hhhh3.ps}}
\put(7,2){\mbox{$
\frac{-ig\lambda m_d}{2m_W}
U^S_{a 3}$}}
\end{picture}
\begin{picture}(12,4)
\put(2.8,0.5){$\tilde{d}_L$}
\put(2.8,3.2){$\tilde{u}_R$}
\put(1.3,0.5){$C^+$}
\put(1.3,3.2){$S_a$}
\put(0,-0.8){\includegraphics{hhhh3.ps}}
\put(7,2){\mbox{$
\frac{-ig\lambda m_u}{2m_W}
U^S_{a 3}$}}
\end{picture}
\end{center}
\caption{Feynman rules for the quartic interactions of one scalar and one
charged Higgs boson and two scalar quarks.}
\label{figscsqsq}
\end{figure}

\begin{figure}[p]
\begin{center}
\begin{picture}(12,4)
\put(2.8,0.5){$\tilde{d}_L$}
\put(2.8,3.2){$\tilde{u}_L$}
\put(1.3,0.5){$C^+$}
\put(1.3,3.2){$P_{\alpha}$}
\put(0,-0.8){\includegraphics{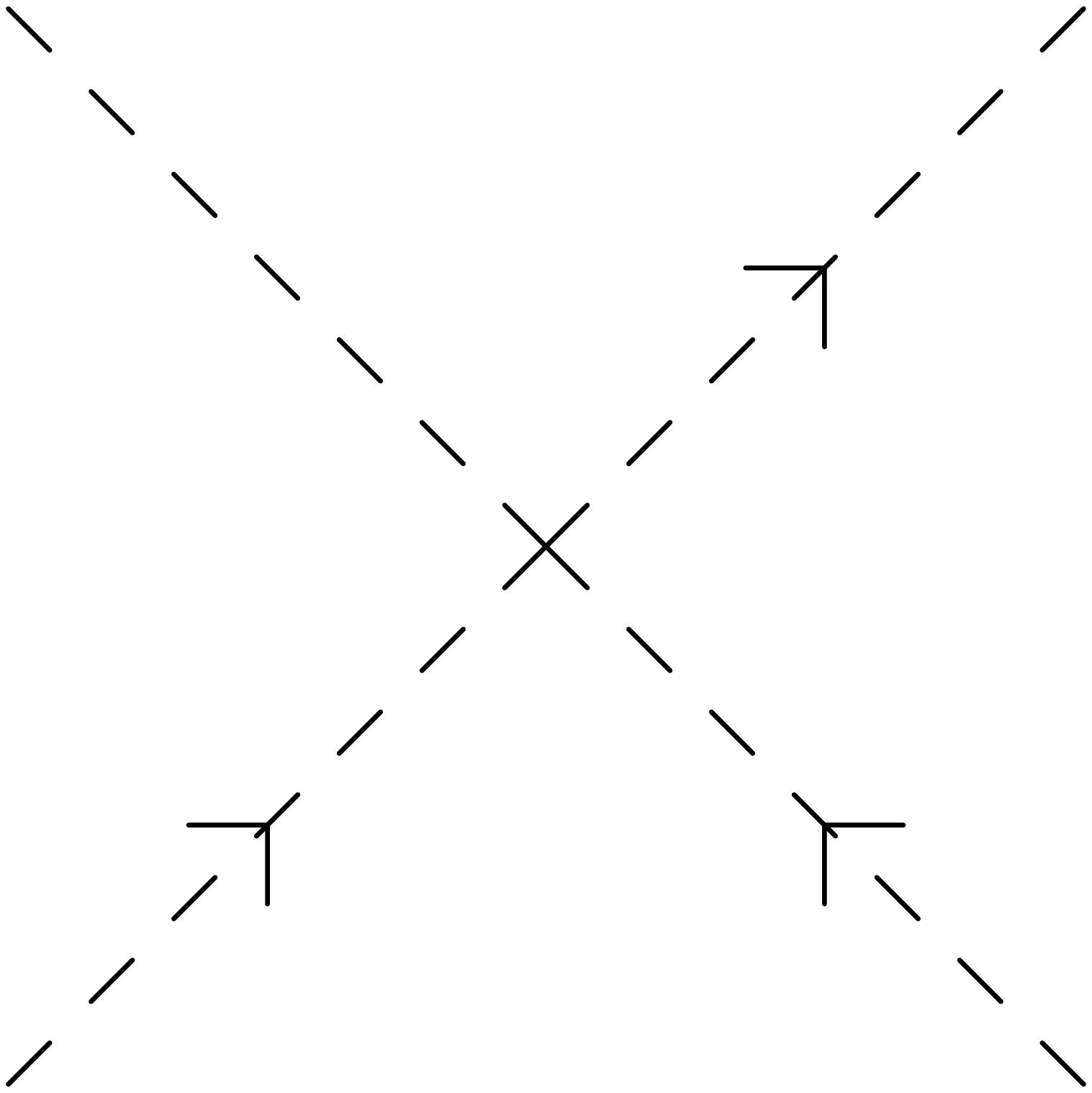}}
\put(7,2){$
\frac{g^2}{2\sqrt{2}} (
U^P_{\alpha 1} \sin\beta - U^P_{\alpha 2} \cos\beta $}
\put(8,1.4){$
+\frac{m_u^2}{m_W^2} \frac{\cos\beta}{\sin^2\beta}
U^P_{\alpha 2}
-\frac{m_d^2}{m_W^2} \frac{\sin\beta}{\cos^2\beta}
U^P_{\alpha 1}) $}
\end{picture}
\begin{picture}(12,4)
\put(2.8,0.5){$\tilde{d}_R$}
\put(2.8,3.2){$\tilde{u}_R$}
\put(1.3,0.5){$C^+$}
\put(1.3,3.2){$P_{\alpha}$}
\put(0,-0.8){\includegraphics{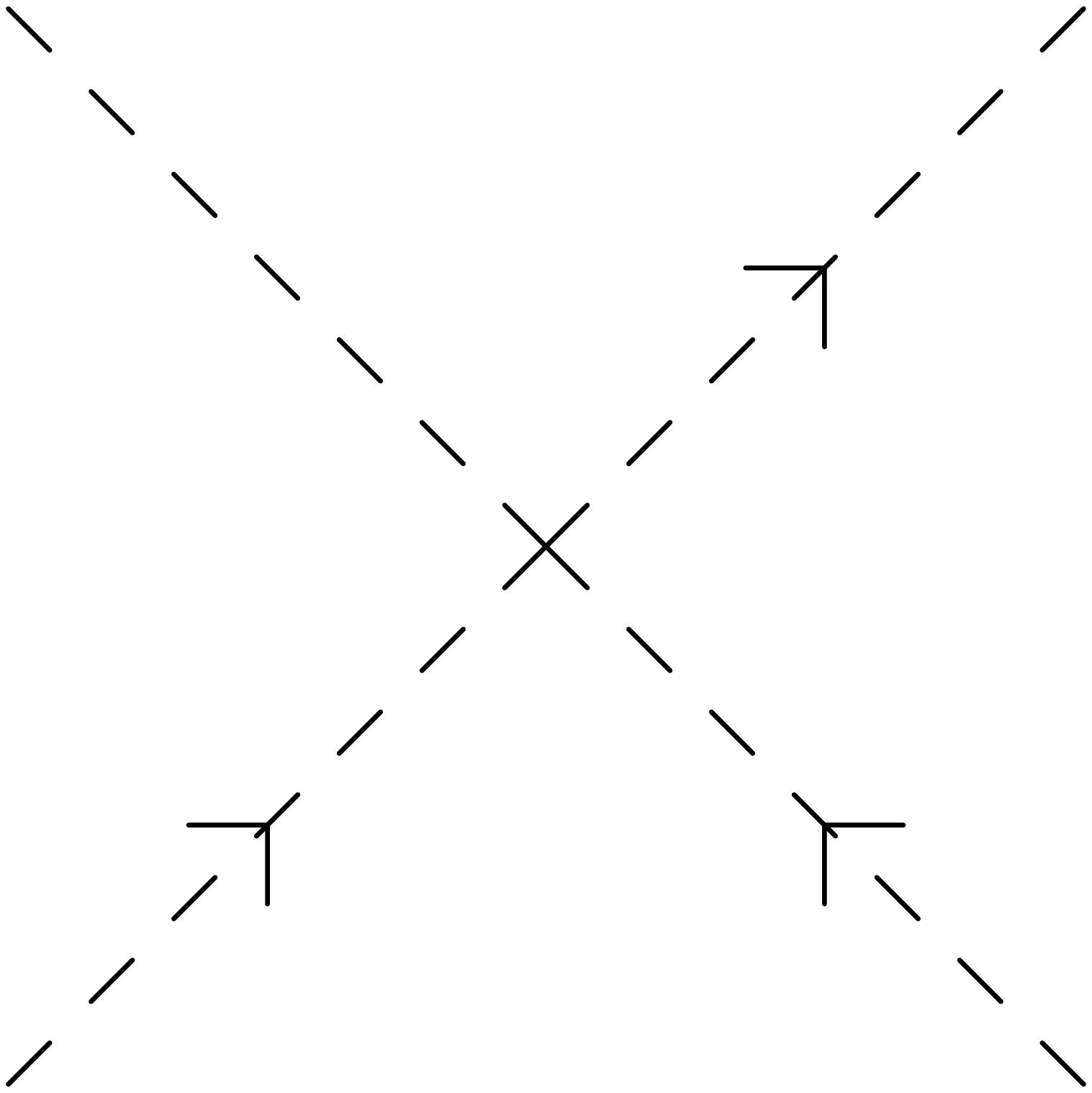}}
\put(7,2){\mbox{$
\frac{g^2m_um_d}{\sqrt{2}m_W^2\sin 2\beta} (
U^P_{\alpha 2} \sin\beta - U^P_{\alpha 1} \cos\beta)$}}
\end{picture}
\begin{picture}(12,4)
\put(2.8,0.5){$\tilde{d}_R$}
\put(2.8,3.2){$\tilde{u}_L$}
\put(1.3,0.5){$C^+$}
\put(1.3,3.2){$P_{\alpha}$}
\put(0,-0.8){\includegraphics{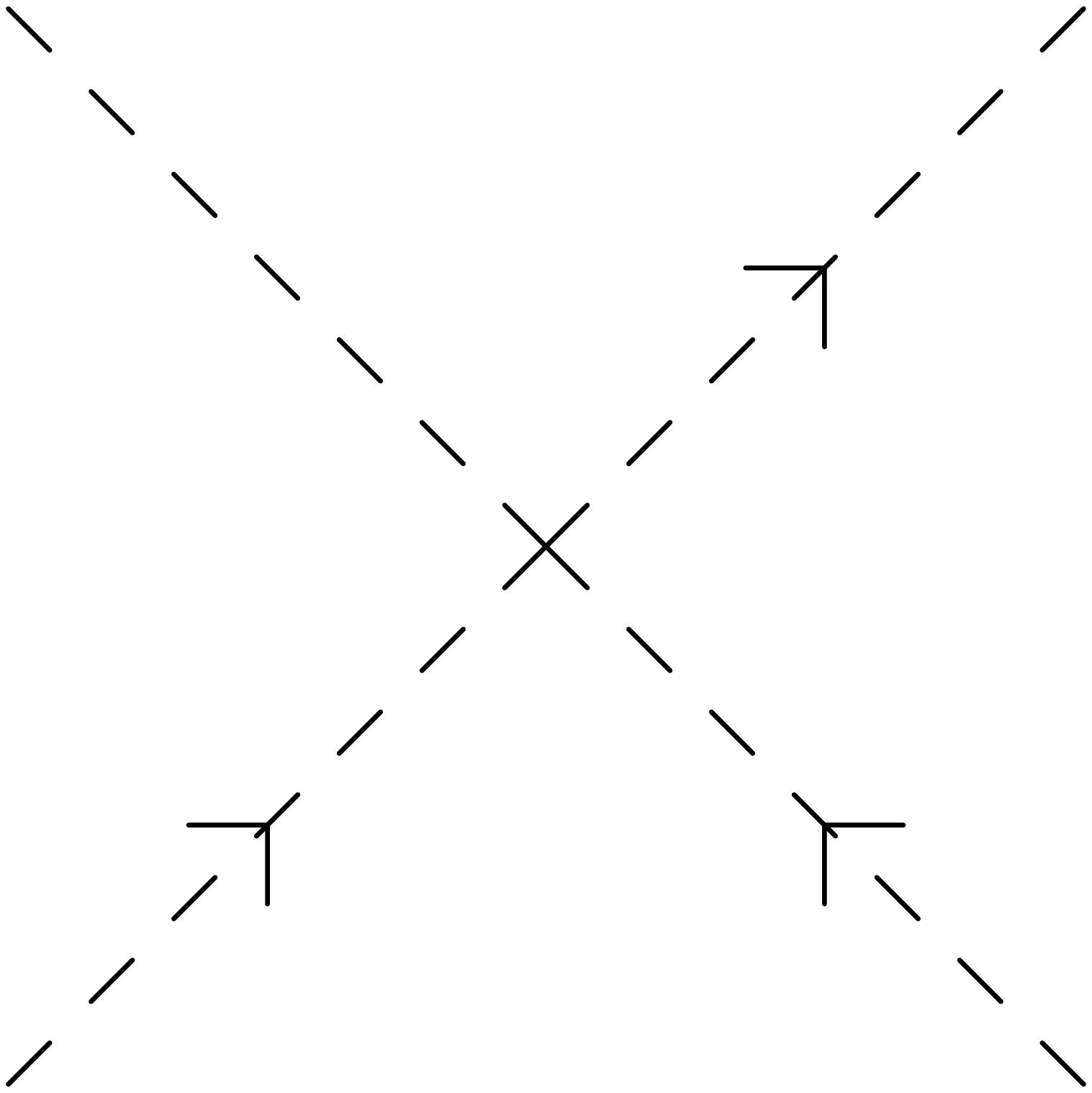}}
\put(7,2){\mbox{$
\frac{g\lambda m_d}{2m_W}
U^P_{\alpha 3}$}}
\end{picture}
\begin{picture}(12,4)
\put(2.8,0.5){$\tilde{d}_L$}
\put(2.8,3.2){$\tilde{u}_R$}
\put(1.3,0.5){$C^+$}
\put(1.3,3.2){$P_{\alpha}$}
\put(0,-0.8){\includegraphics{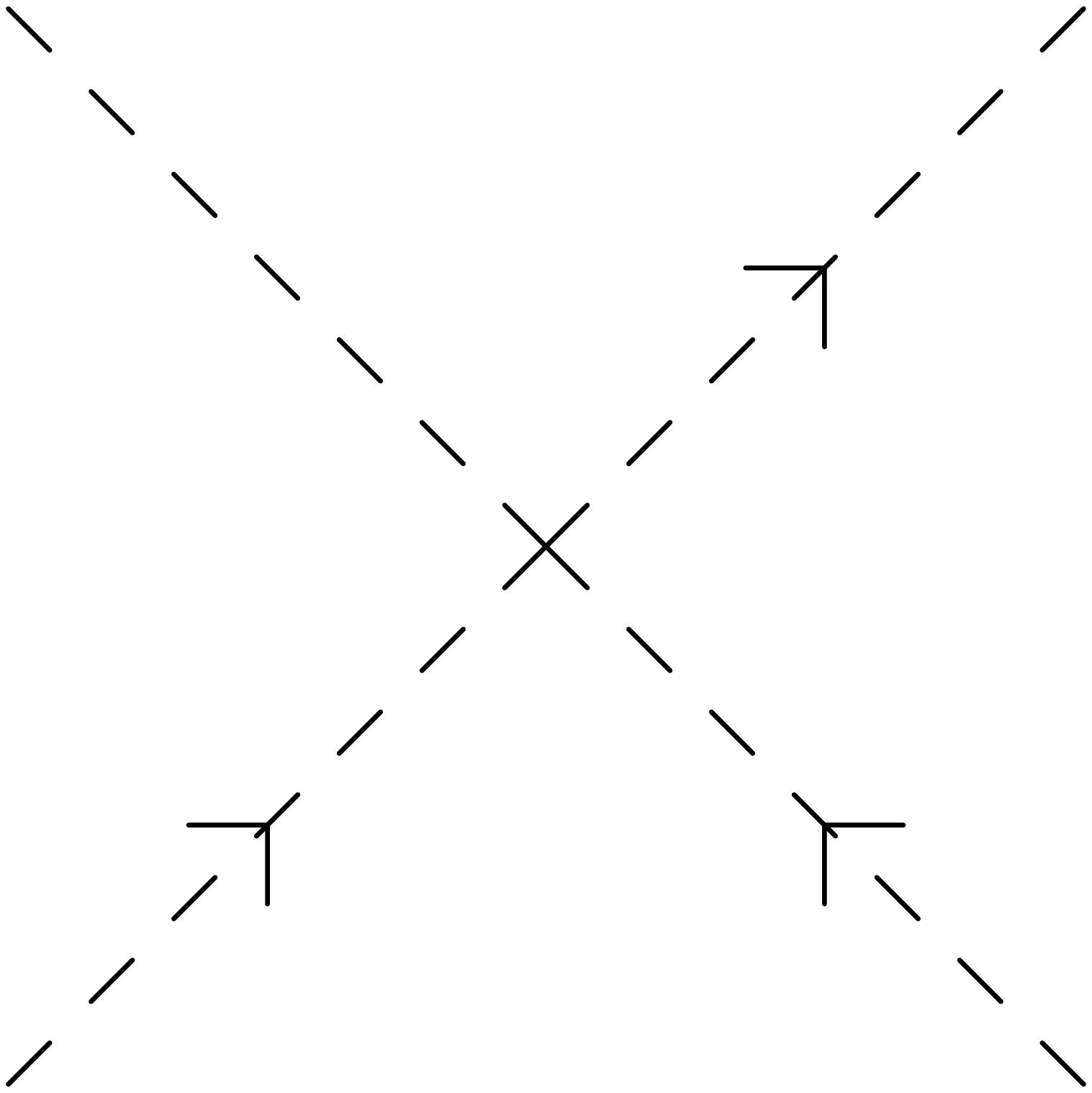}}
\put(7,2){\mbox{$
-\frac{g\lambda m_u}{2m_W}
U^P_{\alpha 3}$}}
\end{picture}
\end{center}
\caption{Feynman rules for the quartic interactions of one pseudoscalar and one
charged Higgs boson and two scalar quarks.}
\label{figpcsqsq}
\end{figure}

In the same way one obtains for the quartic interactions of
\begin{enumerate}
\item two scalar Higgs bosons and two left-handed up-type squarks
\begin{eqnarray}
{\cal L}_{S_aS_b\tilde{u}_L\tilde{u}_L} & = &
\frac{g^2}{8} \Bigg[
\left( \frac{1}{\cos^2\theta_W}-2e_u\tan^2\theta_W \right)
(U^S_{a2}U^S_{b2} - U^S_{a1}U^S_{b1})
\nonumber \\ & & \; \; \; \; \; \; \;
-2\frac{m_u^2}{m_W^2\sin^2\beta} U^S_{a2}U^S_{b2} \Bigg]
(2-\delta_{ab})S_aS_b\tilde{u}^\ast_L \tilde{u}_L ;
\end{eqnarray}
\item two scalar Higgs bosons and two left-handed down-type squarks
\begin{eqnarray}
{\cal L}_{S_aS_b\tilde{d}_L\tilde{d}_L} & = &
-\frac{g^2}{8} \Bigg[
\left( \frac{1}{\cos^2\theta_W}+2e_d\tan^2\theta_W \right)
(U^S_{a2}U^S_{b2} - U^S_{a1}U^S_{b1})
\nonumber \\ & & \; \; \; \; \; \; \;
+2\frac{m_d^2}{m_W^2\cos^2\beta} U^S_{a1}U^S_{b1} \Bigg]
(2-\delta_{ab})S_aS_b\tilde{d}^\ast_L \tilde{d}_L ;
\end{eqnarray}
\item two scalar Higgs bosons and two right-handed up-type squarks
\begin{eqnarray}
{\cal L}_{S_aS_b\tilde{u}_R\tilde{u}_R} & = &
\frac{g^2}{4} \Bigg[
e_u\tan^2\theta_W
(U^S_{a2}U^S_{b2} - U^S_{a1}U^S_{b1})
\nonumber \\ & & \; \; \; \; \; \; \;
-\frac{m_u^2}{m_W^2\sin^2\beta} U^S_{a2}U^S_{b2} \Bigg]
(2-\delta_{ab})S_aS_b\tilde{u}^\ast_R \tilde{u}_R ;
\end{eqnarray}
\item two scalar Higgs bosons and two right-handed down-type squarks
\begin{eqnarray}
{\cal L}_{S_aS_b\tilde{d}_R\tilde{d}_R} & = &
\frac{g^2}{4} \Bigg[
e_d\tan^2\theta_W
(U^S_{a2}U^S_{b2} - U^S_{a1}U^S_{b1})
\nonumber \\ & & \; \; \; \; \; \; \;
-\frac{m_d^2}{m_W^2\cos^2\beta} U^S_{a1}U^S_{b1} \Bigg]
(2-\delta_{ab})S_aS_b\tilde{d}^\ast_R \tilde{d}_R ;
\end{eqnarray}
\item two scalar Higgs bosons and one left-handed and one right-handed
up-type squark
\begin{eqnarray}
\label{ssulur}
{\cal L}_{S_aS_b\tilde{u}_L\tilde{u}_R} & = &
-\lambda \frac{gm_u}{4\sqrt{2}m_W\sin\beta }
(U^S_{a1}U^S_{b3}+U^S_{b1} U^S_{a3})(2-\delta_{ab})
S_a S_b \tilde{u}^\ast_L \tilde{u}_R \nonumber \\ & &
+\mbox{h.c.} \; ;
\end{eqnarray}
\item two scalar Higgs bosons and one left-handed and one right-handed
down-type squark
\begin{eqnarray}
\label{ssdldr}
{\cal L}_{S_aS_b\tilde{d}_L\tilde{d}_R} & = &
-\lambda \frac{gm_d}{4\sqrt{2}m_W\cos\beta }
(U^S_{a2}U^S_{b3}+U^S_{a3} U^S_{b2})
(2-\delta_{ab})S_a S_b \tilde{d}^\ast_L \tilde{d}_R
\nonumber \\ & &
+\mbox{h.c.} \; ;
\end{eqnarray}
\item two pseudoscalar Higgs bosons and two left-handed up-type squarks
\begin{eqnarray}
{\cal L}_{P_\alpha P_\beta\tilde{u}_L\tilde{u}_L} & = &
\frac{g^2}{8} \Bigg[
\left( \frac{1}{\cos^2\theta_W}-2e_u\tan^2\theta_W \right)
(U^P_{\alpha 2}U^P_{\beta 2} - U^P_{\alpha 1}U^P_{\beta 1})
\nonumber \\ & & \; \; \; \; \; \; \;
-2\frac{m_u^2}{m_W^2\sin^2\beta} U^P_{\alpha 2}U^P_{\beta 2} \Bigg]
(2-\delta_{\alpha\beta })P_\alpha P_\beta \tilde{u}^\ast_L \tilde{u}_L ;
\end{eqnarray}
\item two pseudoscalar Higgs bosons and two left-handed down-type squarks
\begin{eqnarray}
{\cal L}_{P_\alpha P_\beta\tilde{d}_L\tilde{d}_L} & = &
\frac{g^2}{8} \Bigg[
\left( \frac{1}{\cos^2\theta_W}+2e_d\tan^2\theta_W \right)
(U^P_{\alpha 2}U^P_{\beta 2} - U^P_{\alpha 1}U^P_{\beta 1})
\nonumber \\ & & \; \; \; \; \; \; \;
-2\frac{m_u^2}{m_W^2\cos^2\beta} U^P_{\alpha 1}U^P_{\beta 1} \Bigg]
(2-\delta_{\alpha\beta })P_\alpha P_\beta \tilde{d}^\ast_L \tilde{d}_L ;
\end{eqnarray}
\item two pseudoscalar Higgs bosons and two right-handed up-type squarks
\begin{eqnarray}
{\cal L}_{P_\alpha P_\beta\tilde{u}_R\tilde{u}_R} & = &
\frac{g^2}{4} \Big[
e_u\tan^2\theta_W
(U^P_{\alpha 2}U^P_{\beta 2} - U^P_{\alpha 1}U^P_{\beta 1})
\nonumber \\ & & \; \; \; \; \; \; \;
-\frac{m_u^2}{m_W^2\sin^2\beta} U^P_{\alpha 2}U^P_{\beta 2} \Big]
(2-\delta_{\alpha\beta })P_\alpha P_\beta \tilde{u}^\ast_R \tilde{u}_R ;
\end{eqnarray}
\item two pseudoscalar Higgs bosons and two right-handed down-type squarks
\begin{eqnarray}
{\cal L}_{P_\alpha P_\beta\tilde{d}_R\tilde{d}_R} & = &
\frac{g^2}{4} \Big[
e_d\tan^2\theta_W
(U^P_{\alpha 2}U^P_{\beta 2} - U^P_{\alpha 1}U^P_{\beta 1})
\nonumber \\ & & \; \; \; \; \; \; \;
-\frac{m_d^2}{m_W^2\cos^2\beta} U^P_{\alpha 1}U^P_{\beta 1} \Big]
(2-\delta_{\alpha\beta })P_\alpha P_\beta \tilde{d}^\ast_R \tilde{d}_R ;
\end{eqnarray}
\item two pseudoscalar Higgs bosons and one left-handed and one
right-handed up-type squark
\begin{eqnarray}
\label{ppulur}
{\cal L}_{P_\alpha P_\beta \tilde{u}_L\tilde{u}_R} & = &
\lambda \frac{gm_u}{4\sqrt{2}m_W\sin\beta }
( U^P_{\alpha 1}U^P_{\beta 3}
+U^P_{\alpha 3}U^P_{\beta 1})(2-\delta_{\alpha\beta})
P_\alpha P_\beta
\tilde{u}^\ast_L \tilde{u}_R  \nonumber \\ & &
+\mbox{h.c.} \; ;
\end{eqnarray}
\item two pseudoscalar Higgs bosons and one left-handed and one
right-handed down-type squark
\begin{eqnarray}
{\cal L}_{P_\alpha P_\beta\tilde{d}_L\tilde{d}_R} & = &
\lambda \frac{gm_d}{4\sqrt{2}m_W\cos\beta }
(U^P_{\alpha 2}U^P_{\beta 3}+U^S_{\alpha 3} U^S_{\beta 2})
(2-\delta_{\alpha\beta})P_\alpha P_\beta
\tilde{d}^\ast_L \tilde{d}_R \nonumber \\ & &
+\mbox{h.c.} \; ;
\label{ppdldr}
\end{eqnarray}
\item one scalar and one pseudoscalar Higgs boson and one left-handed and one
right-handed up-type squark
\begin{eqnarray}
{\cal L}_{S_a P_\alpha \tilde{u}_L \tilde{u}_R} & = &
-i\lambda \frac{gm_u}{2\sqrt{2}m_W\sin\beta}
(U^S_{a1}U^P_{\alpha 3} + U^S_{a3} U^P_{\alpha 1})
S_a P_\alpha \tilde{u}_L^\ast \tilde{u}_R
+ \mbox{h.c.} \; ;
\label{spulur}
\end{eqnarray}
\item one scalar and one pseudoscalar Higgs boson and one left-handed and one
right-handed down-type squark
\begin{eqnarray}
{\cal L}_{S_a P_\alpha \tilde{d}_L \tilde{d}_R} & = &
-i\lambda \frac{gm_d}{2\sqrt{2}m_W\cos\beta}
(U^S_{a2}U^P_{\alpha 3} + U^S_{a3} U^P_{\alpha 2})
S_a P_\alpha \tilde{d}_L^\ast \tilde{d}_R
+ \mbox{h.c.} \; ;
\label{spdldr}
\end{eqnarray}
\item one neutral scalar Higgs boson, one charged Higgs boson, one
left-handed up-type squark and one left-handed down-type squark
\begin{eqnarray}
{\cal L}_{S_aC\tilde{u}_L\tilde{d}_L} & = &
-\frac{g^2}{2\sqrt{2}}
\Big( U^S_{a1} \sin\beta + U^S_{a2} \cos\beta
-\frac{m_u^2}{m_W^2} \frac{\cos\beta}{\sin^2\beta}
U^S_{a2}
\nonumber \\ & & \hspace{1.5cm}
-\frac{m_d^2}{m_W^2} \frac{\sin\beta}{\cos^2\beta}
U^S_{a1}\Big) S_a C^+ \tilde{u}_L^\ast \tilde{d}_L
+\mbox{h.c.} \; ;
\end{eqnarray}
\item one neutral scalar Higgs boson, one charged Higgs boson, one
right-handed up-type squark and one right-handed down-type squark
\begin{eqnarray}
{\cal L}_{S_aC\tilde{u}_R\tilde{d}_R} & = &
\frac{g^2m_um_d}{\sqrt{2}m_W^2\sin 2\beta}
(U^S_{a2} \sin\beta + U^S_{a1} \cos\beta )
S_a C^+ \tilde{u}_R^\ast \tilde{d}_R
+\mbox{h.c.} \; ;
\end{eqnarray}
\item one neutral scalar Higgs boson, one charged Higgs boson, one
left-handed up-type squark and one right-handed down-type squark
\begin{eqnarray}
{\cal L}_{S_a C \tilde{u}_L \tilde{d}_R} & = &
-\lambda \frac{gm_d}{2m_W} U_{a3}^S
S_a C^+ \tilde{u}^\ast_L \tilde{d}_R
+\mbox{h.c.} \; ;
\label{sculdr}
\end{eqnarray}
\item one neutral scalar Higgs boson, one charged Higgs boson, one
left-handed down-type squark and one right-handed up-type squark
\begin{eqnarray}
{\cal L}_{S_a C \tilde{u}_R \tilde{d}_L} & = &
-\lambda \frac{gm_u}{2m_W} U_{a3}^S
S_a C^+ \tilde{u}^\ast_R \tilde{d}_L
+\mbox{h.c.} \; ;
\label{scurdl}
\end{eqnarray}
\item one neutral pseudoscalar Higgs boson, one charged Higgs boson, one
left-handed up-type squark and one left-handed down-type squark
\begin{eqnarray}
{\cal L}_{P_\alpha C\tilde{u}_L\tilde{d}_L} & = &
-\frac{ig^2}{2\sqrt{2}}
\Big( U^P_{\alpha 1} \sin\beta - U^P_{\alpha 2} \cos\beta
+\frac{m_u^2}{m_W^2} \frac{\cos\beta}{\sin^2\beta}
U^P_{\alpha 2}
\nonumber \\ & & \hspace{1.5cm}
-\frac{m_d^2}{m_W^2} \frac{\sin\beta}{\cos^2\beta}
U^P_{\alpha 1}\Big) P_\alpha C^+ \tilde{u}_L^\ast \tilde{d}_L
+\mbox{h.c.} \; ;
\end{eqnarray}
\item one neutral pseudoscalar Higgs boson, one charged Higgs boson, one
right-handed up-type squark and one right-handed down-type squark
\begin{eqnarray}
{\cal L}_{P_\alpha C\tilde{u}_R\tilde{d}_R} & = &
i\frac{g^2m_um_d}{\sqrt{2}m_W^2\sin 2\beta}
(U^P_{\alpha 2} \sin\beta - U^P_{\alpha 1} \cos\beta )
P_\alpha C^+ \tilde{u}_R^\ast \tilde{d}_R
+\mbox{h.c.} \; ;
\end{eqnarray}
\item one neutral pseudoscalar Higgs boson, one charged Higgs boson, one
left-handed up-type squark and one right-handed down-type squark
\begin{eqnarray}
{\cal L}_{P_\alpha C \tilde{u}_L \tilde{d}_R} & = &
-i\lambda \frac{gm_d}{2m_W} U_{\alpha 3}^P
P_\alpha C^+ \tilde{u}^\ast_L \tilde{d}_R
+\mbox{h.c.} \; ;
\label{pculdr}
\end{eqnarray}
\item one neutral pseudoscalar Higgs boson, one charged Higgs boson, one
left-handed down-type squark and one right-handed up-type squark
\begin{eqnarray}
{\cal L}_{P_\alpha C \tilde{u}_R \tilde{d}_L} & = &
i\lambda \frac{gm_u}{2m_W} U_{\alpha 3}^P
P_\alpha C^+ \tilde{u}^\ast_R \tilde{d}_L
+\mbox{h.c.} \; \; .
\label{pcurdl}
\end{eqnarray}
\end{enumerate}
The Feynman rules for these vertices with two Higgs bosons and two
scalar quarks are given in Figs.~\ref{figsssqsq} -- \ref{figpcsqsq}.
Similar as for the trilinear vertex functions, the couplings of
two neutral scalar or pseudoscalar
Higgs bosons to one left-handed and one right-handed scalar quark in
eqs.~(\ref{ssulur}), (\ref{ssdldr}), (\ref{ppulur}) -- (\ref{spdldr})
are explicitly affected by the singlet components of the respective
Higgs particles. Moreover, one left-handed and one right-handed
squark together with a charged Higgs boson couple only to the
singlet component $U^S_{a3}$, $U^P_{\alpha 3}$ of a neutral Higgs boson
in eqs.~(\ref{sculdr}), (\ref{scurdl}), (\ref{pculdr}), (\ref{pcurdl}).
Since these couplings  vanish in the MSSM, their existence could
represent a unique test of the NMSSM.
\subsection{Trilinear self-interaction of Higgs bosons}
The self-interactions of the Higgs bosons are generated by the
following parts of
the $D$ and $F$ terms
and the supersymmetry breaking terms of the scalar potential:
\begin{eqnarray}
\label{ldhhh}
{\cal L}^D_{HHH(H)} & = & -\frac{1}{8}(g^2+{g'}^2)(
H_1^{1\ast}H_1^1H_1^{1\ast}H_1^1+H_2^{2\ast}H_2^2H_2^{2\ast}H_2^2
-2H_1^{1\ast}H_1^1H_2^{2\ast}H_2^2), \\
{\cal L}^F_{HHH(H)} & = &
-\lambda^2H_1^{1\ast}H_1^1H_2^{2\ast}H_2^2
-k^2N^{\ast}NN^{\ast}N
\nonumber \\ & &
+\lambda k (H_1^1H_2^2N^{\ast}N^{\ast}+ \mbox{h.c.} )
\nonumber \\ & &
-\lambda^2NN^{\ast}(H_1^{1\ast}H_1^1
+H_2^{2\ast}H_2^2),
\label{lfhhh} \\
{\cal L}^{\mbox{\scriptsize soft}}_{HHH} & = & \lambda A_{\lambda} H_1^1H_2^2
N+\frac{1}{3}kA_kN^3 + \mbox{h.c.} \hspace{0.5cm} .
\label{lshhh}
\end{eqnarray}
Inserting the mass eigenstates of eqs.~(\ref{shiggs}) --
(\ref{chiggs}) we find for the self-coupling of three
scalar Higgs bosons
\begin{eqnarray}
{\cal L}_{SSS} & = \bigg[ \bigg. &
-\frac{1}{4}\frac{g^2+{g'}^2}{\sqrt{2}}
(v_1 U^S_{a1} U^S_{b1} U^S_{c1} +v_2 U^S_{a2} U^S_{b2} U^S_{c2})
\nonumber \\ & &
+\left(\frac{g^2+{g'}^2}{4\sqrt{2}}-\frac{\lambda^2}{\sqrt{2}}\right)
(v_1
U^S_{a1} U^S_{b2} U^S_{c2}+
v_2
U^S_{a1} U^S_{b1} U^S_{c2})
\nonumber \\ & &
+\frac{1}{\sqrt{2}}(\lambda kv_2-\lambda^2v_1)
U^S_{a1} U^S_{b3} U^S_{c3}
\nonumber \\ & &
+\frac{1}{\sqrt{2}}(\lambda kv_1-\lambda^2v_2)
U^S_{a2} U^S_{b3} U^S_{c3}
\nonumber \\ & &
-\frac{1}{\sqrt{2}}\lambda^2 x
(U^S_{a1} U^S_{b1} U^S_{c3}
+U^S_{a2} U^S_{b2} U^S_{c3} )
\nonumber \\ & &
+\lambda\left(\frac{A_{\lambda}}{\sqrt{2}}+\sqrt{2} kx\right)
U^S_{a1} U^S_{b2} U^S_{c3}
\nonumber \\ & & \bigg.
+(\frac{1}{3\sqrt{2}}kA_k -\sqrt{2}k^2)U^S_{a3} U^S_{b3} U^S_{c3}
\bigg]
S_aS_bS_c \hspace{0.5cm} .
\label{lsss}
\end{eqnarray}
In eq.~(\ref{lsss}) summation over the indices $a$, $b$, and $c$ still
has to be carried out, which is already performed for
the coupling of one scalar with two pseudoscalar Higgs bosons
\begin{eqnarray}
{\cal L}_{S_aP_{\beta}P_{\gamma}} & = \bigg[ \bigg. &
-\frac{g^2+{g'}^2}{4\sqrt{2}}
\left( v_1  U^S_{a1} U^P_{\beta 1} U^P_{\gamma 1} +
v_2 U^S_{a2} U^P_{\beta 2} U^P_{\gamma 2}\right)
\nonumber \\ & &
+\left(\frac{g^2+{g'}^2}{4\sqrt{2}}-\frac{\lambda^2}{\sqrt{2}}\right)
\left( v_1  U^S_{a1} U^P_{\beta 2} U^P_{\gamma 2} +
v_2 U^S_{a2} U^P_{\beta 1} U^P_{\gamma 1}\right)
\nonumber \\ & &
-\frac{1}{\sqrt{2}}(\lambda k v_1 + \lambda^2 v_2)
U^S_{a2} U^P_{\beta 3} U^P_{\gamma 3}
\nonumber \\ & &
-\frac{1}{\sqrt{2}}(\lambda k v_2 + \lambda^2 v_1)
U^S_{a1} U^P_{\beta 3} U^P_{\gamma 3}
\nonumber \\ & &
-\frac{\lambda^2x}{\sqrt{2}} U^S_{a3} (U^P_{\beta 1} U^P_{\gamma 1}+
U^P_{\beta 2} U^P_{\gamma 2})
\nonumber \\ & &
-\left( \sqrt{2}k^2x+\frac{kA_k}{\sqrt{2}}\right)
U^S_{a3} U^P_{\beta 3} U^P_{\gamma 3}
\nonumber \\ & &
+\frac{ \lambda k}{\sqrt{2}} \left( v_1 U_{a3}^S(U^P_{\beta 2} U^P_{\gamma 3} +
U^P_{\beta 3} U^P_{\gamma 2})+v_2 U_{a3}^S(U^P_{\beta 1} U^P_{\gamma 3} +
U^P_{\beta 3} U^P_{\gamma 1})\right)
\nonumber \\ & &
+\left(\frac{\lambda kx}{\sqrt{2}}-\frac{\lambda A_{\lambda}}{2\sqrt{2}}
\right)
\left( U^S_{a1} (U^P_{\beta 2} U^P_{\gamma 3} +
U^P_{\beta 3} U^P_{\gamma 2})
+ U^S_{a2} (U^P_{\beta 1} U^P_{\gamma 3} +
U^P_{\beta 3} U^P_{\gamma 1}\right)
\nonumber \\ & &
-\left(\frac{\lambda kx}{\sqrt{2}}+\frac{\lambda A_{\lambda}}{2\sqrt{2}}
\right)
U^S_{a3} (U^P_{\beta 1} U^P_{\gamma 2} +
U^P_{\beta 2} U^P_{\gamma 1})
\bigg. \bigg]
(2-\delta_{\beta \gamma})S_aP_{\beta}P_{\gamma} \hspace{0.5cm} .
\end{eqnarray}
CP invariance forbids vertices with an odd number of pseudoscalar
Higgs bosons.

The interaction of one neutral Higgs boson with two charged
Higgs particles can be derived from the following parts of the Lagrangian
\begin{eqnarray}
\label{ldhcc}
{\cal L}^D_{H(H)CC} & = & -\frac{g^2}{4} \Big[
(H_2^2H_2^{2\ast}+H_1^1H_1^{1\ast})
(H_2^1H_2^{1\ast}+H_1^2H_1^{2\ast})
\nonumber  \\ & & \hspace*{1.0cm}
+2 (H_1^1H_2^{1\ast}H_1^{2\ast}H_2^2 + \mbox{h.c.}) \Big]
\nonumber \\ [-0.2em] & &
-\frac{{g'}^2}{4}(H_2^2H_2^{2\ast}-H_1^1H_1^{1\ast})
(H_2^1H_2^{1\ast}-H_1^2H_1^{2\ast}), \\ [0.5em]
{\cal L}^F_{H(H)CC} & = & -\lambda k (H_1^2H_2^1N^\ast N^\ast +
\mbox{h.c.})-\lambda^2(H_2^1H_2^{1\ast}+H_1^2H_1^{2\ast})NN^\ast
\nonumber \\ & &
+\lambda^2(H_1^1H_2^2 H_1^{2\ast} H_2^{1\ast} +\mbox{h.c.}),
\label{lfhcc}
\\ [0.5em]
{\cal L}^{\mbox{\scriptsize soft}}_{HCC} & = &
-\lambda A_\lambda (H_1^2H_2^1N + \mbox{h.c.}).
\label{lshcc}
\end{eqnarray}
With eqs.~(\ref{shiggs}) and (\ref{chiggs})
the coupling of a scalar Higgs boson
with two charged Higgs
particles reads
\begin{eqnarray}
{\cal L}_{S_aC^+C^-} & = &
\Big\{ -g m_W ( U_{a1}^S \cos\beta + U_{a2}^S \sin\beta )
\nonumber \\ & & \hspace{0.3cm}
-\frac{gm_Z}{2\cos\theta_W}\left( U_{a2}^S \sin\beta - U_{a1}^S
\cos\beta \right) \cos 2\beta
\nonumber \\ & & \hspace{0.3cm}
+\frac{\lambda^2}{\sqrt{2}} \left( v_1 U_{a2}^S + v_2 U_{a1}^S \right)
\sin 2\beta \nonumber \\ & & \hspace{0.3cm}
-\frac{1}{\sqrt{2}} \lambda U_{a3}^S \left[ (2kx+A_\lambda) \sin 2\beta
+2\lambda x\right]
\Big\} S_aC^+C^- \hspace{0.3cm} .
\end{eqnarray}
The couplings of the pseudoscalar Higgs bosons with two charged
Higgs bosons vanish due to CP conservation.

\begin{figure}[p]
\begin{center}
\begin{picture}(12,8)
\put(-0.0,4.3){$S_a$}
\put(2.0,2.5){$S_b$}
\put(2.0,5.0){$S_c$}
\put(-0.8,1.2){\includegraphics{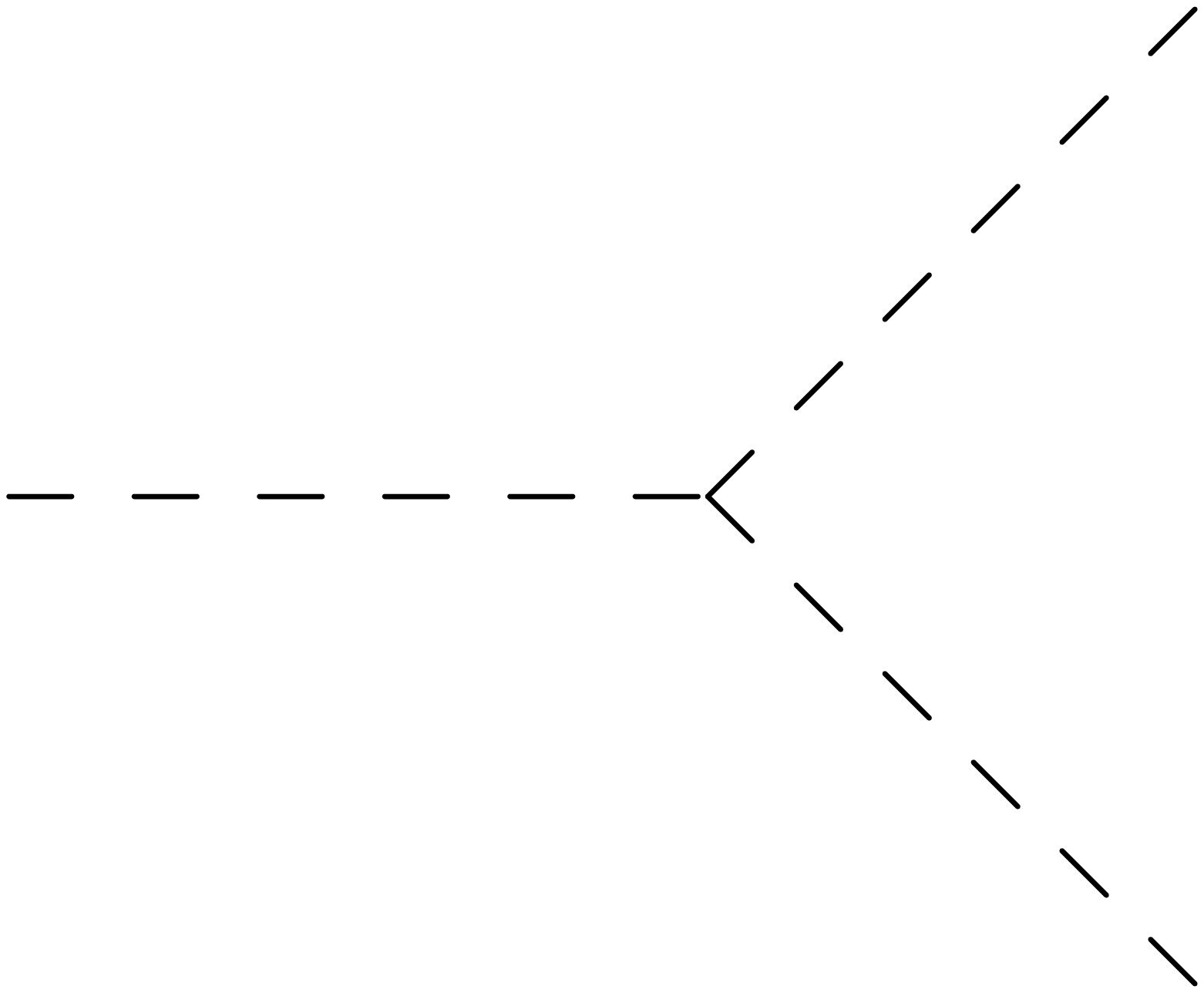}}
\put(4,4){\mbox{$\begin{array}{l}
-\frac{3}{2}i\frac{g^2+{g'}^2}{\sqrt{2}}
(v_1 U^S_{a1} U^S_{b1} U^S_{c1} +v_2 U^S_{a2} U^S_{b2} U^S_{c2})
\\
+i\left(\frac{g^2+{g'}^2}{2\sqrt{2}}-\sqrt{2}\lambda^2\right) v_1
(U^S_{a1} U^S_{b2} U^S_{c2} + U^S_{a2} U^S_{b1} U^S_{c2} +
U^S_{a2} U^S_{b2} U^S_{c1})
\\
+i\left(\frac{g^2+{g'}^2}{2\sqrt{2}}-\sqrt{2}\lambda^2\right) v_2
(U^S_{a1} U^S_{b1} U^S_{c2} +U^S_{a1} U^S_{b2} U^S_{c1} +
U^S_{a2} U^S_{b1} U^S_{c1})
\\
+\sqrt{2}i(\lambda kv_2-\lambda^2v_1)
(U^S_{a1} U^S_{b3} U^S_{c3} + U^S_{a3} U^S_{b1} U^S_{c3} +
U^S_{a3} U^S_{b3} U^S_{c1})
\\
+\sqrt{2}i(\lambda kv_1-\lambda^2v_2)
(U^S_{a2} U^S_{b3} U^S_{c3} + U^S_{a3} U^S_{b2} U^S_{c3} +
U^S_{a3} U^S_{b3} U^S_{c2})
\\
-\sqrt{2}i\lambda^2 x
(U^S_{a1} U^S_{b1} U^S_{c3} + U^S_{a1} U^S_{b3} U^S_{c1} +
U^S_{a3} U^S_{b1} U^S_{c1}
\\
\; \; \; \; \; \; \; \; \; \; \; \; \; \; \; \;
+U^S_{a2} U^S_{b2} U^S_{c3} + U^S_{a2} U^S_{b3} U^S_{c2} +
U^S_{a3} U^S_{b2} U^S_{c2})
\\
+i\lambda\left(\frac{A_{\lambda}}{\sqrt{2}}+\sqrt{2} kx\right)
(U^S_{a1} U^S_{b2} U^S_{c3} + U^S_{a1} U^S_{b3} U^S_{c2} +
U^S_{a2} U^S_{b1}  U^S_{c3}
\\
\;\;\;\;\;\;\;\;\;\;\;\;\;\;\;\;
+ U^S_{a2} U^S_{b3}  U^S_{c1} + U^S_{a3} U^S_{b1} U^S_{c2} +
U^S_{a3} U^S_{b2} U^S_{c1})
\\
+i(\sqrt{2}kA_k -6\sqrt{2}k^2)U^S_{a3} U^S_{b3} U^S_{c3}
\end{array} $}}
\end{picture}
\begin{picture}(12,7)
\put(0.0,4.3){$S_a$}
\put(2.0,2.5){$P_{\beta}$}
\put(2.0,5.0){$P_{\gamma}$}
\put(-0.8,1.2){\includegraphics{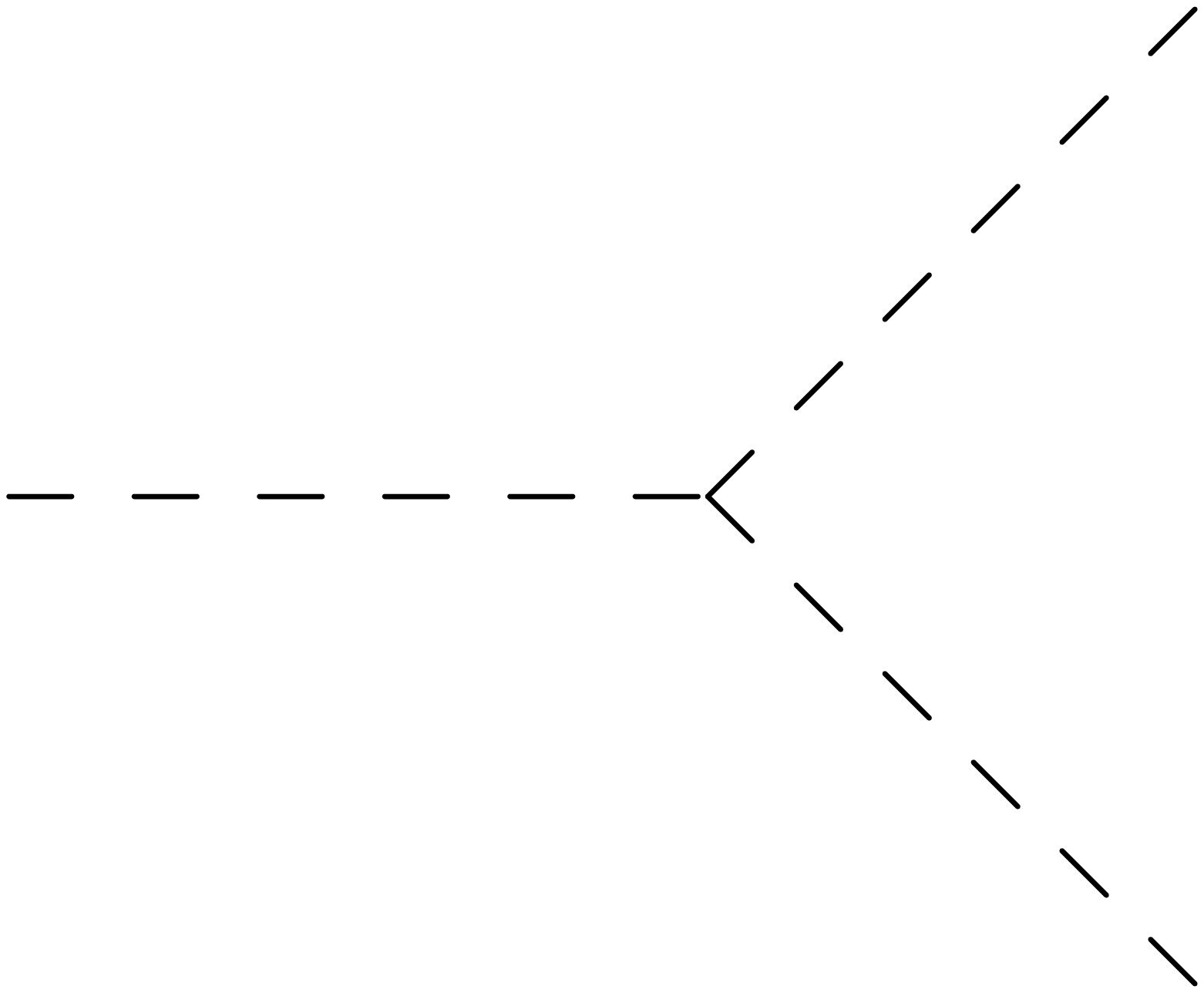}}
\put(4,4){\mbox{$\begin{array}{l}
-i\frac{g^2+{g'}^2}{2\sqrt{2}}
\left( v_1  U^S_{a1} U^P_{\beta 1} U^P_{\gamma 1} +
v_2 U^S_{a2} U^P_{\beta 2} U^P_{\gamma 2}\right) \\
+\left(i\frac{g^2+{g'}^2}{2\sqrt{2}}-\sqrt{2}\lambda^2\right)
\left( v_1  U^S_{a1} U^P_{\beta 2} U^P_{\gamma 2} +
v_2 U^S_{a2} U^P_{\beta 1} U^P_{\gamma 1}\right)  \\
-\sqrt{2}i(\lambda k v_1 + \lambda^2 v_2)
U^S_{a2} U^P_{\beta 3} U^P_{\gamma 3} \\
-\sqrt{2}i(\lambda k v_2 + \lambda^2 v_1)
U^S_{a1} U^P_{\beta 3} U^P_{\gamma 3} \\
-\sqrt{2}i\lambda^2x U^S_{a3} (U^P_{\beta 1} U^P_{\gamma 1}+
U^P_{\beta 2} U^P_{\gamma 2}) \\
-i\left( 2\sqrt{2}k^2x+\sqrt{2} kA_k\right)
U^S_{a3} U^P_{\beta 3} U^P_{\gamma 3} \\
+\sqrt{2}i\lambda k U_{a3}^S \left( v_1 (U^P_{\beta 2} U^P_{\gamma 3} +
U^P_{\beta 3} U^P_{\gamma 2})+v_2 (U^P_{\beta 1} U^P_{\gamma 3} +
U^P_{\beta 3} U^P_{\gamma 1})\right) \\
+i\left( \sqrt{2}\lambda kx-\frac{\lambda A_{\lambda}}{\sqrt{2}}\right)
\left( U^S_{a1} (U^P_{\beta 2} U^P_{\gamma 3} +
U^P_{\beta 3} U^P_{\gamma 2}) \right. \\ \hspace*{4cm}  \left.
+  U^S_{a2} (U^P_{\beta 1} U^P_{\gamma 3} +
U^P_{\beta 3} U^P_{\gamma 1}\right) \\
-i\left( \sqrt{2}\lambda kx+\frac{\lambda A_{\lambda}}{\sqrt{2}}\right)
U^S_{a3} (U^P_{\beta 1} U^P_{\gamma 2} +
U^P_{\beta 2} U^P_{\gamma 1})
\end{array}$}}
\end{picture}
\begin{picture}(12,4)
\put(0.0,2.3){$S_a$}
\put(2.0,0.5){$C^-$}
\put(2.0,3.0){$C^+$}
\put(-0.8,-0.8){\includegraphics{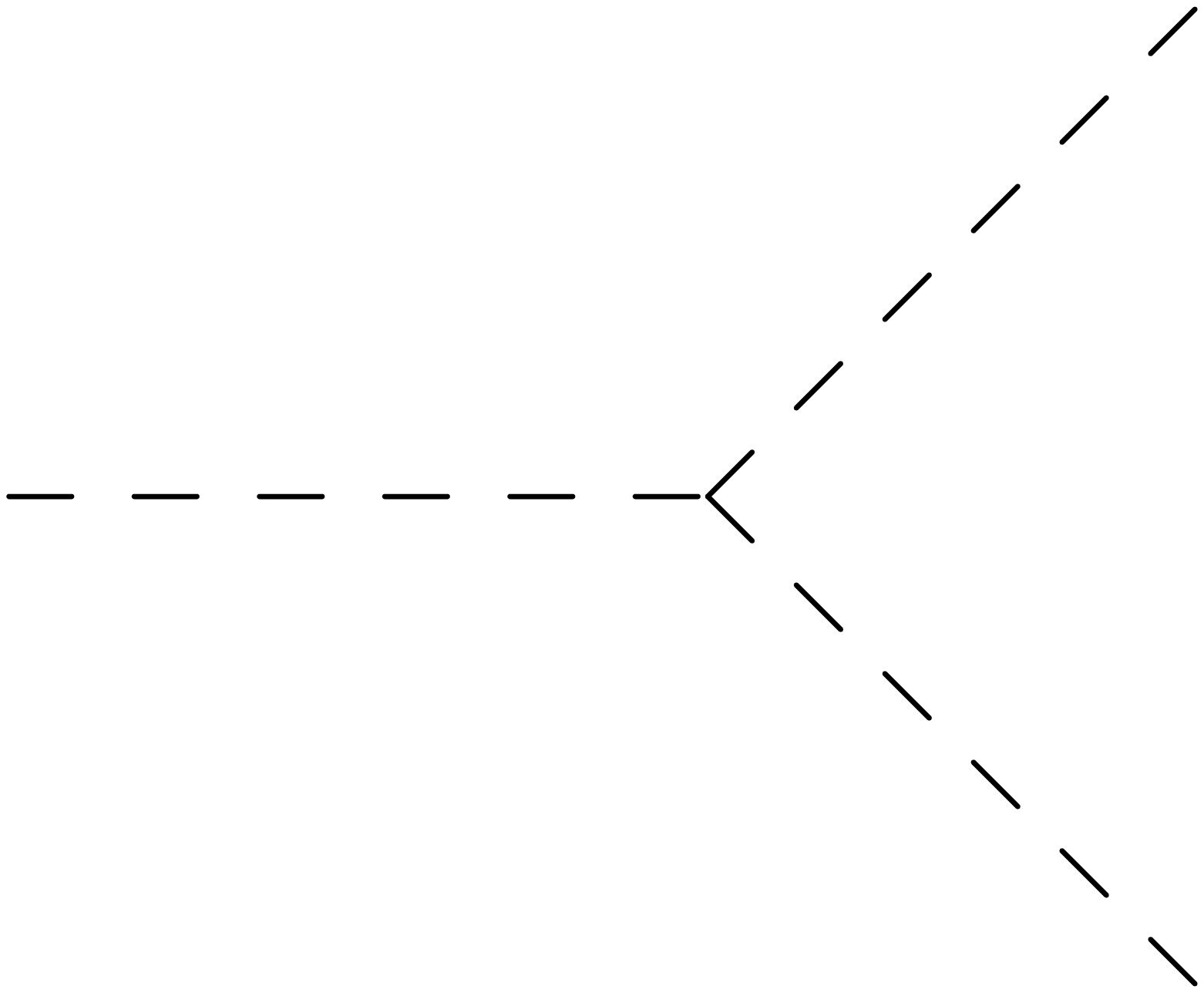}}
\put(4,2){\mbox{$\begin{array}{l}
-ig m_W ( U_{a1}^S \cos\beta + U_{a2}^S \sin\beta )
\\ [0.3ex]
-i\frac{gm_Z}{2\cos\theta_W}\left( U_{a2}^S \sin\beta - U_{a1}^S
\cos\beta \right) \cos 2\beta
\\ [0.7ex]
+i\frac{\lambda^2}{\sqrt{2}} \left( v_1 U_{a2}^S + v_2 U_{a1}^S \right)
\sin 2\beta \\ [0.3ex]
-\frac{i}{\sqrt{2}} \lambda U_{a3}^S \left[ (2kx+A_\lambda) \sin 2\beta
+2\lambda x\right] \end{array} $}}
\end{picture}
\end{center}
\caption{Feynman rules for the trilinear self-interactions of the
Higgs bosons.}
\label{figself}
\end{figure}

The Feynman rules for trilinear Higgs self-couplings of the NMSSM shown in
Fig.~\ref{figself} exhibit significant
differences to their counterparts in the minimal model. They contain
the singlet components $U^S_{a3}$, $U^P_{\alpha 3}$ as well as
terms proportional to the couplings in the superpotential
$\lambda$ and $k$ and to the parameters in the supersymmetry
breaking potential $A_\lambda$ and $A_k$. Therefore, probing the
Higgs self-coupling at high energy colliders probably represents
a highly decisive test in oder to distinguish between NMSSM and MSSM.
\subsection{Quartic self-interaction of Higgs bosons}
The interactions between four neutral Higgs bosons are also contained in
eqs.~(\ref{ldhhh}) and (\ref{lfhhh}).
Since again vertices with an odd number of pseudoscalar Higgs bosons
are forbidden by CP invariance, one obtains for the interactions of
\begin{enumerate}
\item four scalar neutral Higgs bosons
\begin{eqnarray}
{\cal L}_{SSSS} & = & \bigg[ -\frac{1}{32}(g^2+{g'}^2)
(U^S_{a1}U^S_{b1}U^S_{c1}U^S_{d1}+
U^S_{a2}U^S_{b2}U^S_{c2}U^S_{d2}) \bigg.
\nonumber \\ & & \hspace{0.3cm}
+\left( \frac{1}{16}(g^2+{g'}^2)-\frac{\lambda^2}{4} \right)
U^S_{a1}U^S_{b1}U^S_{c2}U^S_{d2}
\nonumber \\ & & \hspace{0.3cm}
-\frac{\lambda^2}{4}(
U^S_{a1}U^S_{b1}U^S_{c3}U^S_{d3}+
U^S_{a2}U^S_{b2}U^S_{c3}U^S_{d3})
\nonumber \\ & & \hspace{0.3cm} \bigg.
-\frac{k^2}{4}
U^S_{a3}U^S_{b3}U^S_{c3}U^S_{d3}-\frac{\lambda k}{2}
U^S_{a1}U^S_{b2}U^S_{c3}U^S_{d3}\bigg]
S_a S_b S_c S_d  \; ;
\label{lssss}
\end{eqnarray}
\item two scalar and two pseudoscalar Higgs bosons
\begin{eqnarray}
{\cal L}_{SSPP} & = & \bigg[ -\frac{1}{16}(g^2+{g'}^2)
(U^S_{a1}U^S_{b1}U^P_{\gamma 1}U^P_{\delta 1}+
U^S_{a2}U^S_{b2}U^P_{\gamma 2}U^P_{\delta 2}) \bigg.
\nonumber \\ & & \hspace{0.3cm}
+\left( \frac{1}{16}(g^2+{g'}^2)-\frac{\lambda^2}{4} \right)
(U^S_{a1}U^S_{b1}U^P_{\gamma 2}U^P_{\gamma 2}+
U^S_{a2}U^S_{b2}U^P_{\gamma 1}U^P_{\delta 1})
\nonumber \\ & & \hspace{0.3cm}
-\frac{\lambda^2}{4}(
U^S_{a1}U^S_{b1}U^P_{\gamma 3}U^P_{\delta 3}+
U^S_{a3}U^S_{b3}U^P_{\gamma 1}U^P_{\delta 1}+
U^S_{a2}U^S_{b2}U^P_{\gamma 3}U^P_{\delta 3}+
U^S_{a3}U^S_{b3}U^P_{\gamma 2}U^P_{\delta 2})
\nonumber \\ & & \hspace{0.3cm}
-\frac{\lambda k}{2}
(U^S_{a1}U^S_{b2}U^P_{\gamma 3}U^P_{\delta 3}+
U^S_{a3}U^S_{b3}U^P_{\gamma 1}U^P_{\delta 2}-2
U^S_{a1}U^S_{b3}U^P_{\gamma 2}U^P_{\delta 3}-2
U^S_{a2}U^S_{b3}U^P_{\gamma 1}U^P_{\delta 3})
\nonumber \\ & & \hspace{0.3cm} \bigg.
-\frac{k^2}{2}
U^S_{a3}U^S_{b3}U^P_{\gamma 3}U^P_{\delta 3}  \bigg]
S_a S_b P_{\gamma} P_{\delta} \; ;
\label{lsspp}
\end{eqnarray}
\item four pseudoscalar Higgs bosons
\begin{eqnarray}
{\cal L}_{PPPP} & = & \bigg[ -\frac{1}{32}(g^2+{g'}^{2})
(U^P_{\alpha 1}U^P_{\beta 1}U^P_{\gamma 1}U^P_{\delta 1}+
U^P_{\alpha 2}U^P_{\beta 2}U^P_{\gamma 2}U^P_{\delta 2}) \bigg.
\nonumber \\ & & \hspace{0.3cm}
+\left( \frac{1}{16}(g^2+{g'}^{2})-\frac{\lambda^2}{4} \right)
U^P_{\alpha 1}U^P_{\beta 1}U^P_{\gamma 2}U^P_{\delta 2}
\nonumber \\ & & \hspace{0.3cm}
-\frac{\lambda^2}{4}(
U^P_{\alpha 1}U^P_{\beta 1}U^P_{\gamma 3}U^P_{\delta 3}+
U^P_{\alpha 2}U^P_{\beta 2}U^P_{\gamma 3}U^P_{\delta 3})
\nonumber \\ & & \hspace{0.3cm} \bigg.
-\frac{k^2}{4}
U^P_{\alpha 3}U^P_{\beta 3}U^P_{\gamma 3}U^P_{\delta 3}
-\frac{\lambda k}{2}
U^P_{\alpha 1}U^P_{\beta 2}U^P_{\gamma 3}U^P_{\delta 3}\bigg]
P_{\alpha} P_{\beta} P_{\gamma} P_{\delta}.
\label{lpppp}
\end{eqnarray}
\end{enumerate}
In eqs.~(\ref{lssss}) -- (\ref{lpppp})
the sum over repeated indices has to be performed.
Latin indices for the scalar Higgs bosons run from 1 to 3,
the greek indices of pseudoscalar Higgs bosons are summed from
1 to 2.

\begin{figure}[p]
\begin{center}
\begin{picture}(12,9)
\put(2.0,2.3){$S_d$}
\put(2.0,5.3){$S_c$}
\put(0.3,2.3){$S_a$}
\put(0.3,5.3){$S_b$}
\put(-0.8,1.2){\includegraphics{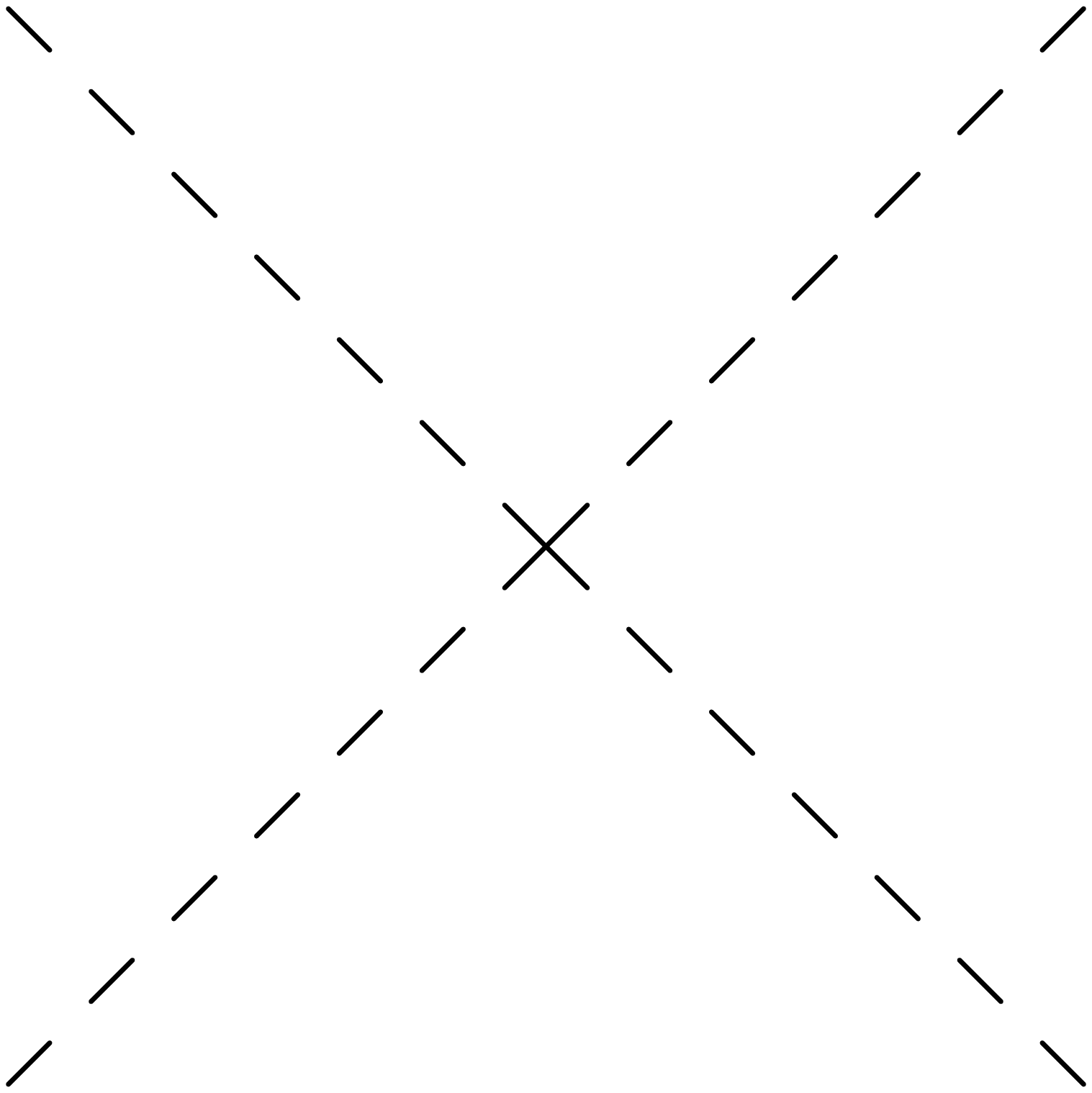}}
\put(3.5,7.0){$
-i\frac{3}{4}(g^2+{g'}^{2})  (
U^S_{a1}U^S_{b1}U^S_{c1}U^S_{d1}+U^S_{a2}U^S_{b2}U^S_{c2}U^S_{d2})$}
\put(3.5,6.4){$
+i\left( \frac{1}{4}(g^2+{g'}^{2})-\lambda^2 \right)$}
\put(4.8,5.8){$(
U^S_{a1}U^S_{b1}U^S_{c2}U^S_{d2}+
U^S_{a1}U^S_{b2}U^S_{c1}U^S_{d2}+
U^S_{a1}U^S_{b2}U^S_{c2}U^S_{d1}$}
\put(5.0,5.2){$ +
U^S_{a2}U^S_{b1}U^S_{c1}U^S_{d2}+
U^S_{a2}U^S_{b1}U^S_{c2}U^S_{d1}+
U^S_{a2}U^S_{b2}U^S_{c1}U^S_{d1})$}
\put(3.5,4.6){$-i\lambda^2  (
U^S_{a1}U^S_{b1}U^S_{c3}U^S_{d3}+
U^S_{a1}U^S_{b3}U^S_{c1}U^S_{d3}+
U^S_{a1}U^S_{b3}U^S_{c3}U^S_{d1}$}
\put(5.0,4.0){$+U^S_{a3}U^S_{b1}U^S_{c1}U^S_{d3}+
U^S_{a3}U^S_{b1}U^S_{c3}U^S_{d1}+
U^S_{a3}U^S_{b3}U^S_{c1}U^S_{d1}$}
\put(5.0,3.4){$+U^S_{a2}U^S_{b2}U^S_{c3}U^S_{d3}+
U^S_{a2}U^S_{b3}U^S_{c2}U^S_{d3}+
U^S_{a2}U^S_{b3}U^S_{c3}U^S_{d2}$}
\put(5.0,2.8){$+U^S_{a3}U^S_{b2}U^S_{c2}U^S_{d3}+
U^S_{a3}U^S_{b2}U^S_{c3}U^S_{d2}+
U^S_{a3}U^S_{b3}U^S_{c2}U^S_{d2})$}
\put(3.5,2.2){$-6ik^2  U^S_{a3}U^S_{b3}U^S_{c3}U^S_{d3}$}
\put(3.5,1.6){$-i \lambda k  (
U^S_{a1}U^S_{b2}U^S_{c3}U^S_{d3}+
U^S_{a1}U^S_{b3}U^S_{c2}U^S_{d3}+
U^S_{a1}U^S_{b3}U^S_{c3}U^S_{d2}$}
\put(5.0,1.0){$+U^S_{a2}U^S_{b1}U^S_{c3}U^S_{d3}+
U^S_{a2}U^S_{b3}U^S_{c1}U^S_{d3}+
U^S_{a2}U^S_{b3}U^S_{c3}U^S_{d1}$}
\put(5.0,0.4){$+U^S_{a3}U^S_{b1}U^S_{c2}U^S_{d3}+
U^S_{a3}U^S_{b1}U^S_{c3}U^S_{d2}+
U^S_{a3}U^S_{b2}U^S_{c1}U^S_{d3}$}
\put(5.0,-0.2){$+U^S_{a3}U^S_{b2}U^S_{c3}U^S_{d1}+
U^S_{a3}U^S_{b3}U^S_{c1}U^S_{d2}+
U^S_{a3}U^S_{b3}U^S_{c2}U^S_{d1}) $}
\end{picture}
\begin{picture}(12,7.5)
\put(2.0,2.3){$P_\delta$}
\put(2.0,5.3){$P_\gamma$}
\put(0.3,2.3){$S_a$}
\put(0.3,5.3){$S_b$}
\put(-0.8,1.2){\includegraphics{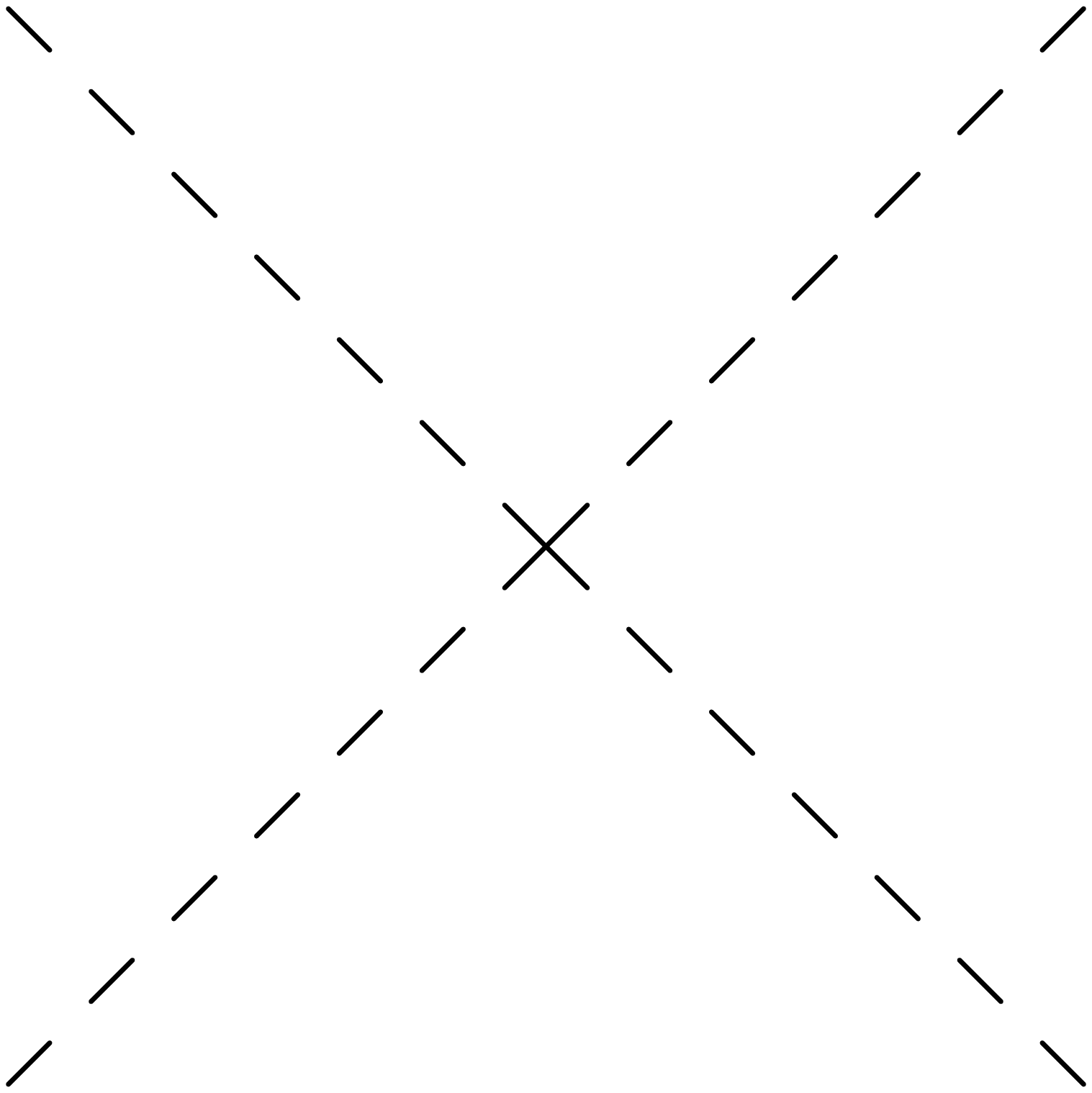}}
\put(3.5,6){$
-i\frac{1}{4}(g^2+{g'}^{2})  (
U^S_{a 1}U^S_{b 1}U^P_{\gamma 1}U^P_{\delta 1}+
U^S_{a 2}U^S_{b 2}U^P_{\gamma 2}U^P_{\delta 2})$}
\put(3.5,5.4){$+i\left( \frac{1}{4}(g^2+{g'}^2)-\lambda^2 \right)  (
U^S_{a 1}U^S_{b 1}U^P_{\gamma 2}U^P_{\delta 2}+
U^S_{a 2}U^S_{b 2}U^P_{\gamma 1}U^P_{\delta 1})$}
\put(3.5,4.8){$-i\lambda^2  (
U^S_{a 1}U^S_{b 1}U^P_{\gamma 3}U^P_{\delta 3}+
U^S_{a 3}U^S_{b 3}U^P_{\gamma 1}U^P_{\delta 1}$}
\put(5.0,4.2){$+U^S_{a 2}U^S_{b 2}U^P_{\gamma 3}U^P_{\delta 3} +
U^S_{a 3}U^S_{b 3}U^P_{\gamma 2}U^P_{\delta 2})$}
\put(3.5,3.6){$-2ik^2  U^S_{a 3}U^S_{b 3}U^P_{\gamma 3}U^S_{\delta 3} $}
\put(3.5,3.0){$-i \lambda k  \Big(
(U^S_{a 1}U^S_{b 2}+
U^S_{a 2}U^S_{b 1})U^P_{\gamma 3}U^P_{\delta 3}
+U^S_{a 3}U^S_{b 3}(U^P_{\gamma 1}U^P_{\delta 2}+
U^P_{\gamma 2}U^P_{\delta 1})$}
\put(5.0,2.4){$-
U^S_{a 1}U^S_{b 3}U^P_{\gamma 2}U^P_{\delta 3} -
U^S_{a 3}U^S_{b 1}U^P_{\gamma 2}U^P_{\delta 3}$}
\put(5.0,1.8){$-U^S_{a 1}U^S_{b 3}U^P_{\gamma 3}U^P_{\delta 2} -
U^S_{a 3}U^S_{b 1}U^P_{\gamma 3}U^P_{\delta 2}$}
\put(5.0,1.2){$-U^S_{a 2}U^S_{b 3}U^P_{\gamma 1}U^P_{\delta 3} -
U^S_{a 3}U^S_{b 2}U^P_{\gamma 1}U^P_{\delta 3}$}
\put(5.0,0.6){$-U^S_{a 2}U^S_{b 3}U^P_{\gamma 3}U^P_{\delta 1} -
U^S_{a 3}U^S_{b 2}U^P_{\gamma 3}U^P_{\delta 1}\Big) $}
\end{picture}
\end{center}
\caption{\label{fig4h1} Feynman rules for the quartic self-interactions
of the Higgs bosons.}
\end{figure}

The interactions with two neutral and two charged Higgs particles
follow directly from eqs.~(\ref{ldhcc}) and (\ref{lfhcc}). They read
for the case of scalar Higgs bosons
\begin{eqnarray}
{\cal L}_{S_aS_bCC} & = & \bigg[ -\frac{1}{8}g^2
(U^S_{a2} U^S_{b2} +
U^S_{a1} U^S_{b1})
\nonumber \\ & & \hspace*{0.3cm}
-\left( \frac{1}{8}g^2 -\frac{\lambda ^2}{4} \right)
(U^S_{a1}U^S_{b2}+U^S_{a2}U^S_{b1}) \sin 2\beta
\nonumber \\ & &    \hspace*{0.3cm}
-\frac{1}{8}{g'}^{2}
(U^S_{a2} U^S_{b2} -
U^S_{a1} U^S_{b1}) \cos 2\beta
\nonumber \\ & & \hspace*{0.3cm}
-\frac{\lambda }{2} (\lambda U^S_{a3} U^S_{b3}+
k U^S_{a3} U^S_{b3} \sin 2\beta  )  \bigg]
(2-\delta_{ab})S_aS_bC^+C^-  ,
\end{eqnarray}
and for pseudoscalar Higgs bosons
\begin{eqnarray}
{\cal L}_{P_\alpha P_\beta CC} & = & \bigg[ -\frac{1}{8}g^2
(U^P_{\alpha 2} U^P_{\beta 2} +
U^P_{\alpha 1} U^P_{\beta1})
\nonumber \\ & & \hspace*{0.3cm}
+\left( \frac{1}{8}g^2 -\frac{\lambda ^2}{4} \right)
(U^P_{\alpha 1}U^P_{\beta 2}+U^P_{\alpha 2}U^P_{\beta 1}) \sin 2\beta
\nonumber \\ & &    \hspace*{0.3cm}
-\frac{1}{8}{g'}^{2}
(U^P_{\alpha 2} U^P_{\beta 2} -
U^P_{\alpha 1} U^P_{\beta 1}) \cos 2\beta
\nonumber \\ & & \hspace*{0.3cm}
-\frac{\lambda }{2} (\lambda U^P_{\alpha 3} U^P_{\beta 3}-
k U^P_{\alpha 3} U^P_{\beta 3} \sin 2\beta  )  \bigg]
(2-\delta_{\alpha\beta})P_\alpha P_\beta C^+C^-  .
\end{eqnarray}
The Feynman rules for the quartic Higgs vertices are displayed in
Fig.~\ref{fig4h1}.
As the trilinear Higgs interactions, also the quartic Higgs self-couplings
of the NMSSM differ significantly from those of the MSSM.
They are, however, of lesser importance for the supersymmetric
processes tested at the particle colliders in the nearer future.
\subsection{Interaction of Higgs bosons with neutralinos and charginos}
\label{sec:hnn}
For these interactions one has to take into account the mass eigenstates
of the Higgs bosons as well as the neutralino/chargino mixing as
described in Sec.~2. In four-component notation
the Lagrangian for the interaction of a neutral Higgs boson with two
charginos reads
\begin{equation}
{\cal L}_{H\tilde{\chi}^+\tilde{\chi}^-}
^{\mbox{\scriptsize int}}  =
-g(H_1^{1\ast} \bar{\tilde{H}}P_L\tilde{W}+
H_2^{2\ast}\bar{\tilde{W}}P_L\tilde{H})
+\lambda N \bar{\tilde{H}}P_L\tilde{H}+\mbox{h.c.} \; \; .
\end{equation}
Substituting the mass eigenstates of the Higgs bosons (eqs.~(\ref{shiggs})
and (\ref{phiggs})) and of the charginos (eq.~(\ref{charmass})) one finds
\begin{eqnarray}
{\cal L}_{H\tilde{\chi}^+\tilde{\chi}^-} & = &
-S_a\bar{\tilde{\chi}}^+_i
\left[ Q^{\ast}_{aij}P_L+Q_{aji}P_R \right]
\tilde{\chi}^+_j  \nonumber \\ & &
-iP_\alpha
\bar{\tilde{\chi}}^+_i
\left[ R^{\ast}_{\alpha ij}P_L-R_{\alpha ji}P_R \right]
\tilde{\chi}^+_j \; \; ,
\end{eqnarray}
where
\begin{eqnarray}
Q_{aij} & = & \frac{g}{\sqrt{2}}
(U^S_{a1}U_{i2}V_{j1}+U^S_{a2}U_{i1}V_{j2})-
\frac{\lambda}{\sqrt{2}} U^{S}_{a3} U_{i2}V_{j2} \; \; ,
\\
R_{\alpha ij} & = &
-\frac{g}{\sqrt{2}}
(U^P_{\alpha 1}U_{i2}V_{j1}+U^P_{\alpha 2}U_{i1}V_{j2})-
\frac{\lambda}{\sqrt{2}} U^{P}_{\alpha 3} U_{i2}V_{j2} \; \; .
\end{eqnarray}
For real matrices $Q$ and $R$ the couplings are indeed scalar for
the $S_a$ and pseudoscalar for the $P_\alpha$.

The source for the coupling of a charged Higgs boson to a neutralino and
a chargino is the Lagrangian
\begin{eqnarray}
{\cal L}_{C^\pm\tilde{\chi}^0\tilde{\chi}^{\pm}}
^{\mbox{\scriptsize int}} & = &
-\frac{g}{\sqrt{2}} \left(
H_2^{1\ast}\left( 2s_W \bar{\tilde{\gamma}}+
\frac{1-2s_W^2}{ c_W}
\bar{\tilde{Z}} \right) P_L \tilde{H}
-H_1^{2\ast} \bar{\tilde{H}} P_L
\left( 2s_W \tilde{\gamma} + \frac{1-2s_W^2}{c_W}
\tilde{Z} \right) \right) \nonumber \\
& & -g \left( H_1^{2\ast} \bar{\tilde{W}} P_L
\left( \tilde{H}_a \cos\beta+\tilde{H}_b \sin\beta
\right) +H_2^{1\ast} \left(
-\bar{\tilde{H}}_a \sin\beta + \bar{\tilde{H}}_b \cos\beta
\right) P_L \tilde{W} \right) \nonumber \\
& & + \lambda  \left( H_1^2 \bar{\tilde{N}} P_L \tilde{H}
+H_2^1 \bar{\tilde{H}} P_L \tilde{N} \right)
+ \mbox{h.c.} \; \; ,
\end{eqnarray}
which leads to the interaction of the mass eigenstates
\begin{eqnarray}
{\cal L}_{C^\pm\tilde{\chi}^0\tilde{\chi}^{\pm}} & = &
-C^- \bar{\tilde{\chi}}^0_i
\left[ Q^{'L \ast}_{ij}P_L+Q^{'R}_{ij}P_R \right]
\tilde{\chi}^+_j +\mbox{h.c.} \; \; ,
\end{eqnarray}
where
\begin{eqnarray}
Q^{'L}_{ij} & = & g\cos\beta \bigg[
\left( -N_{i3}\sin\beta+N_{i4}\cos\beta\right) V_{j1}
\nonumber \\ & & \hspace{2cm}
+ \frac{1}{\sqrt{2}} \left( 2s_WN_{i1}+
(c_W-\frac{s_W^2}{c_W})N_{i2} \right) V_{j2} \bigg]
\nonumber \\ & & -\lambda ^\ast \sin\beta N_{i5}V_{j2} \; \; ,  \\
Q^{'R}_{ij} & = & g\sin\beta \bigg[
\left( N_{i3}\cos\beta+N_{i4}\sin\beta\right) U_{j1}
\nonumber \\ & & \hspace{2cm}
- \frac{1}{\sqrt{2}} \left( 2s_WN_{i1}+
(c_W-\frac{s_W^2}{c_W})N_{i2} \right) U_{j2} \bigg]
\nonumber \\ & & -\lambda ^\ast \cos\beta N_{i5}U_{j2} \; \; .
\end{eqnarray}

Finally, the interaction of a neutral Higgs boson and two neutralinos
arises from
\begin{eqnarray}
{\cal L}^{\mbox{\scriptsize int}}_{H\tilde{\chi}^0\tilde{\chi}^0} & = &
\frac{g}{\sqrt{2}c_W} \left(
H_2^{2\ast} \bar{\tilde{Z}} P_L
\left( \tilde{H}_a \cos\beta +\tilde{H}_b \sin\beta \right)
-H_1^{1\ast} \bar{\tilde{Z}} P_L
\left(-\tilde{H}_a \sin\beta +\tilde{H}_b \cos\beta \right)
\right) \nonumber \\ & &
-\lambda \left( H_1^1 \bar{\tilde{N}} P_L
\left(-\tilde{H}_a \sin\beta +\tilde{H}_b \cos\beta \right)
+H_2^2 \bar{\tilde{N}} P_L
\left( \tilde{H}_a \cos\beta +\tilde{H}_b \sin\beta \right)
\right) \nonumber \\ & &
+2kN\bar{\tilde{N}} P_L \tilde{N} \; \; +\mbox{h.c.} \; \; .
\end{eqnarray}
With the mass eigenstates of the Higgs bosons and
neutralinos one arrives at
\begin{eqnarray}
{\cal L}_{H\tilde{\chi}^0\tilde{\chi}^0} & = &
-\frac{1}{2} S_a \bar{\tilde{\chi}}^0_i
(Q^{L''}_{aij}P_L+Q^{R''}_{aij}P_R)\tilde{\chi}^0_j
\nonumber \\ & &
-\frac{i}{2} P_{\alpha}\bar{\tilde{\chi}}^0_i
(R^{L''}_{\alpha ij}P_L+R^{R''}_{\alpha ij}P_R)\tilde{\chi}^0_j \; \; ,
\end{eqnarray}
where
\begin{eqnarray}
Q^{''L}_{aij} & = &
\frac{1}{2} \Bigg[
(U^S_{a1}\cos\beta+U^S_{a2}\sin\beta)
\nonumber \\ & & \hspace{1.0cm} \times
\left(\frac{g}{c_W}
(N_{i2}N_{j3}^{\ast}+N_{j2}N_{i3}^{\ast}) .
+\sqrt{2}\lambda(N_{i5}N_{j4}^{\ast}+N_{j5}N_{i4}^{\ast})
\right)
\nonumber \\ & & \hspace{0.5cm}
+(U^S_{a1}\sin\beta-U^S_{a2}\cos\beta)
\nonumber \\ & & \hspace{1.0cm} \times
\left(\frac{g}{c_W}
(N_{i2}N_{j4}^{\ast}+N_{j2}N_{i4}^{\ast})
-\sqrt{2}\lambda (N_{i5}N_{j3}^{\ast}+N_{j5}N_{i3}^{\ast})
\right) \Bigg]
\nonumber \\ & &
-\sqrt{2}kU^S_{a3}(N_{i5}N_{j5}^{\ast}+N_{j5}N_{i5}^{\ast}) \; \; ,
\\
Q^{''R}_{aij} & = & Q^{''L\ast}_{aij} \; \; ,
\\
R^{''L}_{\alpha ij} & = &
-\frac{1}{2} \Bigg[
(U^P_{\alpha 1}\cos\beta+U^P_{\alpha 2}\sin\beta)
\nonumber \\ & & \hspace{1.0cm} \times
\left(\frac{g}{c_W}(N_{i2}N_{j3}^{\ast}+N_{j2}N_{i3}^{\ast})
-\sqrt{2}\lambda(N_{i5}N_{j4}^{\ast}+N_{j5}N_{i4}^{\ast})
\right)
\nonumber \\ & & \hspace{0.5cm}
+(U^P_{\alpha 1}\sin\beta-U^P_{\alpha 2}\cos\beta)
\nonumber \\ & & \hspace{1.0cm} \times
\left(\frac{g}{c_W}
(N_{i2}N_{j4}^{\ast}+N_{j2}N_{i4}^{\ast})
+\sqrt{2}\lambda(N_{i5}N_{j3}^{\ast}+N_{j5}N_{i3}^{\ast})
\right) \Bigg]
\nonumber \\ & &
-\sqrt{2}kU^P_{\alpha 3}(N_{i5}N_{j5}^{\ast}+N_{j5}N_{i5}^{\ast}) \; \; ,
\\
R^{''R}_{\alpha ij} & = & -R^{''L\ast}_{\alpha ij} \; \; .
\end{eqnarray}

The corresponding Feynman rules are given in Fig.~\ref{fighneuchar}.
All couplings of Higgs bosons to neutralinos and charginos
derived in this section explicitly contain the singlet components
$U^S_{a3}$, $U^P_{\alpha 3}$ of the neutral Higgs bosons or
$N_{i5}$ of the neutralinos and are therefore obviously different
from the MSSM couplings. They will be discussed in Sec.~\ref{sec:coupl}
together with some
other crucial differences
between the couplings of the
minimal and nonminimal model.

\section{Experimental constraints on the parameter space}
\label{sec:con}
The parameter space of the NMSSM and the masses of the supersymmetric
particles are
constrained by the results from the current high energy colliders
LEP1 at CERN and Tevatron at Fermilab. In this section we reanalyze the
experimental results from the
negative search for neutralinos and Higgs bosons which were
studied in great detail in refs.~\cite{frankeneu,frankehiggs}.
A key role for the production of Higgs bosons at
$e^+e^-$ colliders plays the Higgs coupling to
$Z$ bosons in eqs.~(\ref{frspz}) and (\ref{frszz}), while
neutralino production at LEP1 crucially depends on the
$Z\tilde{\chi}^0\tilde{\chi}^0$ coupling which is formally
identical in NMSSM and MSSM and differs only by the neutralino mixing.
All those couplings are suppressed in the NMSSM if the respective
neutralinos or Higgs bosons have significant singlet components.
Therefore NMSSM neutralino and Higgs mass bounds are much weaker
than in the minimal model. We now discuss these mass bounds and
couplings in detail.

\subsection{Constraints from Higgs search}
First we consider the couplings of the scalar and pseudoscalar Higgs bosons
to $Z$ bosons that are crucial for the Higgs production at future
$e^+e^-$ colliders.
Bounds for these couplings as a function of the Higgs masses
from the experiments at LEP
constrain the masses of the Higgs bosons and the parameters of the
Higgs sector. In Fig.~\ref{xiplots} we compare the couplings of a
scalar Higgs boson with two $Z$ bosons in the MSSM and the NMSSM
for different singlet vacuum expectation values $x$ and parameters $k$
and two values of $\tan\beta$. Radiative corrections to the Higgs masses
due to top/stop loops are included with
$A_t=0$ GeV and
$m_{\tilde{t}_1}=200$ GeV, $m_{\tilde{t}_2}=500$ GeV.
All couplings are normalized with respect to the SM $\Phi ZZ$ coupling.
The solid lines in Fig.~\ref{xiplots} denote the experimental bounds
from the LEP experiments \cite{alephhiggs}.
Here the upper line applies if the Higgs boson decays as in the SM,
the lower curve is valid if it decays invisibly. Also shown
are the range of the NMSSM couplings (dashed lines for
$g_{S_1ZZ}^{NMSSM}/g_{\Phi ZZ}^{SM}$, dotted  for
$g_{S_2ZZ}^{NMSSM}/g_{\Phi ZZ}^{SM}$) scanning over
all values for the parameters $A_\lambda$ and $A_k$,
and the MSSM couplings $g_{hZZ}^{MSSM}/g_{\Phi ZZ}^{SM}$ (double dashed) and
$g_{HZZ}^{MSSM}/g_{\Phi ZZ}^{SM}$ (dashed dotted).

In Fig.~\ref{xiplots} we first choose a singlet vacuum
expectation value $x=200$ GeV near those of the Higgs doublets
$v=\sqrt{v_1^2+v_2^2}=174$ GeV. The couplings in the superpotential
are fixed to typical values $\lambda=0.8$ and $k=0.1$ which allow
light Higgs bosons with a large singlet component.
The dependence on $\tan\beta$ is studied with $\tan\beta=2$ and
$\tan\beta=10$. For the MSSM couplings the $\mu$ parameter which
affects the radiative corrections is set
to $\mu=\lambda x$. Furthermore, we consider in Fig.~\ref{xiplots}
the larger singlet vacuum
expectation value $x=1000$ GeV for the same set of
parameters $\lambda$, $k$ and $\tan\beta$. In this case
the lightest Higgs has only a
small singlet component, so that we also study a smaller
coupling $k=0.01$, where for $\tan\beta=10$ a small Higgs mass is not
experimentally excluded.

First one notes the different mass ranges for the lightest and
next-to-lightest Higgs bosons in the NMSSM and MSSM as well as
for the different parameters sets. For \mbox{$x=200$ GeV}, $\lambda=0.8$,
$k=0.1$, $\tan\beta=2$ and the above described choice of $A_t$ and the
stop masses
the theoretical upper bound for the
lightest NMSSM Higgs scalar $S_1$ is about 110 GeV, while it is increased
to 150 GeV for a large singlet vacuum expectation value \mbox{$x=1000$ GeV}
when the other parameters are unchanged.
A smaller coupling $k$ lets this bounds again decrease.
The second lightest Higgs scalar $S_2$ can accordingly reach
higher mass regions $145\, (108)$ GeV $ < m_{S_2} < \; 269\, (227)$ GeV
for $x=1000$ GeV, $\lambda=0.8$, $k=0.1$, $\tan\beta=2\, (10)$,
while its mass is constrained to a narrow
range in the other considered scenarios.
The larger value $\tan\beta=10$ leads to a decrease of the upper mass bound
for the lightest Higgs scalar of about 50 GeV.
In the MSSM, the mass regions for the Higgs scalars are generally
smaller for the light
Higgs and larger for the heavy scalar Higgs particle.

In the MSSM, the $hZZ$ and $HZZ$ couplings for the light and heavy Higgs
scalars to $Z$ or $W$ bosons are given by $\sin (\alpha - \beta )$ and
$\cos (\alpha- \beta )$ relative to the SM $\Phi ZZ$ coupling, respectively.
Contrary, in the NMSSM these
couplings can become very small if the respective Higgs bosons have a
large singlet component.
Since also the coupling between the lightest scalar and pseudoscalar
Higgs particles and a $Z$ boson is rather weak, as we will discuss in
connection with Fig.~\ref{pseudoplots}, a very light or massless
Higgs boson is experimentally not excluded for $x=200$ GeV.
Here the mixing of the lightest scalar NMSSM Higgs is dominated by its singlet
component, while
the coupling of the next-to-lightest scalar Higgs boson to gauge bosons
is of the same order as in the SM for a Higgs boson of the same mass.
For the larger value $\tan\beta=10$
the squared $S_1ZZ$ coupling can be totally suppressed
compared to the SM in the whole Higgs mass range.

The situation is different for $x=1000$ GeV with $\lambda$, $k$
unchanged. Here the singlet component of the lightest Higgs scalar
with a mass up to about 100 GeV practically vanishes
so that the NMSSM Higgs mass bound becomes
similar to that in the MSSM for $\tan\beta=2$ or even stronger
for $\tan\beta=10$. With decreasing parameter $k$ the possible
singlet component of the lightest Higgs scalar increases, so that
for $k=0.01$ and $\tan\beta=10$ a 10-GeV Higgs scalar is not excluded.

The couplings between a scalar and pseudoscalar Higgs and a $Z$ boson
in the NMSSM are shown in Fig.~\ref{pseudoplots}. Since for
small singlet vacuum expectation values $x$ the lightest Higgs scalar has
a large singlet component, while for large $x$ the light pseudoscalar
Higgs boson is almost a pure singlet,
they are always rather small.
For our representative choice of parameters, their largest squared ratio
relative to the SM $\Phi ZZ$ coupling reaches
about 1 \%  at $x=200$ GeV and
decreases to $10^{-6}$ for $x=1000$ GeV and $\tan\beta =2$. Larger values of
$\tan\beta$ lead to even smaller $S_{1,2}P_1Z$ couplings.
As a consequence the experimental bounds from the direct search for
pseudoscalar Higgs bosons produced together with a Higgs scalar at LEP
do not significantly extend the excluded parameter domain or
raise the Higgs mass bounds.

The constraints for the parameters $A_\lambda$ and $A_k$ of the Higgs sector
and the Higgs masses are summarized in Fig.~\ref{alakplots}.
Here the region above the
$m_{S_1}=0$ contour line is forbidden because the
mass squared would become negative, while the domain beyond the dashed line
is excluded since there exists an alternative lower minimum of the Higgs
potential with vanishing vacuum expectation values.
For $x=200$ GeV a massless Higgs scalar is not excluded by the present
LEP bounds while for $x=1000$ GeV a large value of $\tan\beta$ and
a small parameter $k$ are
necessary for a light scalar Higgs boson. More details for
the interpretation of $(A_\lambda, A_k)$ plots and excluded parameter
regions due to LEP constraints can be found in ref.~\cite{frankehiggs}.

\subsection{Constraints from neutralino search}
Contrary to the MSSM, neutralino and Higgs sectors are strongly
correlated in the NMSSM. In addition to the parameters of the Higgs sector
only the gaugino mass parameters $M$ and $M'$ have to be
fixed in order to determine the masses and mixings of the
neutralinos. So the LEP bounds from neutralino and Higgs searches
have to be combined in order to constrain the NMSSM parameter space
most effectively, e.~g.~small singlet vacuum expectation values
$x<14$ GeV are ruled out for $\tan\beta=2$ \cite{frankehiggs}.

The consequences from the negative neutralino search at LEP
for the parameter space and the neutralino masses have been studied
in ref.~\cite{frankeneu}. We review our previous analysis with the
now slightly improved limits \cite{neubounds}
\begin{enumerate}
\item for new physics contributing to the total $Z$ width
\begin{equation}
\Delta \Gamma _Z < 23.1 \; \mbox{MeV},
\end{equation}
\item for new physics contributing to the invisible $Z$ width
\begin{equation}
\Delta \Gamma _{\mbox{\scriptsize inv}} < 8.4 \; \mbox{MeV},
\end{equation}
\item from the direct neutralino search. Here we extract the following bounds
\begin{eqnarray}
B (Z \rightarrow \tilde{\chi}^0_1 \tilde{\chi}^0_j)
& < & 2 \times 10 ^{-5}, \hspace*{0.5cm} j=2,\ldots,5, \\
B (Z \rightarrow \tilde{\chi}^0_i \tilde{\chi}^0_j)
& < & 5 \times 10 ^{-5}, \hspace*{0.5cm} i,j=2,\ldots,5.
\end{eqnarray}
\end{enumerate}
In Fig.~\ref{lkplots} we show the excluded parameter space due to
the above constraints in the $(\lambda ,k)$ plane for the same
parameters as in the last section, Fig.~\ref{mxplots} depicts the
excluded regions in the $(M,x)$ plane for the same parameters as
in ref.~\cite{frankeneu} in order to study the effects of the
new LEP bounds. For this plots, the usual gaugino mass relation
\begin{equation}
M'  =  \frac{5}{3} \frac{{g'}^2}{g^2} M \simeq 0.5 M
\end{equation}
is employed.

The experimentally excluded
parameter space in Fig.~\ref{lkplots}
generally becomes smaller for larger values of $x$.
For larger $\tan\beta$ the excluded region from the total
$Z$ width measurements also decreases, but on the other hand
now the invisible $Z$ width and direct neutralino search rule out
an increasing domain.

Fig.~\ref{lkplots} shows that the improved LEP bounds only
slightly changed the excluded domain. Generally, the neutralino
mass bounds derived in ref.~\cite{frankeneu} are not affected
by these results, e.~g.~a massless neutralino is still allowed
in the NMSSM. In ref.~\cite{franke4}
it is shown that a very
light NMSSM neutralino cannot even be ruled out at LEP2, so that
its exclusion (or detection) will be one of the challenges at a
future linear $e^+e^-$ collider.

\section{Higgs couplings}
\label{sec:coupl}
In this section we confront some particular couplings of NMSSM  and MSSM,
which are supposed to have some important phenomenological
implications for supersymmetric processes at future particle colliders.
Our analysis includes the Higgs couplings to quarks, which are
generally suppressed in the NMSSM compared to the minimal model, but also
the vertex factors with Higgs bosons and scalar quarks, neutralinos
or charginos that can be suppressed or enhanced according to the
choice of the NMSSM parameters. The Feynman rules with quarks and
squarks are relevant for Higgs production via gluon fusion at proton
colliders or Higgs decay into gluons or photons.
For the Higgs decay into a photon pair also the Higgs-chargino-chargino
couplings may be important. Furthermore, a
strong coupling of the Higgs bosons to two lightest neutralinos may
enhance an invisible supersymmetric decay mode. Finally, the
Higgs self-coupling may be probed at a linear $e^+e^-$ collider.
So all these couplings are suited for decisive tests of the NMSSM.

In all plots in this section, the experimental constraints described
in Sec.~\ref{sec:con} are included.
First we consider the couplings of the scalar neutral Higgs bosons
to quarks.
In the SM as well as in the MSSM and NMSSM, the Higgs-quark-antiquark
couplings are proportional to the quark mass. Both the MSSM and
NMSSM couplings obtain a factor depending on the Higgs mixing
angles which can either suppress or enhance the vertex function.
In Fig.~\ref{figquarkcoupl} the NMSSM and MSSM
couplings of the two light neutral
scalar Higgs bosons with two top and two bottom quarks are plotted
relative to the SM values as a function of the Higgs mass.
For these plots we set $A_t=A_b=0$. The range of the NMSSM
couplings is computed by scanning over all experimentally allowed
parameters $A_\lambda$ and $A_k$.

For all scenarios, the squared MSSM
$ht\bar{t}$ coupling is smaller than the SM value, while the squared
$hb\bar{b}$ coupling is larger. The enhancement of the vertex
functions with bottom quarks and the suppression of those with top
quarks becomes stronger with increasing values of $\tan\beta$.
The NMSSM couplings, however, are always suppressed relative
to the SM vertex functions in our scenarios where the
light Higgs bosons may have significant singlet components.
While the minimum of the ratio for the NMSSM couplings is of the order of
$10^{-1}$ at the lower Higgs mass bound, it significantly decreases
with increasing mass of the light Higgs boson. For a large Higgs mass
range the Higgs couplings to quarks could practically vanish, so that the
loop diagrams with quarks contributing to the Higgs-gluon-gluon
and Higgs-$\gamma$-$\gamma$ couplings can be neglected. Since, however,
the quark loops are generally dominant for typical scenarios with
nearly degenerate squark masses \cite{higgs}, also the production of the
lightest NMSSM Higgs boson
via gluon fusion is heavily suppressed in this case. This disadvantage for
the Higgs search may be partially balanced by the fact, that
the quark couplings
of the second lightest Higgs boson may become rather large.

We now turn to the squark couplings. As already mentioned in
Sec.~\ref{sec:hsqsq}, the vertex functions for one Higgs boson and
one left-handed and and right-handed squark explicitly depend on
the Higgs singlet component, so that they may be significantly
enhanced or suppressed. In Fig.~\ref{figsquarkcoupl} we plot the "average"
Higgs couplings to squarks
\begin{equation}
\label{squarkratio}
\sum_{i,j=L,R} \frac{g^2_{S_a \tilde{u}_i\tilde{u}_j}+
g^2_{S_a \tilde{d}_i\tilde{d}_j} }{8g^2m_W^2} =
\frac{<g^2_{S_a \tilde{q}\tilde{q}}>}{g^2m_W^2}
\end{equation}
for the lightest and next-to-lightest Higgs scalar in the NMSSM and
the MSSM. Again these couplings are plotted
for third generation squarks, the range for the NMSSM couplings is
obtained by scanning over the allowed
$(A_\lambda, A_k)$-plane with $A_t=A_b =0$.
For $x=200$ GeV, the NMSSM couplings for the lightest Higgs scalar are
weakly suppressed relative to the MSSM vertex functions, while the
$S_2$ couplings are slightly enhanced. For $x=1000$ GeV the NMSSM couplings
may become rather weak in the case of small parameters $k$ or
comparable to the MSSM couplings for some Higgs mass regions
if the value for $k$ is increased.

The couplings of the neutral Higgs bosons to neutralinos and
charginos are important in order to determine the branching ratios for
the supersymmetric decay channels. The vertex factors for
charginos also affect the loop decay into two photons. The
NMSSM couplings explicitly differ from the MSSM vertex functions
since they
depend on the singlet components of the Higgs bosons as well as of the
neutralinos. Fig.~\ref{figcharneucoupl} shows the squared SUSY
Higgs couplings to neutralino and charginos relative to $g^2$
for the gaugino mass parameters $M=100$ GeV, $M'=0.5M$.
These vertex functions crucially depend on the
mixing of the Higgs bosons as well as on the mixing of the neutralinos and
charginos. Due to the strong correlation of neutralino and Higgs sectors
in the NMSSM, the possiblities for neutralino mixing are restricted
if the Higgs parameters are fixed.
While in the scenarios for the figures
the light Higgs bosons have significant singlet components,
the light neutralinos are not dominantly singlets.
For $k=0.1$ the singlet component of the
lightest neutralino is smaller than 10 \% , it increases with
decreasing $k$ values. For a detailed discussion of the neutralino
singlet component as a function of the NMSSM parameters see
ref.~\cite{franke4}.
Due to the neutralino and chargino mixing the NMSSM couplings of
the lightest neutral scalar Higgs boson
to neutralinos and charginos are generally somewhat smaller in our
scenarios than the MSSM couplings.
As for the previously discussed
couplings, there again exists a Higgs mass range where the NMSSM
couplings could be heavily suppressed.

Finally we compare the trilinear Higgs self-couplings of NMSSM, MSSM
and SM. In the SM the vertex function is directly proportional
to the squared Higgs mass $m_{\Phi}^2$
\begin{equation}
g_{\scriptsize \Phi \Phi \Phi}=\frac{3}{2} \frac{ m_{\Phi}^2}{m_W},
\end {equation}
while the MSSM couplings
are suppressed \cite{ellis,hunter}:
\begin{eqnarray}
g_{\scriptsize hhh} & = & \frac{-3}{2} \frac{gm_z}{\cos\theta_W}
\cos 2\alpha \sin
(\alpha + \beta ) ,\\
g_{\scriptsize HHH} & = & \frac{-3}{2} \frac{gm_z}{\cos\theta_W}
\cos 2\alpha \cos
(\alpha + \beta ) ,\\
g_{\scriptsize hHH} & = & \frac{g}{2} \frac{m_z}{\cos\theta_W}
\left( 2\sin 2\alpha \cos
(\alpha + \beta ) + \cos 2\alpha \sin (\alpha + \beta ) \right) ,\\
g_{\scriptsize Hhh} & = & \frac{g}{2} \frac{m_z}{\cos\theta_W} \left(
2\sin 2\alpha \sin
(\alpha + \beta ) - \cos 2\alpha \cos (\alpha + \beta ) \right) ,\\
g_{\scriptsize hAA} & = & \frac{-g}{2} \frac{m_z}{\cos\theta_W}
\cos 2\beta \sin
(\alpha + \beta ) ,\\
g_{\scriptsize HAA} & = & \frac{-g}{2} \frac{m_z}{\cos\theta_W}
\cos 2\beta \cos
(\alpha + \beta ) .
\end{eqnarray}
In the NMSSM, however, the trilinear Higgs self-couplings
contain terms proportional to the
singlet vacuum expectation value $x$ and the parameters $A_\lambda$ and
$A_k$. Since these parameters can in principle become as large as
some TeV even for a very small Higgs mass, the Higgs self-couplings
may become very strong.
We show in Fig.~\ref{selfplots} the
trilinear self-coupling of the lightest Higgs scalar and also the coupling
between
two lightest pseudoscalar Higgs particles and the lightest scalar
Higgs boson in the NMSSM and MSSM. Again, the couplings are
normalized with respect to the self-coupling in the SM.
While the MSSM self-couplings are always smaller than the
SM couplings with a Higgs boson of the same mass, the situation is
completely different in the NMSSM. Here large $A_\lambda$ and $A_k$ values
lead to an increase of the squared self-couplings by a factor up
to $10^5$. This fact could be a crucial key for distinguishing
between minimal and nonminimal model, if one probes the trilinear
Higgs vertex e.~g.~in processes as $e^+e^- \rightarrow S_1S_1Z$ at a
future linear collider \cite{ilyin}.

\section{Conclusion}
We have presented a complete set of Feynman rules in the NMSSM which
can be used as starting point for all calculations
of NMSSM processes. The couplings of NMSSM and MSSM
mainly differ by the mixing of the neutralinos and Higgs bosons,
which contain an additional singlet component in the NMSSM.
The following NMSSM vertex functions depend only on
the doublet components of the Higgs bosons and are therefore
generally suppressed in the NMSSM compared to the minimal model:
\begin{itemize}
\item the trilinear and quartic Higgs couplings to gauge bosons.
Contrary to the MSSM, however, the vertex functions with two
different neutral scalar Higgs bosons and two gauge bosons
do not vanish in the NMSSM.
\item the trilinear Higgs couplings to quarks and leptons.
\item the trilinear or quartic Higgs couplings to two left-handed or
two right-handed
scalar quarks or leptons.
\end{itemize}
Moreover, some Feynman rules depend explicitly on the
singlet components of the Higgs bosons. These NMSSM vertex functions
which may be suppressed or enhanced compared to the MSSM are
\begin{itemize}
\item the trilinear and quartic Higgs couplings to one left-handed and
one right-handed scalar quark. The quartic Higgs couplings to one
left-handed and one right-handed scalar quark or lepton even vanish
in the MSSM.
\item the trilinear and quartic Higgs self-couplings. These vertex functions
probably exhibit the most significant differences between the minimal and
nonminimal supersymmetric model. Unfortunately, they cannot be fully tested
until new powerful colliders start operating.
\item the trilinear coupling of one Higgs boson to neutralinos
or charginos.
\end{itemize}

We have demonstrated some fundamental differences between the SM, the
minimal and the nonminimal supersymmetric model by comparing
the Higgs couplings to gauge bosons, quarks, scalar quarks,
neutralinos and charginos as well as the
trilinear Higgs self-couplings.
The vertex functions with gauge bosons
are generally reduced in both supersymmetric
models compared to the SM. The MSSM quark coupling may be suppressed or
enhanced
relative to the SM, but is always enhanced compared to the NMSSM.
The ratio between the couplings with scalar quarks, neutralinos,
charginos as well as
the Higgs self-couplings of MSSM and NMSSM is not determined, it depends
on the choice for the parameters and therefore on the Higgs and neutralino
masses and mixings.

We also have reanalyzed the excluded parameter space by applying
the derived Feynman rules to
the experimental bounds from the so far negative search for Higgs bosons
and neutralinos. Especially very light neutralinos and Higgs bosons
are not excluded in the NMSSM.

Until now, there is no experimental evidence against supersymmetry --
the by far largest part of the parameter space of NMSSM and MSSM
is compatible with all experimental results.
In order to reveal the nature of new physics
all phenomenological implications of the different supersymmetric
models have to be studied by computing relevant cross section and
decay rates and
by providing Monte-Carlo Simulations for
the present detectors.
The discussion of concrete supersymmetric processes
was beyond the scope of this paper.
The derived Feynman rules of the NMSSM
represent an indispensable prerequisite for this task clearing the way for
further efforts
needed to verify or to exclude supersymmetry at the next generation
of particle colliders.
\section*{Acknowledgements}
We would like to thank S.~Hesselbach for many helpful comments on the
manuscript.
This work was supported by the Deutsche Forschungsgemeinschaft under contract
no.~FR 1064/2-1.

\newpage
\clearpage
\newpage
\newpage
\newpage
\newpage
\newpage
\newpage
\begin{figure}[p]
\begin{center}
\begin{picture}(12,9)
\put(2.0,2.3){$P_\delta$}
\put(2.0,5.3){$P_\gamma$}
\put(0.3,2.3){$P_\alpha$}
\put(0.3,5.3){$P_\beta$}
\put(-0.8,1.2){\includegraphics{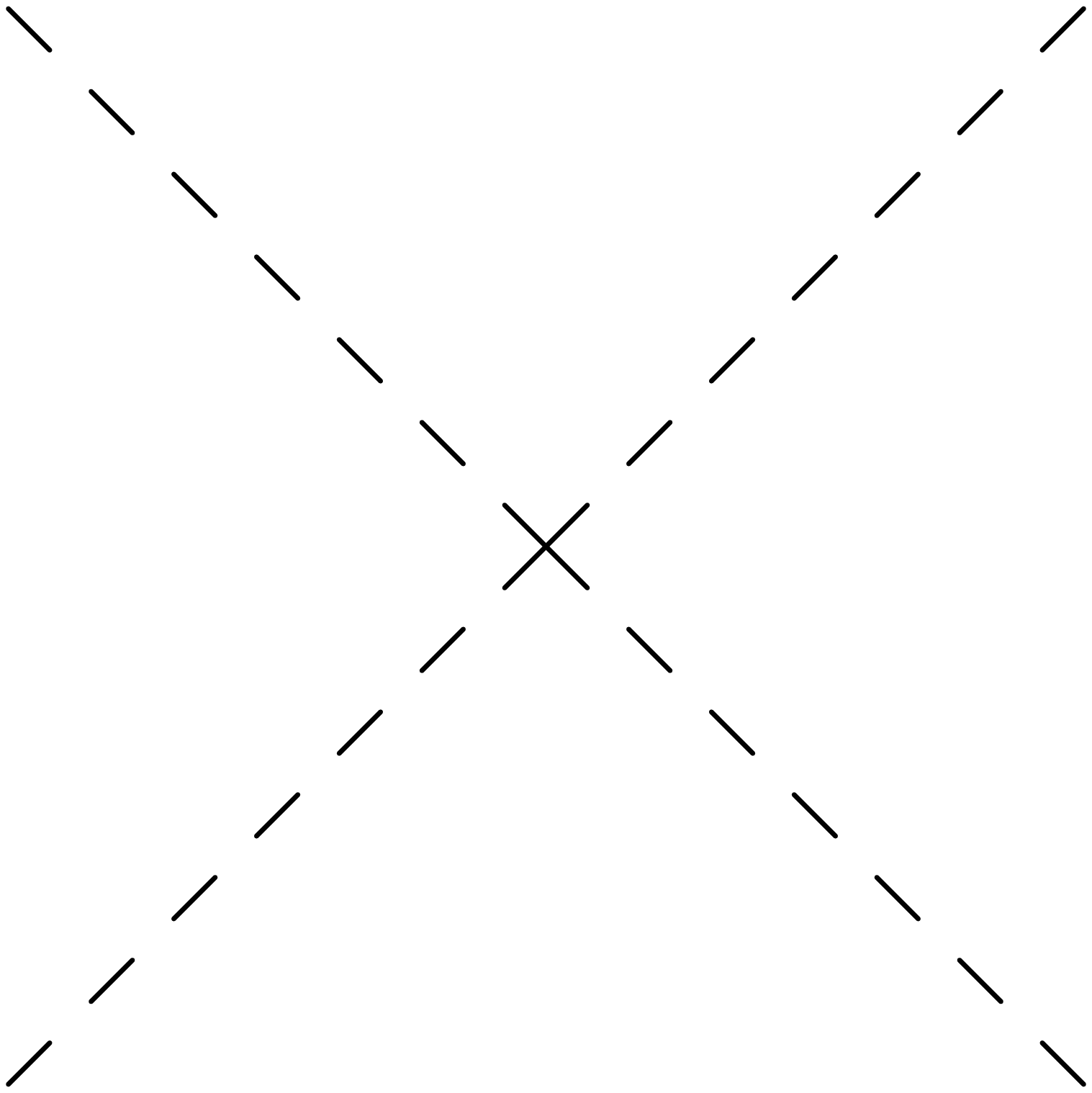}}
\put(3.5,7){$
-i\frac{3}{4}(g^2+{g'}^{2})  (
U^P_{\alpha 1}U^P_{\beta 1}U^P_{\gamma 1}U^P_{\delta 1}+
U^P_{\alpha 2}U^P_{\beta 2}U^P_{\gamma 2}U^P_{\delta 2})$}
\put(3.5,6.4){$+i\left( \frac{1}{4}(g^2+{g'}^{2})-\lambda^2 \right)$}
\put(4.8,5.8){$(U^P_{\alpha 1}U^P_{\beta 1}U^P_{\gamma 2}U^P_{\delta 2}+
U^P_{\alpha 1}U^P_{\beta 2}U^P_{\gamma 1}U^P_{\delta 2}+
U^P_{\alpha 1}U^P_{\beta 2}U^P_{\gamma 2}U^P_{\delta 1}$}
\put(5.0,5.2){$+U^P_{\alpha 2}U^P_{\beta 1}U^P_{\gamma 1}U^P_{\delta 2}+
U^P_{\alpha 2}U^P_{\beta 1}U^P_{\gamma 2}U^P_{\delta 1}+
U^P_{\alpha 2}U^P_{\beta 2}U^P_{\gamma 1}U^P_{\delta 1})$}
\put(3.5,4.6){$-i\lambda^2  (
U^P_{\alpha 1}U^P_{\beta 1}U^P_{\gamma 3}U^P_{\delta 3}+
U^P_{\alpha 1}U^P_{\beta 3}U^P_{\gamma 1}U^P_{\delta 3}+
U^P_{\alpha 1}U^P_{\beta 3}U^P_{\gamma 3}U^P_{\delta 1}$}
\put(5.0,4.0){$+U^P_{\alpha 3}U^P_{\beta 1}U^P_{\gamma 1}U^P_{\delta 3}+
U^P_{\alpha 3}U^P_{\beta 1}U^P_{\gamma 3}U^P_{\delta 1}+
U^P_{\alpha 3}U^P_{\beta 3}U^P_{\gamma 1}U^P_{\delta 1}$}
\put(5.0,3.4){$+U^P_{\alpha 2}U^P_{\beta 2}U^P_{\gamma 3}U^P_{\delta 3}+
U^P_{\alpha 2}U^P_{\beta 3}U^P_{\gamma 2}U^P_{\delta 3}+
U^P_{\alpha 2}U^P_{\beta 3}U^P_{\gamma 3}U^P_{\delta 2}$}
\put(5.0,2.8){$+U^P_{\alpha 3}U^P_{\beta 2}U^P_{\gamma 2}U^P_{\delta 3}+
U^P_{\alpha 3}U^P_{\beta 2}U^P_{\gamma 3}U^P_{\delta 2}+
U^P_{\alpha 3}U^P_{\beta 3}U^P_{\gamma 2}U^P_{\delta 2})$}
\put(3.5,2.2){$-6ik^2 U^P_{\alpha 3}U^P_{\beta 3}U^P_{\gamma 3}U^P_{\delta 3}$}
\put(3.5,1.6){$-i \lambda k  (
U^P_{\alpha 1}U^P_{\beta 2}U^P_{\gamma 3}U^P_{\delta 3}+
U^P_{\alpha 1}U^P_{\beta 3}U^P_{\gamma 2}U^P_{\delta 3}+
U^P_{\alpha 1}U^P_{\beta 3}U^P_{\gamma 3}U^P_{\delta 2}$}
\put(5.0,1.0){$+U^P_{\alpha 2}U^P_{\beta 1}U^P_{\gamma 3}U^P_{\delta 3}+
U^P_{\alpha 2}U^P_{\beta 3}U^P_{\gamma 1}U^P_{\delta 3}+
U^P_{\alpha 2}U^P_{\beta 3}U^P_{\gamma 3}U^P_{\delta 1}$}
\put(5.0,0.4){$+U^P_{\alpha 3}U^P_{\beta 1}U^P_{\gamma 2}U^P_{\delta 3}+
U^P_{\alpha 3}U^P_{\beta 1}U^P_{\gamma 3}U^P_{\delta 2}+
U^P_{\alpha 3}U^P_{\beta 2}U^P_{\gamma 1}U^P_{\delta 3}$}
\put(5.0,-0.2){$+U^P_{\alpha 3}U^P_{\beta 2}U^P_{\gamma 3}U^P_{\delta 1}+
U^P_{\alpha 3}U^P_{\beta 3}U^P_{\gamma 1}U^P_{\delta 2}+
U^P_{\alpha 3}U^P_{\beta 3}U^P_{\gamma 2}U^P_{\delta 1})$}
\end{picture}
\begin{picture}(12,6)
\put(2.0,1.3){$C^-$}
\put(2.0,4.3){$C^+$}
\put(0.3,1.3){$S_b$}
\put(0.3,4.3){$S_a$}
\put(-0.8,0.2){\includegraphics{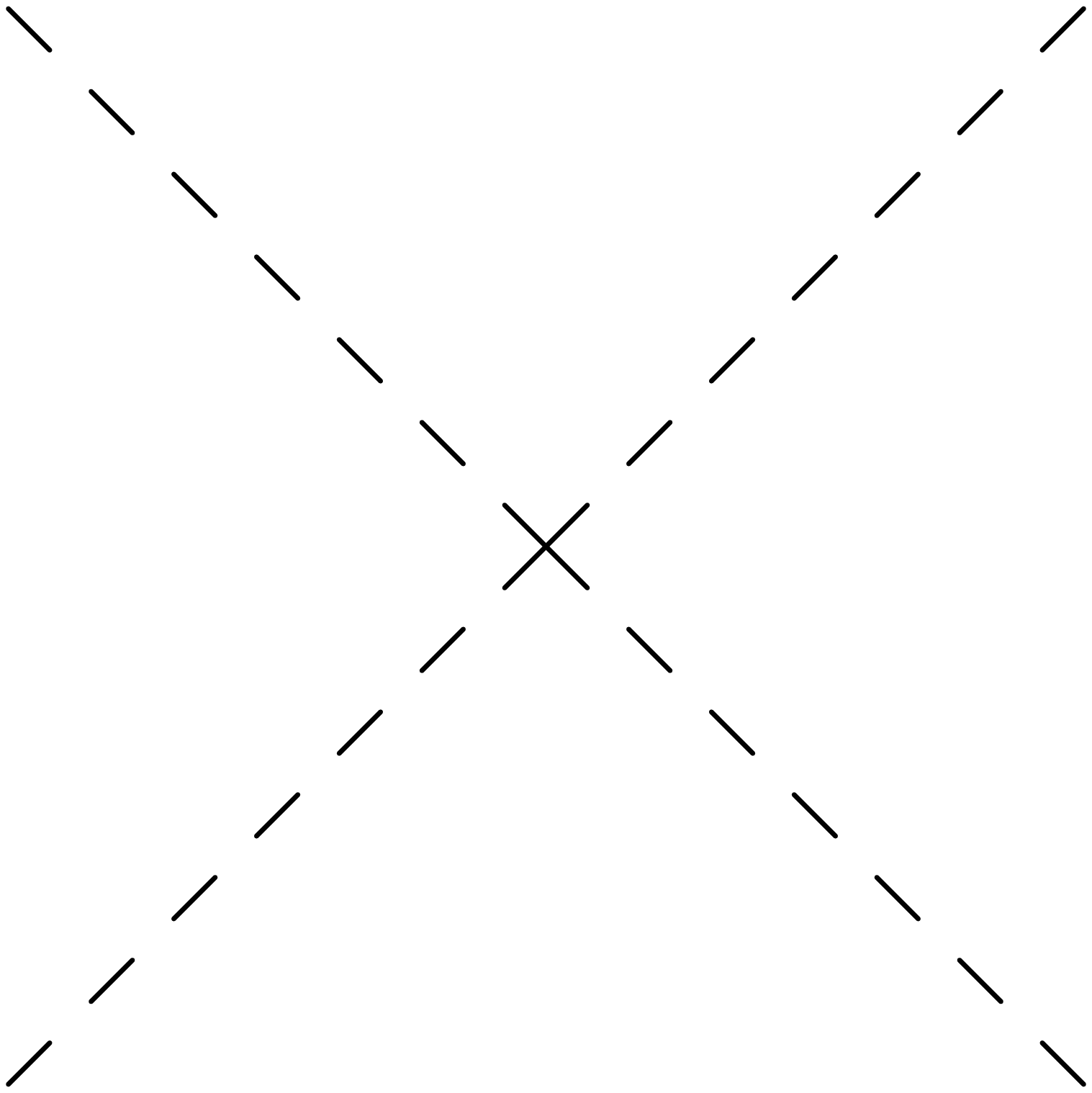}}
\put(5.5,3){\mbox{$\begin{array}{l}
-\frac{1}{4}ig^2
(U^S_{a2} U^S_{b2} +
U^S_{a1} U^S_{b1})
\\
-i\left( \frac{1}{4}g^2 -\frac{\lambda ^2}{2} \right)
(U^S_{a1}U^S_{b2}+U^S_{a2}U^S_{b1}) \sin 2\beta
\\
-\frac{1}{4}i{g'}^{2}
(U^S_{a2} U^S_{b2} -
U^S_{a1} U^S_{b1}) \cos 2\beta
\\
-i\lambda  (\lambda U^S_{a3} U^S_{b3}+
k U^S_{a3} U^S_{b3} \sin 2\beta  )
\end{array}$}}
\end{picture}
\begin{picture}(12,5)
\put(2.0,1.3){$C^-$}
\put(2.0,4.3){$C^+$}
\put(0.3,1.3){$P_\beta$}
\put(0.3,4.3){$P_\alpha$}
\put(-0.8,0.2){\includegraphics{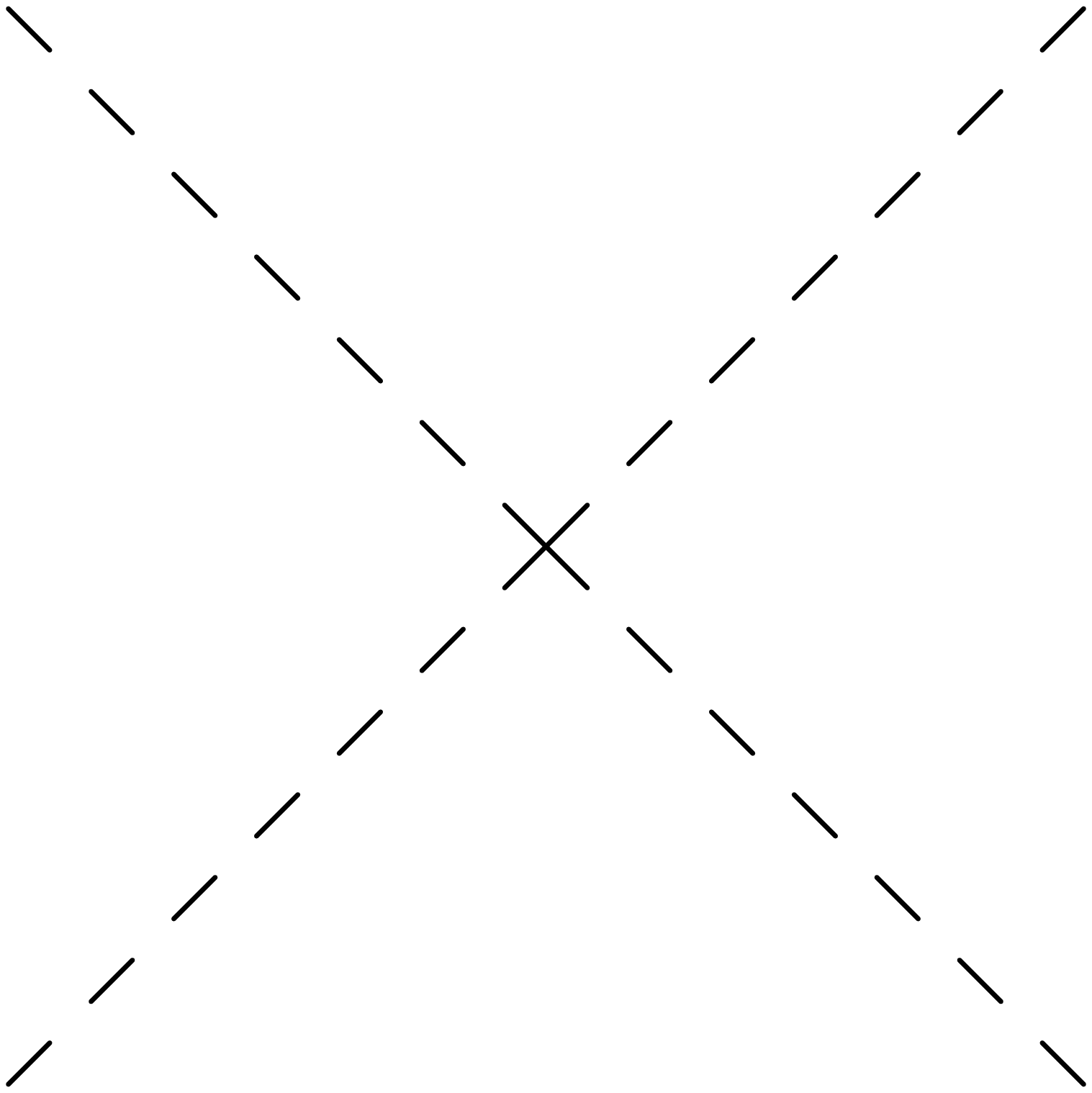}}
\put(5.5,3){\mbox{$\begin{array}{l}
-\frac{1}{4}ig^2
(U^P_{\alpha 2} U^P_{\beta 2} +
U^P_{\alpha 1} U^P_{\beta1})
\\
+i\left( \frac{1}{4}g^2 -\frac{\lambda ^2}{2} \right)
(U^P_{\alpha 1}U^P_{\beta 2}+U^P_{\alpha 2}U^P_{\beta 1}) \sin 2\beta
\\
-\frac{1}{4}i{g'}^{2}
(U^P_{\alpha 2} U^P_{\beta 2} -
U^P_{\alpha 1} U^P_{\beta 1}) \cos 2\beta
\\
-i\lambda  (\lambda U^P_{\alpha 3} U^P_{\beta 3}-
k U^P_{\alpha 3} U^P_{\beta 3} \sin 2\beta  )
\end{array}$}}
\put(-2.2,-1.0){Figure \ref{fig4h1} (continued): Feynman rules for the quartic
self-interactions of the Higgs bosons.}
\end{picture}
\end{center}
\end{figure}
\clearpage
\begin{figure}[p]
\begin{center}
\begin{picture}(12,4)
\put(0.8,2.3){$S_a$}
\put(2.8,0.5){$\tilde{\chi}^+_j$}
\put(2.8,3.0){$\tilde{\chi}^+_i$}
\put(0,-0.8){\includegraphics{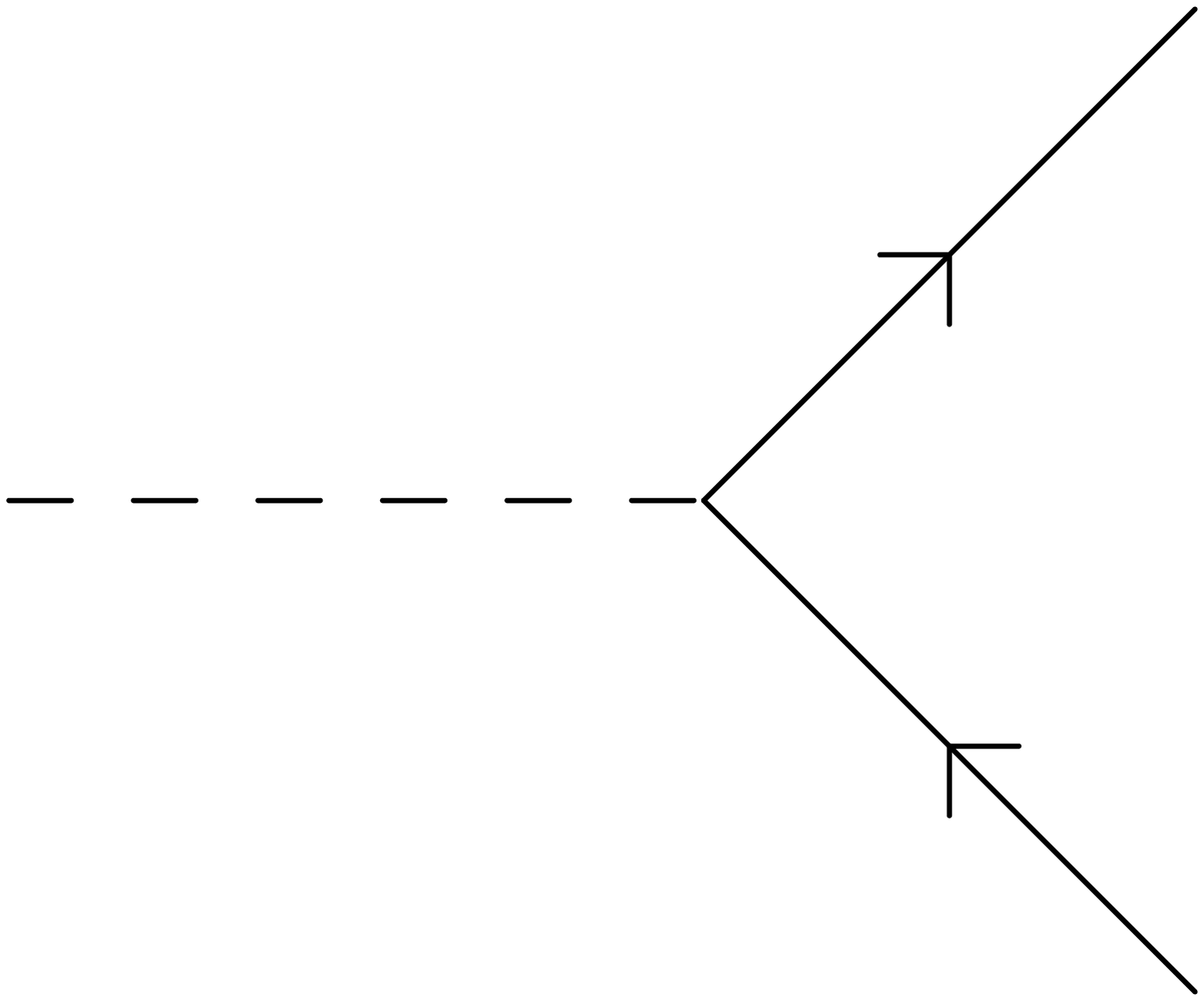}}
\put(7,2){\mbox{$-i(Q_{aij}^{\ast}P_L+Q_{aij}P_R)$}}
\end{picture}
\begin{picture}(12,4)
\put(0.8,2.3){$P_{\alpha}$}
\put(2.8,0.5){$\tilde{\chi}^+_j$}
\put(2.8,3.0){$\tilde{\chi}^+_i$}
\put(0,-0.8){\includegraphics{hqq.ps}}
\put(7,2){\mbox{$R_{\alpha ij}^{\ast}P_L-R_{\alpha ij}P_R$}}
\end{picture}
\begin{picture}(12,4)
\put(0.8,2.3){$C^-$}
\put(2.8,0.5){$\tilde{\chi}^+_j$}
\put(2.8,3.0){$\tilde{\chi}^0_i$}
\put(0,-0.8){\includegraphics{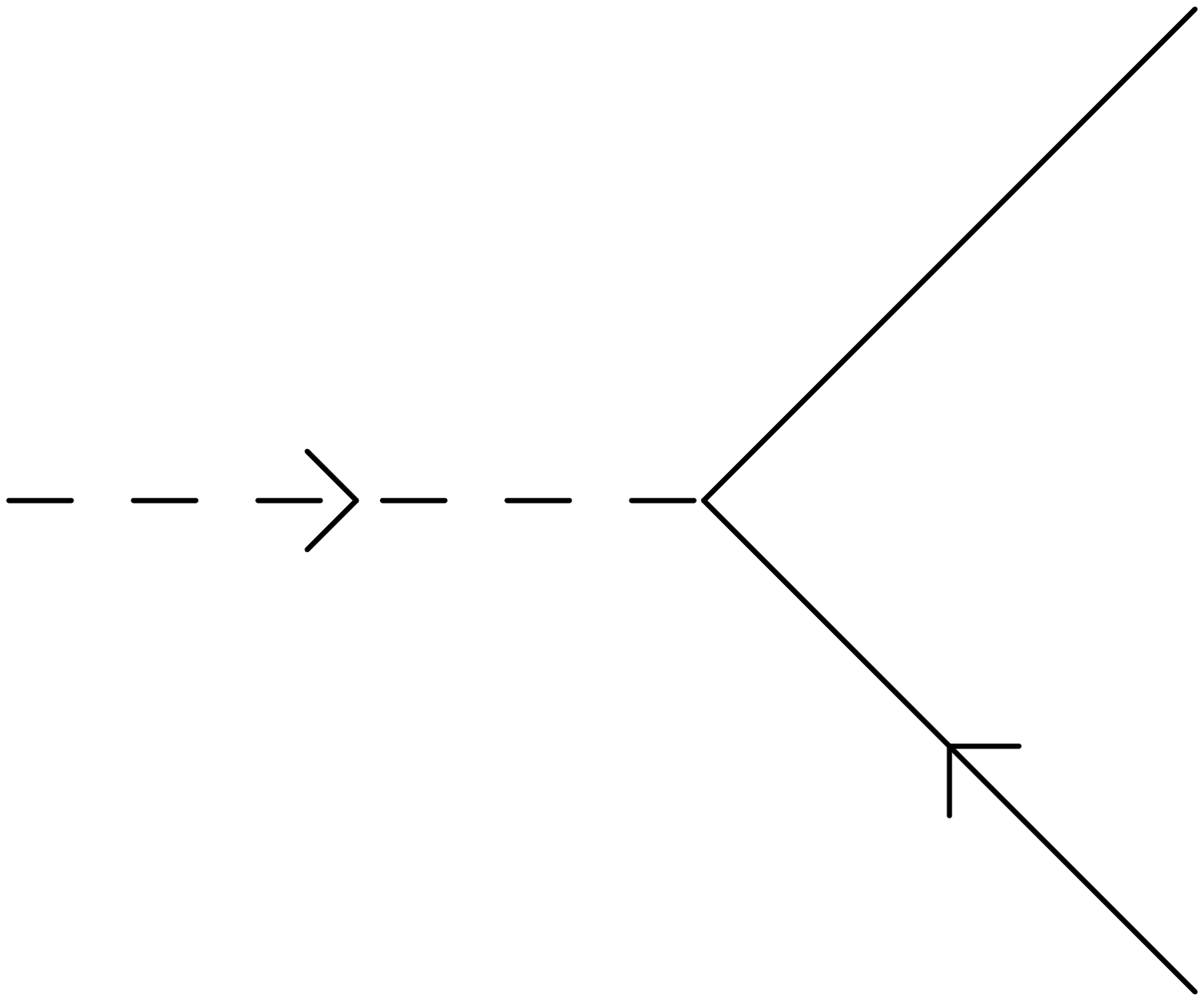}}
\put(7,2){\mbox{$i(Q_{ij}^{'L\ast}P_L+Q_{ij}^{'R}P_R)$}}
\end{picture}
\begin{picture}(12,4)
\put(0.8,2.3){$S_a$}
\put(2.8,0.5){$\tilde{\chi}^0_j$}
\put(2.8,3.0){$\tilde{\chi}^0_i$}
\put(0,-0.8){\includegraphics{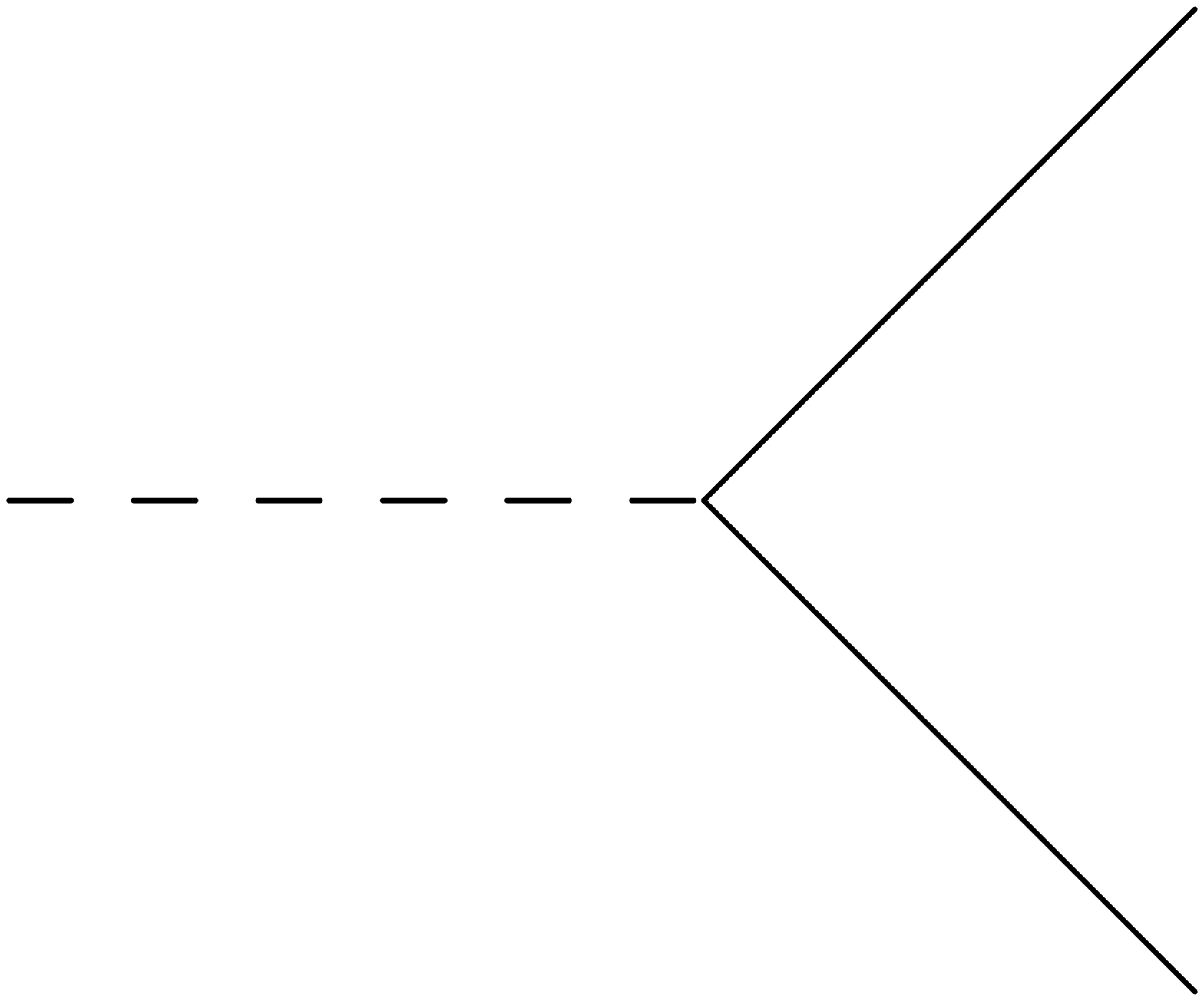}}
\put(7,2){\mbox{$-i(Q_{aij}^{L''}P_L+Q_{aij}^{R''}P_R)$}}
\end{picture}
\begin{picture}(12,4)
\put(0.8,2.3){$P_{\alpha}$}
\put(2.8,0.5){$\tilde{\chi}^0_j$}
\put(2.8,3.0){$\tilde{\chi}^0_i$}
\put(0,-0.8){\includegraphics{hqqohne.ps}}
\put(7,2){\mbox{$R_{\alpha ij}^{L''}P_L+R_{\alpha ij}^{R''}P_R$}}
\end{picture}
\end{center}
\caption{Feynman rules for the couplings of Higgs bosons to neutralinos
and charginos. The relevant factors $Q$ and $R$ are given in the text.}
\label{fighneuchar}
\end{figure}
\clearpage
\begin{figure}[p]
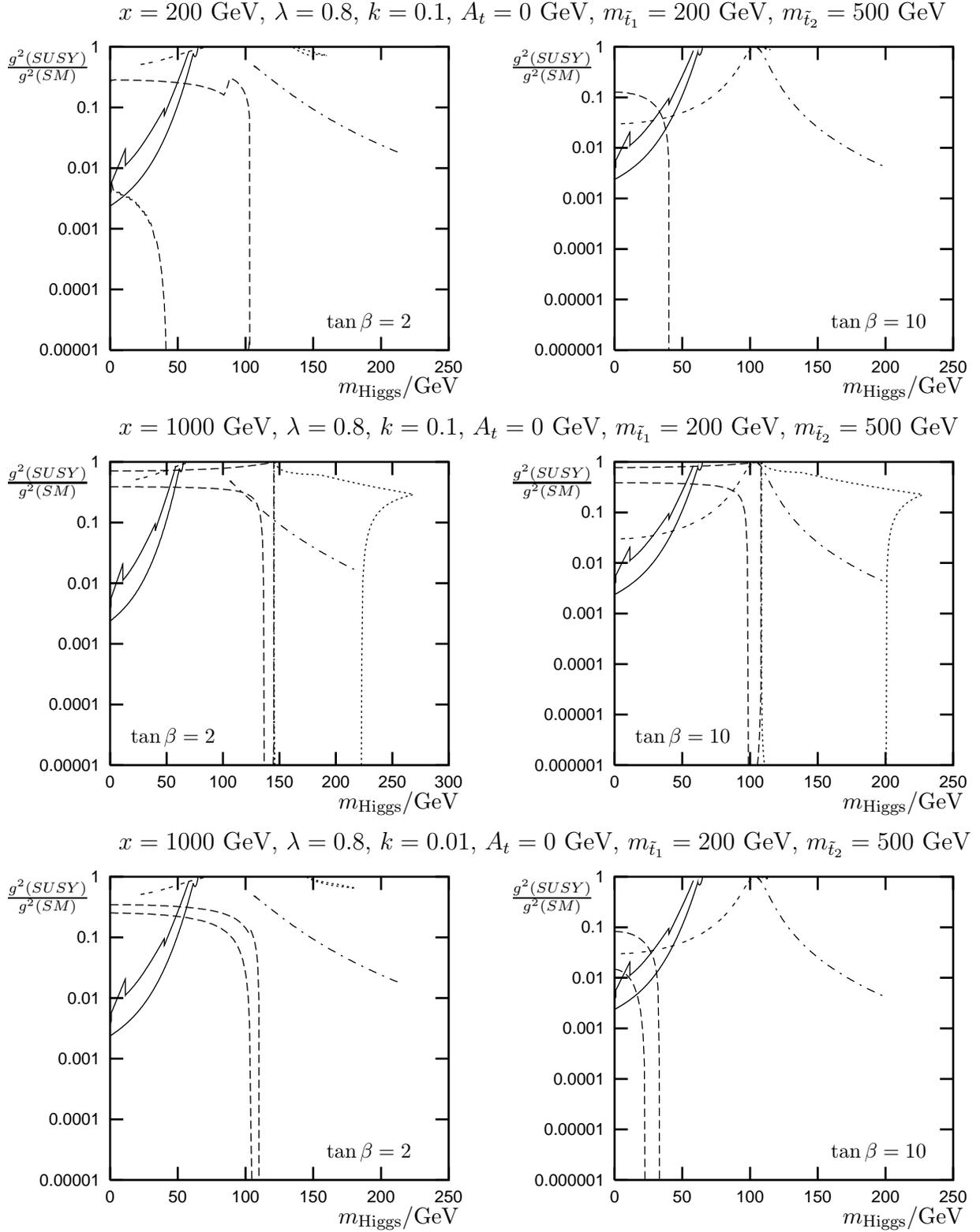

\begin{center}
\begin{picture}(16,18)
\put(0.5,14){\includegraphics{xi200_2.ps}}
\put(9.0,14){\includegraphics{xi200_10.ps}}
\put(0.5,7){\includegraphics{xi1000a_2.ps}}
\put(9.0,7){\includegraphics{xi1000a_10.ps}}
\put(0.5,0){\includegraphics{xi1000b_2.ps}}
\put(9.0,0){\includegraphics{xi1000b_10.ps}}
\put(2.0,20.0){$x=200$ GeV, $\lambda=0.8$, $k=0.1$, $A_t=0$ GeV,
$m_{\tilde{t}_1}=200$ GeV, $m_{\tilde{t}_2}=500$ GeV}
\put(2.0,13.0){$x=1000$ GeV, $\lambda=0.8$, $k=0.1$, $A_t=0$ GeV,
$m_{\tilde{t}_1}=200$ GeV, $m_{\tilde{t}_2}=500$ GeV}
\put(2.0,6.0){$x=1000$ GeV, $\lambda=0.8$, $k=0.01$, $A_t=0$ GeV,
$m_{\tilde{t}_1}=200$ GeV, $m_{\tilde{t}_2}=500$ GeV}
\put(5.5,14.8){\footnotesize $\tan\beta=2$}
\put(14.0,14.8){\footnotesize $\tan\beta=10$}
\put(2.2,7.8){\footnotesize $\tan\beta=2$}
\put(10.7,7.8){\footnotesize $\tan\beta=10$}
\put(5.5,0.8){\footnotesize $\tan\beta=2$}
\put(14.0,0.8){\footnotesize $\tan\beta=10$}
\put(5.7,13.7){\small $m_{\mbox{\scriptsize Higgs}}$/GeV}
\put(14.2,13.7){\small $m_{\mbox{\scriptsize Higgs}}$/GeV}
\put(5.7,6.7){\small $m_{\mbox{\scriptsize Higgs}}$/GeV}
\put(14.2,6.7){\small $m_{\mbox{\scriptsize Higgs}}$/GeV}
\put(5.7,-0.3){\small $m_{\mbox{\scriptsize Higgs}}$/GeV}
\put(14.2,-0.3){\small $m_{\mbox{\scriptsize Higgs}}$/GeV}
\put(0.1,19.1){\footnotesize $\frac{g^2(SUSY)}{g^2(SM)}$}
\put(8.6,19.1){\footnotesize $\frac{g^2(SUSY)}{g^2(SM)}$}
\put(0.1,12.1){\footnotesize $\frac{g^2(SUSY)}{g^2(SM)}$}
\put(8.6,12.1){\footnotesize $\frac{g^2(SUSY)}{g^2(SM)}$}
\put(0.1,5.1){\footnotesize $\frac{g^2(SUSY)}{g^2(SM)}$}
\put(8.6,5.1){\footnotesize $\frac{g^2(SUSY)}{g^2(SM)}$}
\end{picture}
\end{center}
\caption{The ratios of the SUSY Higgs couplings to two $Z$ bosons
relative to those of the SM.
Shown are the range for the NMSSM couplings
$g_{\scriptsize S_1ZZ}^2(NMSSM)/g_{\scriptsize \Phi ZZ}^2(SM)$ (dashed),
$g_{\scriptsize S_2ZZ}^2(NMSSM)/g_{\scriptsize \Phi ZZ}^2(SM)$ (dotted),
and the MSSM couplings
$g_{\scriptsize hZZ}^2(MSSM)/g_{\scriptsize \Phi ZZ}^2(SM)$ (double dashed),
$g_{\scriptsize HZZ}^2(MSSM)/g_{\scriptsize \Phi ZZ}^2(SM)$ (dashed-dotted).
The solid lines denote the experimental bounds from LEP1 for a visibly
and invisibly decaying Higgs boson.}
\label{xiplots}
\end{figure}
\clearpage
\begin{figure}[p]
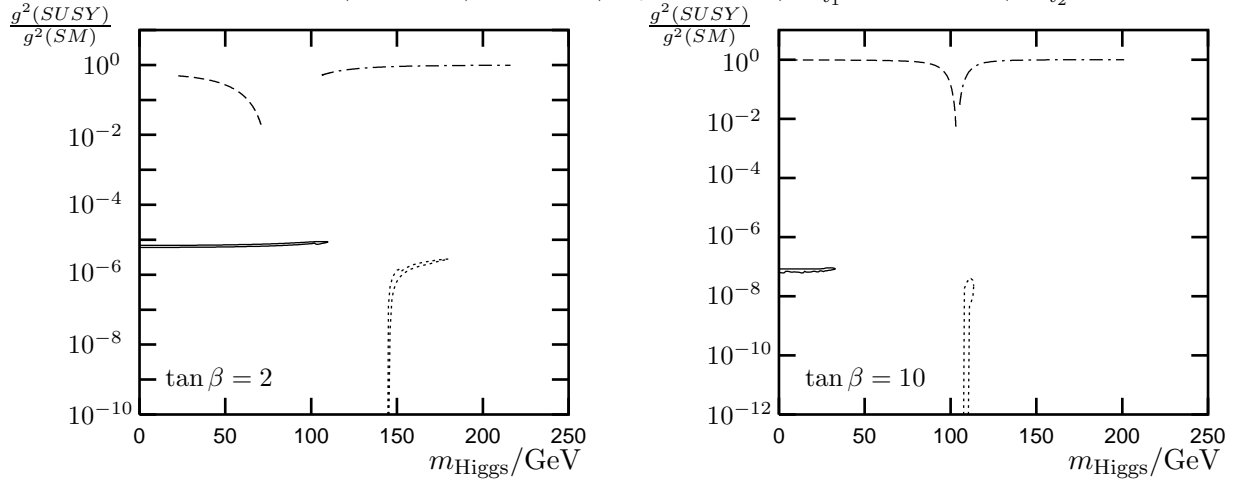

\begin{center}
\begin{picture}(16,18)
\put(0.5,14){\includegraphics{pseudo200_2.ps}}
\put(9.0,14){\includegraphics{pseudo200_10.ps}}
\put(0.5,7){\includegraphics{pseudo1000a_2.ps}}
\put(9.0,7){\includegraphics{pseudo1000a_10.ps}}
\put(0.5,0){\includegraphics{pseudo1000b_2.ps}}
\put(9.0,0){\includegraphics{pseudo1000b_10.ps}}
\put(2.0,20.0){$x=200$ GeV, $\lambda=0.8$, $k=0.1$, $A_t=0$ GeV,
$m_{\tilde{t}_1}=200$ GeV, $m_{\tilde{t}_2}=500$ GeV}
\put(2.0,13.0){$x=1000$ GeV, $\lambda=0.8$, $k=0.1$, $A_t=0$ GeV,
$m_{\tilde{t}_1}=200$ GeV, $m_{\tilde{t}_2}=500$ GeV}
\put(2.0,6.0){$x=1000$ GeV, $\lambda=0.8$, $k=0.01$, $A_t=0$ GeV,
$m_{\tilde{t}_1}=200$ GeV, $m_{\tilde{t}_2}=500$ GeV}
\put(2.2,14.8){\footnotesize $\tan\beta=2$}
\put(10.7,14.8){\footnotesize $\tan\beta=10$}
\put(2.2,7.8){\footnotesize $\tan\beta=2$}
\put(10.7,7.8){\footnotesize $\tan\beta=10$}
\put(2.2,0.8){\footnotesize $\tan\beta=2$}
\put(10.7,0.8){\footnotesize $\tan\beta=10$}
\put(5.7,13.7){\small $m_{\mbox{\scriptsize Higgs}}$/GeV}
\put(14.2,13.7){\small $m_{\mbox{\scriptsize Higgs}}$/GeV}
\put(5.7,6.7){\small $m_{\mbox{\scriptsize Higgs}}$/GeV}
\put(14.2,6.7){\small $m_{\mbox{\scriptsize Higgs}}$/GeV}
\put(5.7,-0.3){\small $m_{\mbox{\scriptsize Higgs}}$/GeV}
\put(14.2,-0.3){\small $m_{\mbox{\scriptsize Higgs}}$/GeV}
\put(0.1,19.1){\footnotesize $\frac{g^2(SUSY)}{g^2(SM)}$}
\put(8.6,19.1){\footnotesize $\frac{g^2(SUSY)}{g^2(SM)}$}
\put(0.1,12.5){\footnotesize $\frac{g^2(SUSY)}{g^2(SM)}$}
\put(8.6,12.5){\footnotesize $\frac{g^2(SUSY)}{g^2(SM)}$}
\put(0.1,5.5){\footnotesize $\frac{g^2(SUSY)}{g^2(SM)}$}
\put(8.6,5.5){\footnotesize $\frac{g^2(SUSY)}{g^2(SM)}$}
\put(0.9,0.3){\footnotesize $10^{-10}$}
\put(0.9,1.25){\footnotesize $10^{-8}$}
\put(0.9,2.2){\footnotesize $10^{-6}$}
\put(0.9,3.1){\footnotesize $10^{-4}$}
\put(0.9,4.0){\footnotesize $10^{-2}$}
\put(1.1,4.95){\footnotesize $10^{0}$}
\put(9.4,0.3){\footnotesize $10^{-12}$}
\put(9.4,1.1){\footnotesize $10^{-10}$}
\put(9.4,1.9){\footnotesize $10^{-8}$}
\put(9.4,2.7){\footnotesize $10^{-6}$}
\put(9.4,3.5){\footnotesize $10^{-4}$}
\put(9.4,4.2){\footnotesize $10^{-2}$}
\put(9.6,5.0){\footnotesize $10^{0}$}
\put(0.9,7.3){\footnotesize $10^{-10}$}
\put(0.9,8.25){\footnotesize $10^{-8}$}
\put(0.9,9.2){\footnotesize $10^{-6}$}
\put(0.9,10.1){\footnotesize $10^{-4}$}
\put(0.9,11.0){\footnotesize $10^{-2}$}
\put(1.1,11.95){\footnotesize $10^{0}$}
\put(9.4,7.3){\footnotesize $10^{-12}$}
\put(9.4,8.1){\footnotesize $10^{-10}$}
\put(9.4,8.9){\footnotesize $10^{-8}$}
\put(9.4,9.7){\footnotesize $10^{-6}$}
\put(9.4,10.5){\footnotesize $10^{-4}$}
\put(9.4,11.2){\footnotesize $10^{-2}$}
\put(9.6,12.0){\footnotesize $10^{0}$}
\end{picture}
\end{center}
\caption{The squared couplings between a scalar and pseudoscalar Higgs boson
and a $Z$ boson as a function of the scalar Higgs mass. Shown are the range
for the NMSSM couplings $g_{\scriptsize S_1P_1Z}^2(NMSSM)/
g_{\scriptsize \Phi ZZ}^2(SM)$
(solid line), $g_{\scriptsize S_2P_1Z}^2(NMSSM)/
g_{\scriptsize \Phi ZZ}^2(SM)$
(dotted) and the MSSM couplings
$g_{\scriptsize hAZ}^2(MSSM)/g_{\scriptsize \Phi ZZ}^2(SM)$ (dashed),
$g_{\scriptsize HAZ}^2(MSSM)/g_{\scriptsize \Phi ZZ}^2(SM)$ (dashed dotted).}
\label{pseudoplots}
\end{figure}
\clearpage
\begin{figure}[p]
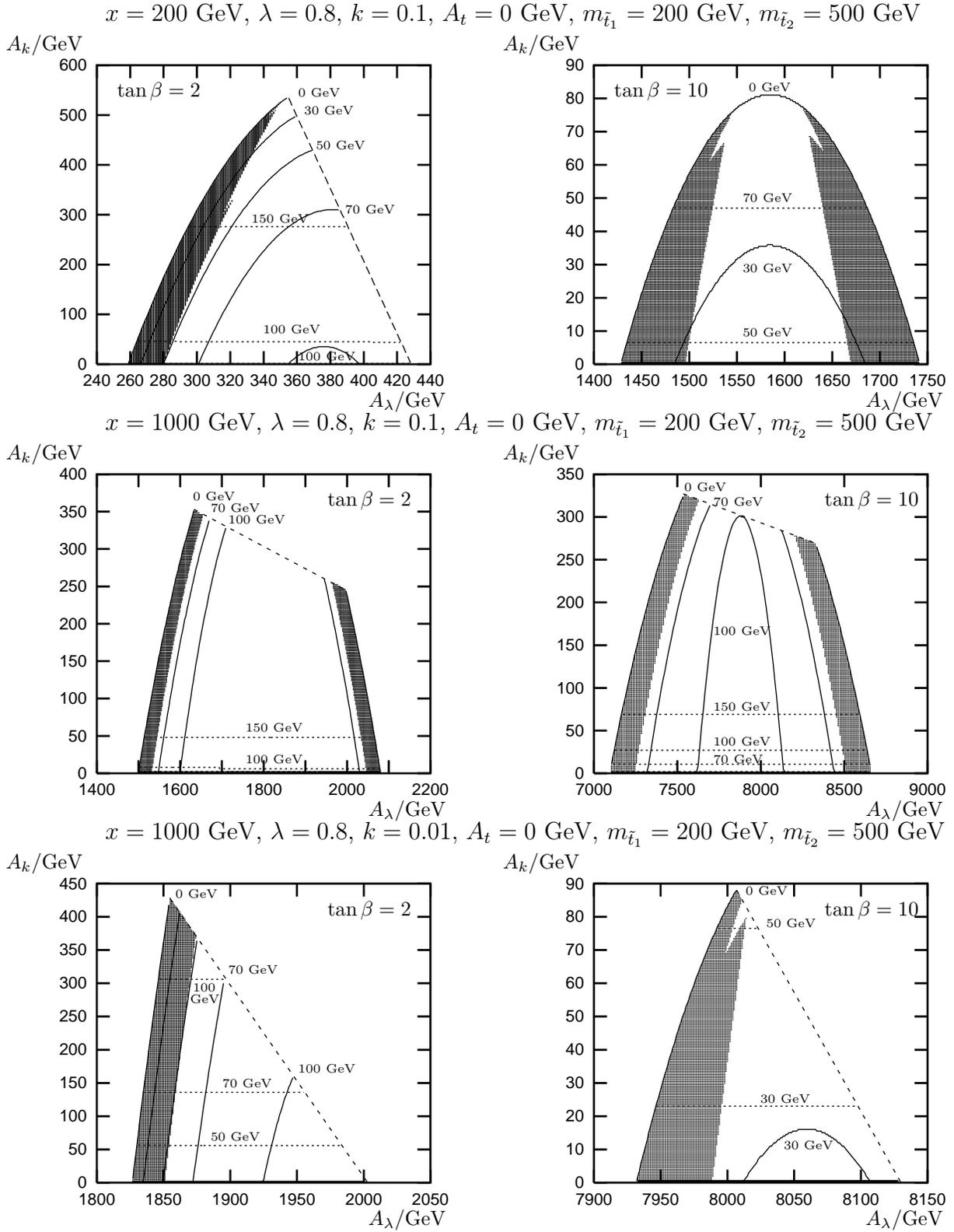

\begin{center}
\begin{picture}(16,18)
\put(0.5,14){\includegraphics{alak200_2.ps}}
\put(9.0,14){\includegraphics{alak200_10.ps}}
\put(0.5,7){\includegraphics{alak1000a_2.ps}}
\put(9.0,7){\includegraphics{alak1000a_10.ps}}
\put(0.5,0){\includegraphics{alak1000b_2.ps}}
\put(9.0,0){\includegraphics{alak1000b_10.ps}}
\put(2.0,20.3){$x=200$ GeV, $\lambda=0.8$, $k=0.1$, $A_t=0$ GeV,
$m_{\tilde{t}_1}=200$ GeV, $m_{\tilde{t}_2}=500$ GeV}
\put(2.0,13.3){$x=1000$ GeV, $\lambda=0.8$, $k=0.1$, $A_t=0$ GeV,
$m_{\tilde{t}_1}=200$ GeV, $m_{\tilde{t}_2}=500$ GeV}
\put(2.0,6.3){$x=1000$ GeV, $\lambda=0.8$, $k=0.01$, $A_t=0$ GeV,
$m_{\tilde{t}_1}=200$ GeV, $m_{\tilde{t}_2}=500$ GeV}
\put(2.2,19.0){\footnotesize $\tan\beta=2$}
\put(10.7,19.0){\footnotesize $\tan\beta=10$}
\put(5.8,12.0){\footnotesize $\tan\beta=2$}
\put(14.3,12.0){\footnotesize $\tan\beta=10$}
\put(5.8,5.0){\footnotesize $\tan\beta=2$}
\put(14.3,5.0){\footnotesize $\tan\beta=10$}
\put(6.5,13.7){\footnotesize $A_{\lambda }/$GeV}
\put(15.0,13.7){\footnotesize $A_{\lambda }/$GeV}
\put(6.5,6.7){\footnotesize $A_{\lambda }/$GeV}
\put(15.0,6.7){\footnotesize $A_{\lambda }/$GeV}
\put(6.5,-0.3){\footnotesize $A_{\lambda }/$GeV}
\put(15.0,-0.3){\footnotesize $A_{\lambda }/$GeV}
\put(0.3,19.8){\footnotesize $A_k/$GeV}
\put(8.8,19.8){\footnotesize $A_k/$GeV}
\put(0.3,12.8){\footnotesize $A_k/$GeV}
\put(8.8,12.8){\footnotesize $A_k/$GeV}
\put(0.3,5.8){\footnotesize $A_k/$GeV}
\put(8.8,5.8){\footnotesize $A_k/$GeV}
\put(5.3,19.0){\tiny 0 GeV}
\put(5.4,18.7){\tiny 30 GeV}
\put(5.6,18.1){\tiny 50 GeV}
\put(6.1,17.0){\tiny 70 GeV}
\put(5.3,14.5){\tiny 100 GeV}
\put(4.7,14.95){\tiny 100 GeV}
\put(4.5,16.85){\tiny 150 GeV}
\put(13.0,19.1){\tiny 0 GeV}
\put(12.9,17.2){\tiny 70 GeV}
\put(12.9,16.0){\tiny 30 GeV}
\put(12.9,14.9){\tiny 50 GeV}
\put(3.5,12.1){\tiny 0 GeV}
\put(3.8,11.9){\tiny 70 GeV}
\put(4.1,11.7){\tiny 100 GeV}
\put(4.4,8.15){\tiny 150 GeV}
\put(4.4,7.6){\tiny 100 GeV}
\put(11.9,12.25){\tiny 0 GeV}
\put(12.4,12.0){\tiny 70 GeV}
\put(12.4,9.8){\tiny 100 GeV}
\put(12.4,8.5){\tiny 150 GeV}
\put(12.4,7.9){\tiny 100 GeV}
\put(12.4,7.6){\tiny 70 GeV}
\put(3.2,5.3){\tiny 0 GeV}
\put(4.1,4.0){\tiny 70 GeV}
\put(3.5,3.7){\tiny 100}
\put(3.45,3.5){\tiny GeV}
\put(5.3,2.3){\tiny 100 GeV}
\put(4.0,2.05){\tiny 70 GeV}
\put(3.8,1.15){\tiny 50 GeV}
\put(12.95,5.35){\tiny 0 GeV}
\put(13.6,1.0){\tiny 30 GeV}
\put(13.2,1.8){\tiny 30 GeV}
\put(13.3,4.8){\tiny 50 GeV}
\end{picture}
\end{center}
\caption{The experimentally excluded parameter space in the
$(A_\lambda,A_k)$ plane (dark area). Also shown are
the contour lines for the mass of the lightest
Higgs scalar (solid) and of the light pseudoscalar Higgs boson (dotted).
The region beyond
the $m_{S_1}=0$-GeV line and the dashed line is theoretically excluded
as explained in the text.}
\label{alakplots}
\end{figure}
\clearpage
\begin{figure}[p]
\begin{center}
\begin{picture}(16,18)
\put(0.5,14){\includegraphics{lk100200_2.ps}}
\put(9.0,14){\includegraphics{lk100200_10.ps}}
\put(0.5,7){\includegraphics{lk200200_2.ps}}
\put(9.0,7){\includegraphics{lk200200_10.ps}}
\put(0.5,0){\includegraphics{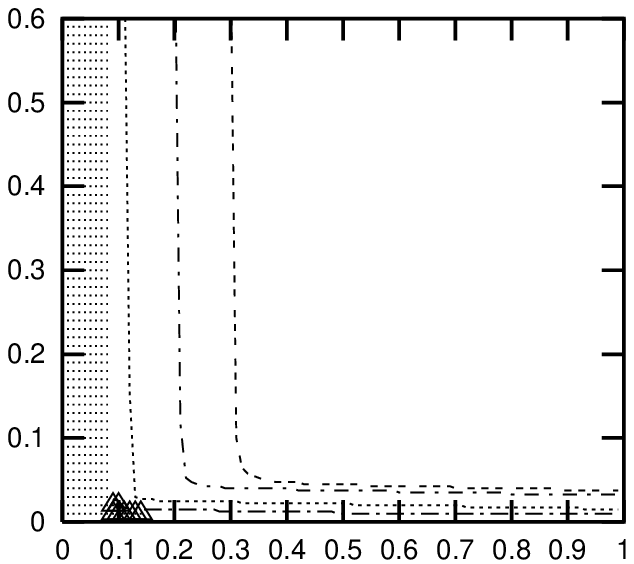}}
\put(9.0,0){\includegraphics{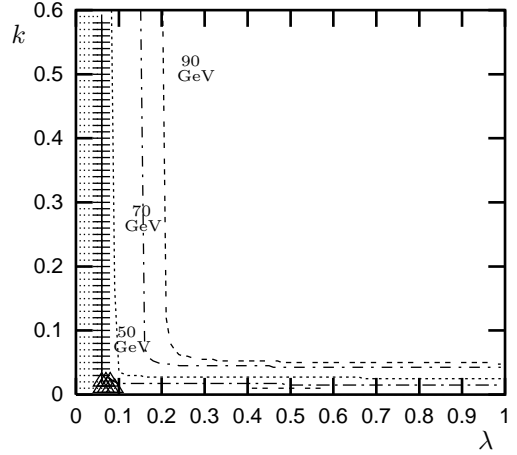}}
\put(1.0,20.0){$M=100$ GeV, $x=200$ GeV, $\tan\beta=2$}
\put(1.0,13.0){$M=200$ GeV, $x=200$ GeV, $\tan\beta=2$}
\put(1.0,6.0){$M=200$ GeV, $x=1000$ GeV, $\tan\beta=2$}
\put(9.5,20.0){$M=100$ GeV, $x=200$ GeV, $\tan\beta=10$}
\put(9.5,13.0){$M=200$ GeV, $x=200$ GeV, $\tan\beta=10$}
\put(9.5,6.0){$M=200$ GeV, $x=1000$ GeV, $\tan\beta=10$}
\put(1.0,19.1){\footnotesize $k$}
\put(7.2,13.7){\footnotesize $\lambda $}
\put(9.5,19.1){\footnotesize $k$}
\put(15.7,13.7){\footnotesize $\lambda $}
\put(1.0,12.1){\footnotesize $k$}
\put(7.2,6.7){\footnotesize $\lambda $}
\put(9.5,12.1){\footnotesize $k$}
\put(15.7,6.7){\footnotesize $\lambda $}
\put(1.0,5.1){\footnotesize $k$}
\put(7.2,-0.3){\footnotesize $\lambda $}
\put(9.5,5.1){\footnotesize $k$}
\put(15.7,-0.3){\footnotesize $\lambda $}
\put(6.2,17.0){\tiny 15 }
\put(6.15,16.8){\tiny GeV}
\put(6.7,18.7){\tiny 20 }
\put(6.65,18.5){\tiny GeV}
\put(13.2,17.7){\tiny 10 }
\put(13.15,17.5){\tiny GeV}
\put(13.5,18.4){\tiny 15 }
\put(13.45,18.2){\tiny GeV}
\put(14.2,19.0){\tiny 20 }
\put(14.15,18.8){\tiny GeV}
\put(4.5,8.5){\tiny 20 }
\put(4.45,8.3){\tiny GeV}
\put(5.0,9.4){\tiny 30 }
\put(4.95,9.2){\tiny GeV}
\put(6.1,10.3){\tiny 50 }
\put(6.05,10.1){\tiny GeV}
\put(6.8,12.0){\tiny 70 }
\put(6.75,11.8){\tiny GeV}
\put(12.4,9.1){\tiny 10 }
\put(12.35,8.9){\tiny GeV}
\put(13.0,11.0){\tiny 30 }
\put(12.95,10.8){\tiny GeV}
\put(14.5,11.7){\tiny 50 }
\put(14.45,11.5){\tiny GeV}
\put(2.6,1.1){\tiny 50 }
\put(2.55,0.9){\tiny GeV}
\put(3.1,2.8){\tiny 70 }
\put(3.05,2.6){\tiny GeV}
\put(3.8,4.8){\tiny 90 }
\put(3.75,4.6){\tiny GeV}
\put(10.9,1.2){\tiny 50 }
\put(10.85,1.0){\tiny GeV}
\put(11.1,2.8){\tiny 70 }
\put(11.00,2.6){\tiny GeV}
\put(11.75,4.8){\tiny 90 }
\put(11.70,4.6){\tiny GeV}
\end{picture}
\end{center}
\caption{The excluded parameter space from neutralino search at LEP
in the $(\lambda ,k)$ plane for various values of $M$, $x$ and
$\tan\beta$: from total $Z$ width measurements (dotted area),
invisible $Z$ width measurements (checkered area) and direct neutralino
search (dark area). Also shown are the mass contour lines of the
lightest neutralino.}
\label{lkplots}
\end{figure}
\clearpage
\begin{figure}[p]
\begin{center}
\begin{picture}(16,15)
\put(0.5,7){\includegraphics{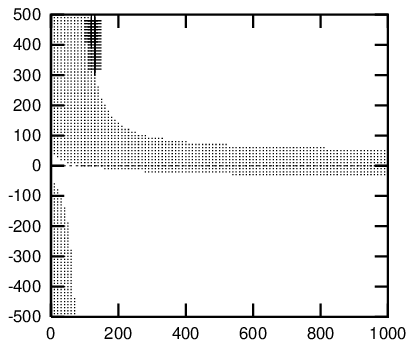}}
\put(9.0,7){\includegraphics{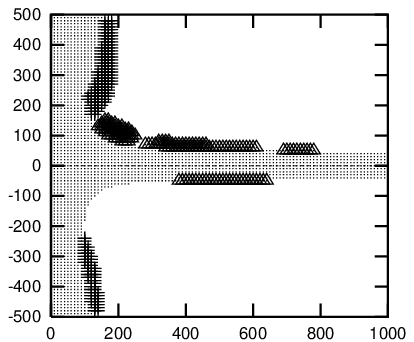}}
\put(0.5,0){\includegraphics{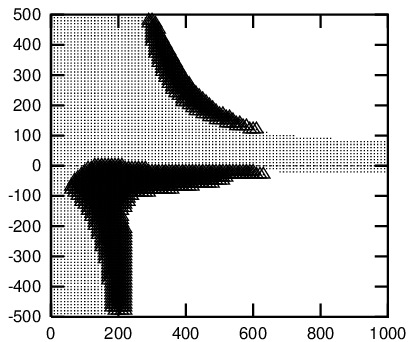}}
\put(9.0,0){\includegraphics{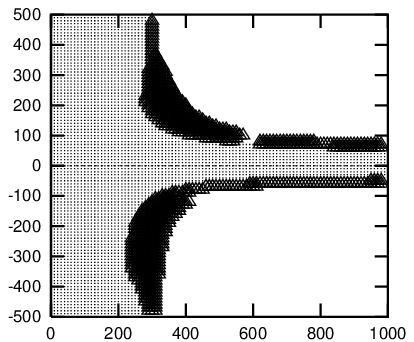}}
\put(2.0,13.0){$\lambda=0.5$, $k=0.2$, $\tan\beta=2$}
\put(2.0,6.0){$\lambda=0.2$, $k=0.05$, $\tan\beta=2$}
\put(10.5,13.0){$\lambda=0.5$, $k=0.2$, $\tan\beta=20$}
\put(10.5,6.0){$\lambda=0.2$, $k=0.05$, $\tan\beta=20$}
\put(-0.0,9.9){\footnotesize $M/$GeV}
\put(6.5,6.6){\footnotesize $x/$GeV}
\put(8.5,9.9){\footnotesize $M/$GeV}
\put(15.,6.6){\footnotesize $x/$GeV}
\put(-0.0,2.9){\footnotesize $M/$GeV}
\put(6.5,-0.4){\footnotesize $x/$GeV}
\put(8.5,2.9){\footnotesize $M/$GeV}
\put(15.,-0.4){\footnotesize $x/$GeV}
\end{picture}
\end{center}
\caption{The excluded parameter space from neutralino search at LEP
in the $(M ,x)$ plane for various values of $\lambda$, $k$ and
$\tan\beta$: from total $Z$ width measurements (dotted area),
invisible $Z$ width measurements (checkered area) and direct neutralino
search (dark area).}
\label{mxplots}
\end{figure}
\clearpage
\begin{figure}[p]
\begin{center}
\begin{picture}(16,18)
\put(0.5,14){\includegraphics{quark2a.ps}}
\put(9.0,14){\includegraphics{quark10a.ps}}
\put(0.5,7){\includegraphics{quark2b.ps}}
\put(9.0,7){\includegraphics{quark10b.ps}}
\put(0.5,0){\includegraphics{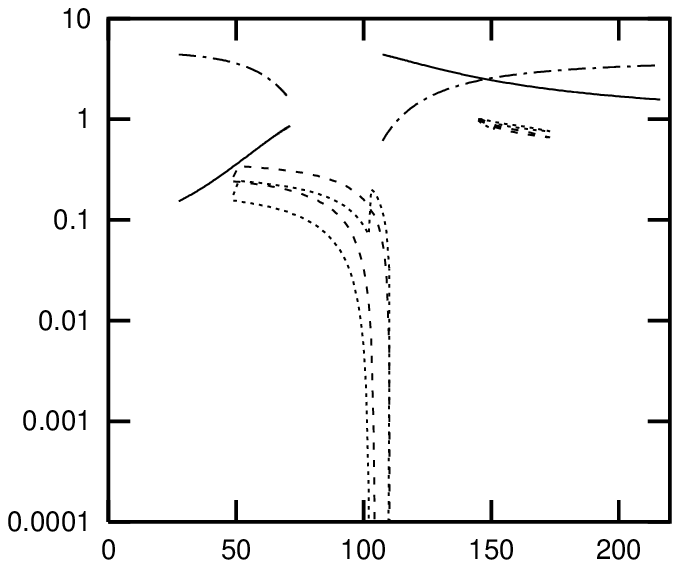}}
\put(9.0,0){\includegraphics{quark10c.ps}}
\put(2.0,20.0){$x=200$ GeV, $\lambda=0.8$, $k=0.1$, $A_t=0$ GeV,
$m_{\tilde{t}_1}=200$ GeV, $m_{\tilde{t}_2}=500$ GeV}
\put(2.0,13.0){$x=1000$ GeV, $\lambda=0.8$, $k=0.1$, $A_t=0$ GeV,
$m_{\tilde{t}_1}=200$ GeV, $m_{\tilde{t}_2}=500$ GeV}
\put(2.0,6.0){$x=1000$ GeV, $\lambda=0.8$, $k=0.01$, $A_t=0$ GeV,
$m_{\tilde{t}_1}=200$ GeV, $m_{\tilde{t}_2}=500$ GeV}
\put(5.5,14.8){\footnotesize $\tan\beta=2$}
\put(14.0,14.8){\footnotesize $\tan\beta=10$}
\put(2.2,7.8){\footnotesize $\tan\beta=2$}
\put(10.7,7.8){\footnotesize $\tan\beta=10$}
\put(5.5,0.8){\footnotesize $\tan\beta=2$}
\put(14.0,0.8){\footnotesize $\tan\beta=10$}
\put(5.7,13.7){\small $m_{\mbox{\scriptsize Higgs}}$/GeV}
\put(14.2,13.7){\small $m_{\mbox{\scriptsize Higgs}}$/GeV}
\put(5.7,6.7){\small $m_{\mbox{\scriptsize Higgs}}$/GeV}
\put(14.2,6.7){\small $m_{\mbox{\scriptsize Higgs}}$/GeV}
\put(5.7,-0.3){\small $m_{\mbox{\scriptsize Higgs}}$/GeV}
\put(14.2,-0.3){\small $m_{\mbox{\scriptsize Higgs}}$/GeV}
\put(0.1,19.0){\footnotesize $\frac{g^2(SUSY)}{g^2(SM)}$}
\put(8.3,19.1){\footnotesize $\frac{g^2(SUSY)}{g^2(SM)}$}
\put(0.1,12.0){\footnotesize $\frac{g^2(SUSY)}{g^2(SM)}$}
\put(8.3,12.1){\footnotesize $\frac{g^2(SUSY)}{g^2(SM)}$}
\put(0.1,5.0){\footnotesize $\frac{g^2(SUSY)}{g^2(SM)}$}
\put(8.3,5.1){\footnotesize $\frac{g^2(SUSY)}{g^2(SM)}$}
\end{picture}
\end{center}
\caption{The squared ratios of the SUSY Higgs-quark-antiquark-couplings
relative to the corresponding SM couplings. Shown are the MSSM couplings
of the
scalar Higgs bosons to top quarks (solid) and bottom quarks (dashed-dotted)
as well as the range of the NMSSM couplings to top quarks (dashed) and
bottom quarks (dotted).}
\label{figquarkcoupl}
\end{figure}
\clearpage
\begin{figure}[p]
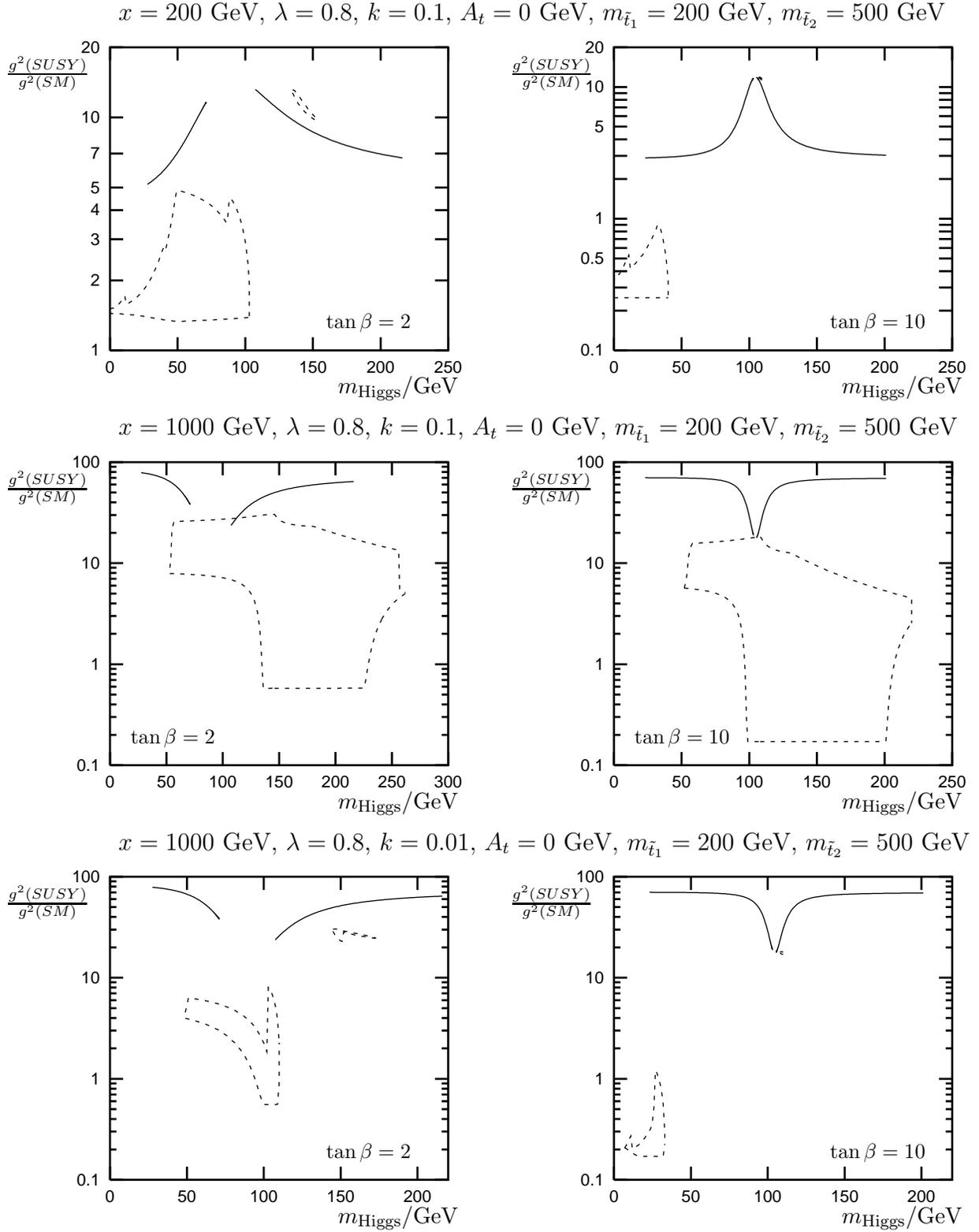

\begin{center}
\begin{picture}(16,18)
\put(0.5,14){\includegraphics{squark2a.ps}}
\put(9.0,14){\includegraphics{squark10a.ps}}
\put(0.5,7){\includegraphics{squark2b.ps}}
\put(9.0,7){\includegraphics{squark10b.ps}}
\put(0.5,0){\includegraphics{squark2c.ps}}
\put(9.0,0){\includegraphics{squark10c.ps}}
\put(2.0,20.0){$x=200$ GeV, $\lambda=0.8$, $k=0.1$, $A_t=0$ GeV,
$m_{\tilde{t}_1}=200$ GeV, $m_{\tilde{t}_2}=500$ GeV}
\put(2.0,13.0){$x=1000$ GeV, $\lambda=0.8$, $k=0.1$, $A_t=0$ GeV,
$m_{\tilde{t}_1}=200$ GeV, $m_{\tilde{t}_2}=500$ GeV}
\put(2.0,6.0){$x=1000$ GeV, $\lambda=0.8$, $k=0.01$, $A_t=0$ GeV,
$m_{\tilde{t}_1}=200$ GeV, $m_{\tilde{t}_2}=500$ GeV}
\put(5.5,14.8){\footnotesize $\tan\beta=2$}
\put(14.0,14.8){\footnotesize $\tan\beta=10$}
\put(2.2,7.8){\footnotesize $\tan\beta=2$}
\put(10.7,7.8){\footnotesize $\tan\beta=10$}
\put(5.5,0.8){\footnotesize $\tan\beta=2$}
\put(14.0,0.8){\footnotesize $\tan\beta=10$}
\put(5.7,13.7){\small $m_{\mbox{\scriptsize Higgs}}$/GeV}
\put(14.2,13.7){\small $m_{\mbox{\scriptsize Higgs}}$/GeV}
\put(5.7,6.7){\small $m_{\mbox{\scriptsize Higgs}}$/GeV}
\put(14.2,6.7){\small $m_{\mbox{\scriptsize Higgs}}$/GeV}
\put(5.7,-0.3){\small $m_{\mbox{\scriptsize Higgs}}$/GeV}
\put(14.2,-0.3){\small $m_{\mbox{\scriptsize Higgs}}$/GeV}
\put(0.1,19.0){\footnotesize $\frac{g^2(SUSY)}{g^2(SM)}$}
\put(8.6,19.0){\footnotesize $\frac{g^2(SUSY)}{g^2(SM)}$}
\put(0.1,12.0){\footnotesize $\frac{g^2(SUSY)}{g^2(SM)}$}
\put(8.6,12.0){\footnotesize $\frac{g^2(SUSY)}{g^2(SM)}$}
\put(0.1,5.0){\footnotesize $\frac{g^2(SUSY)}{g^2(SM)}$}
\put(8.6,5.0){\footnotesize $\frac{g^2(SUSY)}{g^2(SM)}$}
\end{picture}
\end{center}
\caption{The SUSY Higgs couplings to two squarks as
defined in eq.~(\ref{squarkratio}). Shown are the MSSM couplings
(solid)
as well as the range of the NMSSM couplings (dashed).}
\label{figsquarkcoupl}
\end{figure}
\clearpage
\begin{figure}[p]
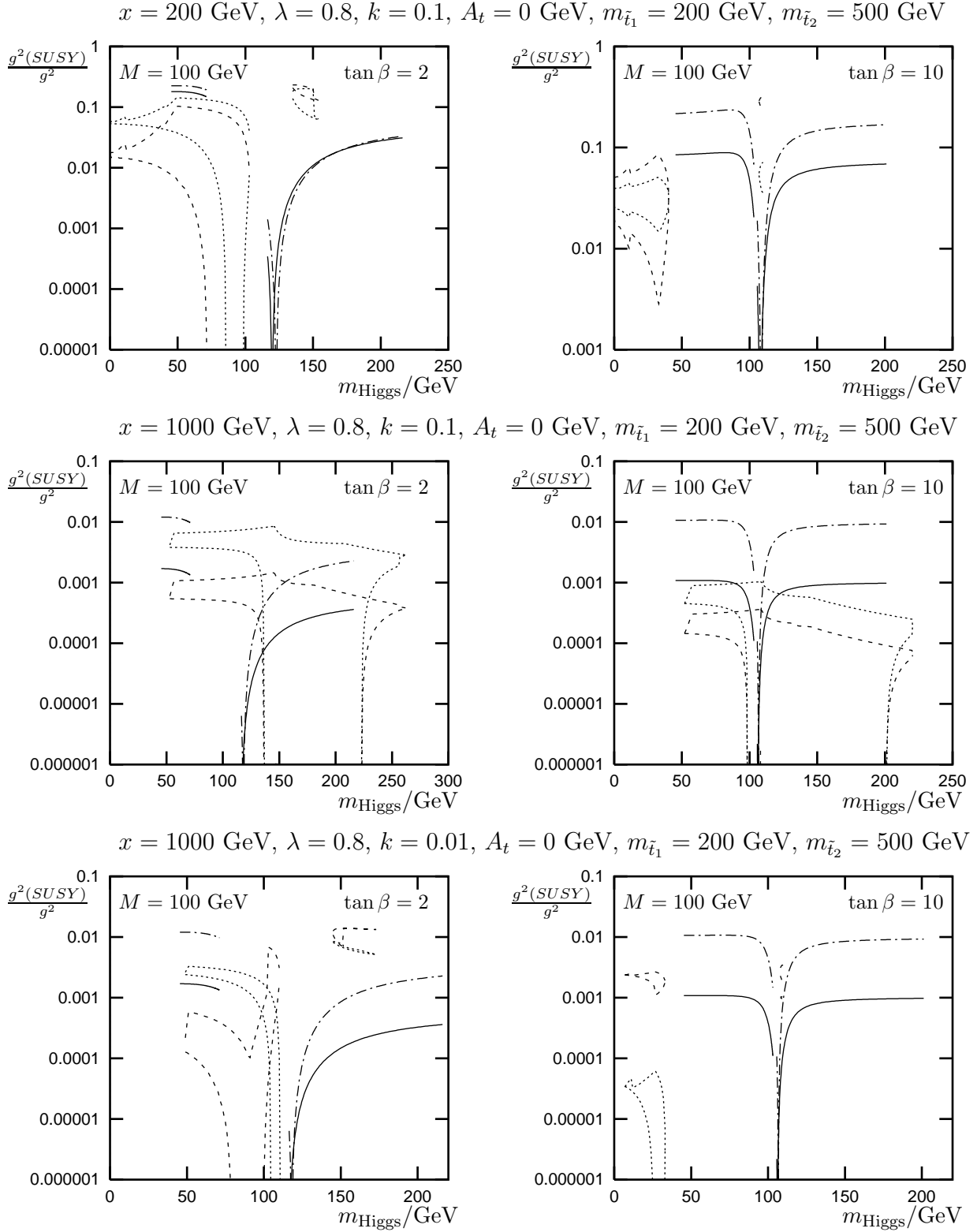

\begin{center}
\begin{picture}(16,18)
\put(0.5,14){\includegraphics{charneu2a.ps}}
\put(9.0,14){\includegraphics{charneu10a.ps}}
\put(0.5,7){\includegraphics{charneu2b.ps}}
\put(9.0,7){\includegraphics{charneu10b.ps}}
\put(0.5,0){\includegraphics{charneu2c.ps}}
\put(9.0,0){\includegraphics{charneu10c.ps}}
\put(2.0,20.0){$x=200$ GeV, $\lambda=0.8$, $k=0.1$, $A_t=0$ GeV,
$m_{\tilde{t}_1}=200$ GeV, $m_{\tilde{t}_2}=500$ GeV}
\put(2.0,13.0){$x=1000$ GeV, $\lambda=0.8$, $k=0.1$, $A_t=0$ GeV,
$m_{\tilde{t}_1}=200$ GeV, $m_{\tilde{t}_2}=500$ GeV}
\put(2.0,6.0){$x=1000$ GeV, $\lambda=0.8$, $k=0.01$, $A_t=0$ GeV,
$m_{\tilde{t}_1}=200$ GeV, $m_{\tilde{t}_2}=500$ GeV}
\put(2.0,19.0){\footnotesize $M=100$ GeV}
\put(10.5,19.0){\footnotesize $M=100$ GeV}
\put(2.0,12.0){\footnotesize $M=100$ GeV}
\put(10.5,12.0){\footnotesize $M=100$ GeV}
\put(2.0,5.0){\footnotesize $M=100$ GeV}
\put(10.5,5.0){\footnotesize $M=100$ GeV}
\put(5.8,19.0){\footnotesize $\tan\beta=2$}
\put(14.3,19.0){\footnotesize $\tan\beta=10$}
\put(5.8,12.0){\footnotesize $\tan\beta=2$}
\put(14.3,12.0){\footnotesize $\tan\beta=10$}
\put(5.8,5.0){\footnotesize $\tan\beta=2$}
\put(14.3,5.0){\footnotesize $\tan\beta=10$}
\put(5.7,13.7){\small $m_{\mbox{\scriptsize Higgs}}$/GeV}
\put(14.2,13.7){\small $m_{\mbox{\scriptsize Higgs}}$/GeV}
\put(5.7,6.7){\small $m_{\mbox{\scriptsize Higgs}}$/GeV}
\put(14.2,6.7){\small $m_{\mbox{\scriptsize Higgs}}$/GeV}
\put(5.7,-0.3){\small $m_{\mbox{\scriptsize Higgs}}$/GeV}
\put(14.2,-0.3){\small $m_{\mbox{\scriptsize Higgs}}$/GeV}
\put(0.1,19.1){\footnotesize $\frac{g^2(SUSY)}{g^2}$}
\put(8.6,19.1){\footnotesize $\frac{g^2(SUSY)}{g^2}$}
\put(0.1,12.0){\footnotesize $\frac{g^2(SUSY)}{g^2}$}
\put(8.6,12.0){\footnotesize $\frac{g^2(SUSY)}{g^2}$}
\put(0.1,5.0){\footnotesize $\frac{g^2(SUSY)}{g^2}$}
\put(8.6,5.0){\footnotesize $\frac{g^2(SUSY)}{g^2}$}
\end{picture}
\end{center}
\caption{The squared SUSY Higgs couplings to neutralinos and
charginos relative to $g^2$. Shown are the MSSM couplings of the
scalar Higgs bosons to lightest neutralinos (solid) and light charginos
(dashed-dotted) as well as the range of the NMSSM couplings to lightest
neutralinos (dashed) and light charginos (dotted).}
\label{figcharneucoupl}
\end{figure}
\clearpage
\begin{figure}[p]
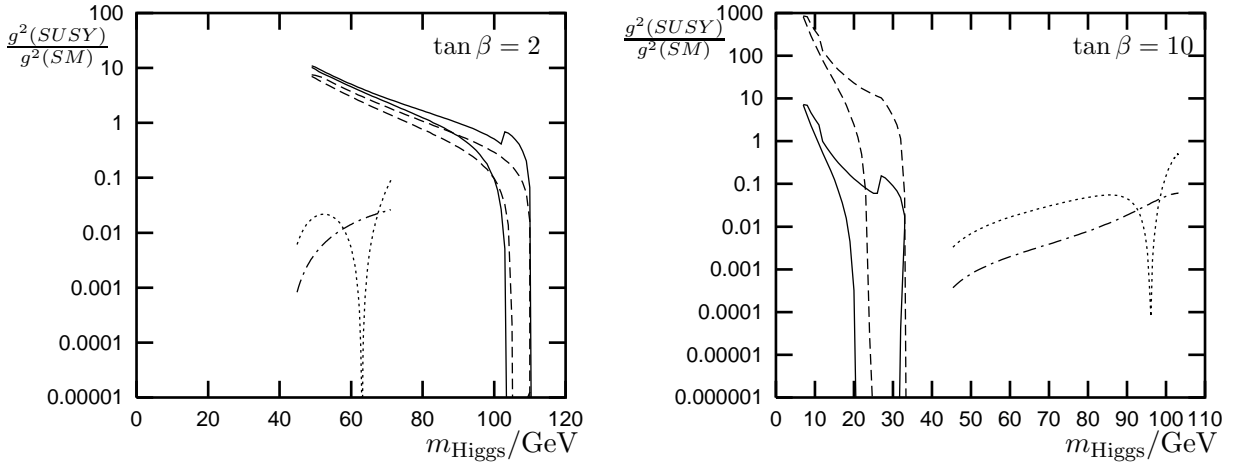

\begin{center}
\begin{picture}(16,18)
\put(0.5,14){\includegraphics{self200_2.ps}}
\put(9.0,14){\includegraphics{self200_10.ps}}
\put(0.5,7){\includegraphics{self1000a_2.ps}}
\put(9.0,7){\includegraphics{self1000a_10.ps}}
\put(0.5,0){\includegraphics{self1000b_2.ps}}
\put(9.0,0){\includegraphics{self1000b_10.ps}}
\put(2.0,20.0){$x=200$ GeV, $\lambda=0.8$, $k=0.1$, $A_t=0$ GeV,
$m_{\tilde{t}_1}=200$ GeV, $m_{\tilde{t}_2}=500$ GeV}
\put(2.0,13.0){$x=1000$ GeV, $\lambda=0.8$, $k=0.1$, $A_t=0$ GeV,
$m_{\tilde{t}_1}=200$ GeV, $m_{\tilde{t}_2}=500$ GeV}
\put(2.0,6.0){$x=1000$ GeV, $\lambda=0.8$, $k=0.01$, $A_t=0$ GeV,
$m_{\tilde{t}_1}=200$ GeV, $m_{\tilde{t}_2}=500$ GeV}
\put(5.8,19.0){\footnotesize $\tan\beta=2$}
\put(14.3,19.0){\footnotesize $\tan\beta=10$}
\put(5.8,12.0){\footnotesize $\tan\beta=2$}
\put(14.3,12.0){\footnotesize $\tan\beta=10$}
\put(5.8,5.0){\footnotesize $\tan\beta=2$}
\put(14.3,5.0){\footnotesize $\tan\beta=10$}
\put(5.7,13.7){\small $m_{\mbox{\scriptsize Higgs}}$/GeV}
\put(14.2,13.7){\small $m_{\mbox{\scriptsize Higgs}}$/GeV}
\put(5.7,6.7){\small $m_{\mbox{\scriptsize Higgs}}$/GeV}
\put(14.2,6.7){\small $m_{\mbox{\scriptsize Higgs}}$/GeV}
\put(5.7,-0.3){\small $m_{\mbox{\scriptsize Higgs}}$/GeV}
\put(14.2,-0.3){\small $m_{\mbox{\scriptsize Higgs}}$/GeV}
\put(-0.4,19.1){\footnotesize $\frac{g^2(SUSY)}{g^2(SM)}$}
\put(8.1,19.1){\footnotesize $\frac{g^2(SUSY)}{g^2(SM)}$}
\put(0.1,12.0){\footnotesize $\frac{g^2(SUSY)}{g^2(SM)}$}
\put(8.6,12.0){\footnotesize $\frac{g^2(SUSY)}{g^2(SM)}$}
\put(0.1,5.0){\footnotesize $\frac{g^2(SUSY)}{g^2(SM)}$}
\put(8.3,5.1){\footnotesize $\frac{g^2(SUSY)}{g^2(SM)}$}
\end{picture}
\end{center}
\caption{The squared ratios of the trilinear SUSY Higgs self-couplings
relative to those of the SM. Shown are the range of the NMSSM couplings
$g_{S_1S_1S_1}^2(NMSSM)/g_{\Phi \Phi \Phi}^2(SM)$ (solid),
$g_{\scriptsize S_1P_1P_1}^2(NMSSM)/g_{\scriptsize \Phi \Phi \Phi}^2(SM)$
(dotted) and the MSSM couplings $g_{\scriptsize hhh}^2(MSSM)/
g_{\scriptsize \Phi \Phi \Phi}^2(SM)$ (double dashed),
$g_{\scriptsize hAA}^2(MSSM)/g_{\scriptsize \Phi \Phi \Phi}^2(SM)$
(dashed dotted).}
\label{selfplots}
\end{figure}

\begin{thebibliography}{99}
\bibitem{cdf} CDF Collaboration, F. Abe et al., Phys. Rev. Lett.
{\bf 74} (1995) 2626; \\
D0 Collaboration, S. Abachi et al., Phys. Rev. Lett.
{\bf 74} (1995) 2632
\bibitem{sm} S.L. Glashow, Nucl. Phys. {\bf B 22} (1961) 579; \\
S. Weinberg, Phys. Rev. Lett. {\bf 19} (1967) 1264; \\
A. Salam, Proc. 8th Nobel Symposium, Stockholm 1968, ed. N. Svartholm,
Almquist and Wiksells, p. 367
\bibitem{kibble} P.W. Higgs, Phys. Lett. {\bf 12} (1964) 132;
Phys. Rev. Lett. {\bf 13} (1964) 508;
Phys. Rev. {\bf 145} (1966) 1156; \\
R. Brout and F. Englert, Phys. Rev. Lett. {\bf 13} (1964) 321; \\
T.W.B. Kibble, Phys. Rev. {\bf 155} (1967) 1554
\bibitem{unitarity} B.W. Lee, C. Quigg and G.B. Thacker, Phys. Rev. Lett.
{\bf 38} (1977) 883; Phys. Rev. {\bf D 16}(1977) 1519
\bibitem{alephsm} P. Bagnaia, presented at the third meeting of the
LEP2 workshop, Geneva, November 1995
\bibitem{amaldi} U. Amaldi, W. de Boer and H. F\"urstenau,
Phys. Lett. {\bf B 260} (1991) 447
\bibitem{hier} E. Gildener, Phys. Rev. {\bf D 14} (1976) 1667; \\
S. Weinberg, Phys. Lett. {\bf B 82} (1979) 387; \\
L. Susskind, Phys. Rep. {\bf 104} (1984) 181
\bibitem{fine} S. Weinberg, Phys. Rev. {\bf D 13} (1976) 974; \\
L. Susskind, Phys. Rev. {\bf D 20} (1979) 2619
\bibitem{haka} H.E. Haber and G.L. Kane, Phys. Rep. {\bf 117} (1985) 75
\bibitem{drees} M. Drees, Int. J. of Mod. Phys. {\bf A4} (1989) 3635
\bibitem{ellis} J. Ellis, J.F. Gunion, H.E. Haber, L. Roszkowski
and F. Zwirner, Phys. Rev. {\bf D 39} (1989) 844
\bibitem{barr} S.M. Barr, Phys. Lett. {\bf B 112} (1982) 219
\bibitem{nilsredwy} H.P. Nilles, M. Srednicki and D. Wyler,
Phys. Lett. {\bf B 120} (1983) 346
\bibitem{derendinger} J.-P. Derendinger and C.A. Savoy,
Nucl. Phys. {\bf B 237} (1984) 307
\bibitem{kim} J.E. Kim and H.P. Nilles, Phys. Lett. {\bf B 138} (1984) 150
\bibitem{alephneu} ALEPH Collaboration, D. Decamp et al.,
Phys. Rep. {\bf 216} (1992) 253; \\
L3 Collaboration, O. Adriani et al., Phys. Rep.
{\bf 236} (1993) 1
\bibitem{alephhiggs} ALEPH Collaboration, D. Buskulic et al.,
Phys. Lett. {\bf B 313} (1993) 312; \\
OPAL Collaboration, R. Akers et al., Z. Phys. {\bf C 64} (1994) 1
\bibitem{frankeneu} F. Franke, H. Fraas and A. Bartl, Phys. Lett.
{\bf B 336} (1994) 415
\bibitem{frankehiggs} F. Franke and H. Fraas, Phys. Lett.
{\bf B 353} (1995) 234
\bibitem{rad} J. Ellis, G. Ridolfi and F. Zwirner, Phys. Lett.
{\bf B 257} (1991) 83; \\
J.L. Lopez and D. Nanopoulos, Phys. Lett. {\bf B 266} (1991) 397
\bibitem{ellwanger} U. Ellwanger, Phys. Lett. {\bf B 303} (1993) 271
\bibitem{elliott} T. Elliott, S.F. King and P.L. White,
Phys. Lett. {\bf B 314} (1993) 56; Phys. Rev. {\bf D 49} (1994) 2435
\bibitem{pandita} P.N. Pandita, Z. Phys. {\bf C 59} (1993) 575
\bibitem{ellrs} U. Ellwanger, M. Rausch de Traubenberg and C.A. Savoy,
Phys. Lett. {\bf B 315} (1993) 331; LPTHE Orsay 95-04, LPT Strasbourg 95-01,
SPhT Saclay T95/04, hep-ph/9502206
\bibitem{kimproc} B.R. Kim, S.K. Oh and A. Stephan,
Proceedings of the Workshop $e^+e^-$ Collisions at 500 GeV.
The Physics Potential, Munich,
Annecy, Hamburg, Ed. P. Zerwas, DESY 92-123B
(1992) 697; DESY 93-123C (1993) 491
\bibitem{kimneu} B.R. Kim, S.K. Oh and A. Stephan, Phys. Lett.
{\bf B 336} (1994) 200
\bibitem{higgs} J.F. Gunion and H.E. Haber, Nucl. Phys. {\bf B 272} (1986) 1;
Nucl. Phys. {\bf B 278} (1986) 449; Nucl. Phys. {\bf B 307} (1988) 445
\bibitem{hunter} J.F. Gunion, H.E. Haber, G. Kane and S. Dawson,
{\em The Higgs Hunter's Guide}, Addison-Wesley, Redwood City 1990
\bibitem{whitehiggs} S.F. King and P.L. White, SHEP 95-27,
OUTP-95-31-P, hep-ph/9508346
\bibitem{franke4} F. Franke and H. Fraas, WUE-ITP-95-021, hep-ph/9511275
\bibitem{ilyin} V.A. Ilyin, A.E.Pukhov, Y. Kurihara, Y. Shimizu and
T. Kaneko, INP MSU Preprint-95-16/380, KEK CP-030, hep-ph/9506326
\bibitem{boudjema} F. Boudjema and E. Chopin, ENSLAPP-A-534/95,
hep-ph/9507396
\bibitem{gira} L. Girardello and M.T. Grisaru, Nucl. Phys. {\bf B 194}
(1982) 65
\bibitem{renor} J.-P- Derendinger and C.A. Savoy,
Nucl. Phys. {\bf B 237} (1984) 307; \\
K. Inoue, A. Kakuto, H. Komatsu and S. Takeshito,
Prog. Theor. Phys. {\bf 67} (1982) 1859;
Prog. Theor. Phys. {\bf 68} (1982) 927
\bibitem{ekwuni} T. Elliott, S.F. King and P.L. White, Phys. Lett.
{\bf B 351} (1995) 213
\bibitem{kwconstr} S.F. King and P.L. White, Phys. Rev. {\bf D 52} (1995)
4183
\bibitem{bartlneu} A. Bartl, H. Fraas, W. Majerotto and
N. Oshimo, Phys. Rev. {\bf D 40} (1989) 1594
\bibitem{bartlchar} A. Bartl, H. Fraas, W. Majerotto and
B. M\"osslacher, Z. Phys. {\bf C 55} (1992) 257
\bibitem{neubounds} L3 Collaboration, M. Acciarri et al.,
Phys. Lett. {\bf B 350} (1995) 109
\end{thebibliography}
\end{document}